\documentclass[rmp]{revtex4}\usepackage{graphicx}
\usepackage{dcolumn}
\usepackage{bm}

\newcommand{\nablab}{\mbox{\boldmath $\nabla $}}

\newcommand{\sigmab}{\mbox{\boldmath $\sigma $}}

\newcommand{\br}{\mbox{\boldmath $r $}}
\newcommand{\bA}{\mbox{\boldmath $A $}}
\newcommand{\bS}{\mbox{\boldmath $S $}}
\newcommand{\bz}{\mbox{\boldmath $z $}}

\newcommand{\bd}{\mbox{\boldmath $d $}}
\newcommand{\bp}{\mbox{\boldmath $p $}}
\newcommand{\bi}{\mbox{\boldmath $i $}}
\newcommand{\bj}{\mbox{\boldmath $j $}}
\newcommand{\bs}{\mbox{\boldmath $s$}}

\newcommand{\bq}{\mbox{\boldmath $q $}}
\newcommand{\bk}{\mbox{\boldmath $k$}}

\newcommand{\bP}{\mbox{\boldmath $P $}}
\newcommand{\bR}{\mbox{\boldmath $R $}}
\newcommand{\bK}{\mbox{\boldmath $K $}}
\newcommand{\bv}{\mbox{\boldmath $v $}}
\newcommand{\bh}{\mbox{\boldmath $h $}}
\newcommand{\be}{\mbox{\boldmath $e $}}
\newcommand{\bE}{\mbox{\boldmath $E $}}

\newcommand{\bB}{\mbox{\boldmath $B $}}

\def\bG{{\mathbf G}}
\def\bl{{\mathbf l}}
\def\G{{\Gamma}}

\def\half{{1\over 2}}

\def\a{{\alpha}}
\def\g{{\gamma}}
\def\d{{\delta}}
\def\D{{\Delta}}
\def\b{{\beta}}
\def\l{{\lambda}}
\def\L{{\Lambda}}
\def\t{{\theta}}
\def\T{{\Theta}}
\def\s{{\sigma}}

\makeindex
\newcommand{\beq}{\begin{equation}}
\newcommand{\eeq}{\end{equation}}
\newcommand{\beqr}{\begin{eqnarray}}
\newcommand{\eeqr}{\end{eqnarray}}
\newcommand{\lt}{\left[}
\newcommand{\rt}{\right]}

\newcommand{\e}{\varepsilon}

\newcommand{\p}{\partial}

\newtheorem{ex}{Exercise}[section]
\begin{document}


\title{Topological Insulators- A review}

\author{R.Shankar}
\email{r.shankar@yale.edu}
 \homepage{http://www.yale.edu/~r.shankar}
\affiliation{ Sloane Physics Lab\\ Yale University  New Haven CT
06520 }

\date{\today}
             \begin{abstract}

Central message: 
{\em Do not stand on a block of topological insulator to change  a light bulb.}

\vspace{1in}

These lecture notes were prepared for a mixed audience of students, postdocs and faculty from the Indian Institute of Technology Madras, India and neighboring institutions, particularly the Institute of Mathematical Sciences. I am not an expert on the subject and during the few years I spent working on  the Quantum Hall effect, I had not fully appreciated that it was part of a family of topological insulators. It was a pleasure to dig a little deeper into this subject and to share its wonders with others. In preparing these lectures I relied heavily on the help of Ganpathy Murthy (UKy) and a very helpful conversation with Steve Kivelson (Stanford.) I am of course responsible any errors despite their efforts. I also relied on some excellent  Powerpoint slides of various talks. I have furnished a few choice references at the end and very few references to original papers. I cover only $d=1$ and $d=2$. 
\end{abstract}
\maketitle
\tableofcontents

\section{Scalar and vector potentials}
\label{chap:potentials}

\subsection{Introducing  $\bA$ and $\phi$ in classical electrodynamics}

The following two Maxwell equations 
\beqr
\nablab \cdot \bB&=&0\\
\nablab \times \bE + { \partial \bB\over \p t}&=&0
\eeqr
can be satisfied as identities by setting 
\beqr
\bE &=& - \nablab \phi - { \p \bA \over \p t}\\
\bB&=& \nablab \times \bA
\eeqr
where $\bA$ and $\phi$ are the {\em vector and scalar potential} \index{vector and scalar potential $\bA$, \$\phi$} respectively. 

By writing the other two Maxwell equations involving the current density $\bj$,  and  the charge density $\rho$,

\beqr
\nablab \cdot \bE&=&{\rho \over \e_0}\\
\nablab \times \bB &=& \mu_0 \bj+\mu_0 \e_0 {\p \bE \over \p t} 
\eeqr
 in terms of $(\bA, \phi)$ we can obtain equations that determine $(\bA, \phi)$ in terms of $(\bj, \rho )$.  

In that process we  invoke  {\em gauge freedom}, \index{ gauge freedom}  which refers to the fact that $(\bA, \phi)$ can be traded for $\bA'$ and $\phi'$ related by a {\em gauge transformation}:
\beqr
\bA' &=& \bA + \nablab \chi\\
\phi'&=& \phi -{\p \chi \over \p t}
\eeqr
without changing $(\bE, \bB)$. 
\begin{ex} Verify this claim.
\end{ex}

Even thought $\bA$ is gauge-dependent, its line integral around a closed loop $C$ which is the boundary of a surface $S$ (i.e., $C = \p S$)  is gauge invariant:
\beq
\oint_{C= \p S}\bA \cdot d\br = \int_S (\nablab \times \bA) \cdot d \bS= \mbox{gauge invariant}. 
\eeq

In classical physics we can work  with either $(\bE, \bB)$ or $(\bA, \phi)$. Each version has its advantages. For example in electrostatics it is easier to solve for the scalar function $\phi$ rather than the vector field $\bE$ in terms of $\rho$. On the other hand one has the lingering feeling that the potentials are unphysical because they can be altered (by a gauge transformation) without changing the fields which are physical and  can be measured by probes.

\subsection{$\bA$ and $\phi$ in  quantum theory}
By "in quantum theory" I mean the particles are treated quantum mechanically, in contrast to quantum electrodynamics where even the electromagnetic  fields are treated quantum mechanically. 

Quantum theory gives us no choice: we have to work with the potentials from the outset. 

In Feynman's sum over paths each path is weighted by 
\beq e^{(i/\hbar)\int {\cal L}dt}
\eeq
 where the Lagrangian for a particle of mass $m$ and charge $e$ in the presence of  an electromagnetic  field is expressed in terms of  $(\bA, \phi)$:
\beq {\cal L}(\br,\bv) =\half mv^2 +e \bv \cdot \bA- e \phi.\label{vdota}
\eeq

The same is true of the  Hamiltonian approach where we have 
\beq
 H = {(\bp -e\bA)^2 \over 2m}+ e \phi\label{hamA}.
\eeq
Despite the appearance of potentials one can show that physical answers (say energy levels and probabilities) are gauge invariant.

Here is an example of gauge freedom and gauge invariance  in the Hamiltonian formalism. Consider the eigenvalue equation
\beq
H \psi = {(\bp -e\bA)^2 \over 2m}\psi = E \psi.
\eeq
It is readily verified that 
\beqr
{(\bp -e\bA')^2 \over 2m}\psi' &=& E \psi'\ \ \ \ \mbox{where}\\
\psi' (\br)&=& \exp \left[ {ie\over \hbar} \chi (\br)\right]\psi (\br)\\
\bA'&=& \bA + \nablab \chi.
\eeqr
In other words, a change of gauge and a corresponding change of phase of $\psi$ leave physical quantities like $E$ invariant. 

In short, you have to chose some gauge to do the calculation but the physical quantities  (e.g., $E$) will not depend on the choice. The wave function itself will change but probabilities or densities will be unaffected.

\subsection{Aharanov-Bohm  experiment}

\begin{figure}
    \centering
\includegraphics[width=3in]{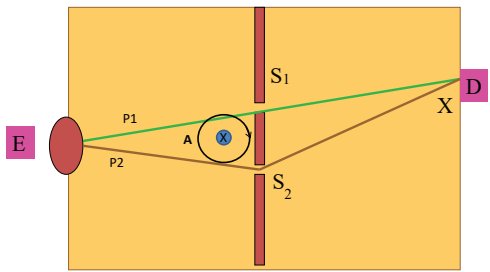}
    \vspace{1cm}
    \caption{The double slit experiment with electrons.  Between the source of electrons $E$ and the screen is a solenoid  carrying flux $\Phi$ going into the page.
The interference pattern responds to the flux inside the solenoid.  There is no $\bB$ outside the solenoid, but there is a concentric $\bA$ everywhere, shown by a circle. }
    \label{dse}
  \end{figure}

So far it looks like we are to  work with $\bA$ but obtain  results that depend only on $\bB$. But there is a celebrated example in which there is no $\bB$ acting on the particle  and its behavior is modified by just $\bA$.  Consider the double-slit experiment in Figure \ref{dse}. Between the source of electrons and the screen is an impenetrable  solenoid  carrying flux  $\Phi$ into the page.
The interference pattern responds to the flux inside the solenoid.  There is no $\bB$ outside the solenoid, but there is an $\bA$ everywhere.  A classical particle excluded from the solenoid will not change its motion because $\bB= \nablab \times \bA=0$ wherever it goes.

How does the electron respond to $\bA$  and how does  the response manage  to be gauge-invariant? One way to see this is in the path integral approach  where  the electron goes along paths on either side of the solenoid. 

Suppose without the $\bA$  we had at a point $\br$ on the screen 
\beq
\psi(\br) = \psi_{P1}(\br)+ \psi_{P2}(\br)
\eeq
 where the two contributions come from classical  paths $P1$ and $P2$ and their neighbors with nearly the same action (i.e., within $\hbar$). Because of the $e\bv\cdot \bA$ term in Eqn. \ref{vdota}, the contributions from the two classical paths $P1$ and $P2$  now get modified to yield 
\beqr
\psi(\br)  &=& \psi_{P1}(\br) \exp\left[ {ie \over \hbar}\int_{P1} \bA \cdot d \br\right]+ \psi_{P2}(\br) \exp\left[ {ie \over \hbar}\int_{P2} \bA \cdot d \br\right]\nonumber \\
| \psi(\br ) |^2 &=& \left| \exp \left[ {ie\over \hbar} \oint_{C=P1-P2}\bA \cdot d\br\right]\psi_{P1}+\psi_{P2} \right|^2.
\eeqr
Thus the usual phase difference $\Delta \phi_{12}$ between the two paths is compounded by the 
line integral of the flux penetrating the closed loop made of path $P1$ and the reverse of path $P2$.

By going on both sides of the solenoid and comparing the phase difference, the electron is able to respond to the flux in the solenoid and respond   gauge invariantly. 

For future use note that if $\Phi$, the enclosed flux in the solenoid, obeys 
\beqr
  \exp  {\lt{i e  \over \hbar}\oint \bA\cdot d {\bl}\rt} &=&\exp \lt{{i e \Phi \over \hbar}}\rt \\
  &=& \exp \lt {2 \pi i m}\rt \ \ m=0, \pm 1, .. \ \ \ \ \mbox{or}\\
\Phi &=& m \Phi_0 \equiv m {2 \pi \hbar \over e}\ \ \ \mbox{where}\\
\Phi_0 &\equiv& \mbox{the flux quantum},
\eeqr
 the solenoid is unobservable. 

\subsection{Monopole problem}
Imagine a particle moving in the field of a monopole of charge $g$,
\beq
\bB= {g \be_r\over r^2}.
\eeq
which  emits a total magnetic flux $4 \pi g$.
{\em We cannot describe this $\bB$ by a non-singular $\bA$.} To see this consider a closed surface, which I take to be a sphere for convenience,  enclosing the monopole and a closed loop $C$ surrounding the north pole in the sense of increasing azimuthal angle $\phi$.  By Stokes' theorem 
\beq
\oint_{C} \bA \cdot d \br = \Phi_{enc}
\eeq
where $\Phi_{enc}$ is the flux enclosed in the  region bounded by $C$. Now slowly increase the size of $C$ till it passes the equator and then starts to shrink and ends up as  a point at the south pole. The flux enclosed now becomes 
$4\pi g$. On the other hand, it is not possible for an infinitesimal loop to enclose a finite amount of flux unless $\bA$ is singular. 

Here is a concrete example
\beq
 \bA = \be_{\phi}{g (1 - \cos \theta) \over r \sin \theta}
\eeq
where $\be_{\phi}$ is a unit vector along the azimuthal direction. If we integrate this $\bA$ around a circle at fixed latitude $\theta$ we find
\beq
\oint \bA \cdot d \br =\int_{0}^{2 \pi} A_{\phi} r  sin \t  d\phi = 2 \pi g (1 -\cos \t)
\eeq
 which is indeed the enclosed flux, which grows from $0$ at $\t =0$ to $4 \pi g$ at $\t = \pi$.
Notice that this $\bA$ is singular at $\t = \pi$. We shall refer to it as $\bA_+$ because it good in the upper part of the sphere that excludes the south pole. Likewise 
\beq
\bA_- = -\be_{\phi}{g (1 + \cos \theta) \over r \sin \theta}
\eeq
is good everywhere except the north pole and yields the same $\bB$ because 
\beqr \bA_+-\bA_-&=& {2g \over r \sin \t}\be_{\phi}\\
&=&\nablab \chi, \ \ \ \mbox{where}\  \chi (r, \t, \phi)= 2 g \phi.
\eeqr

\begin{figure}
    \centering
\includegraphics[width=5in]{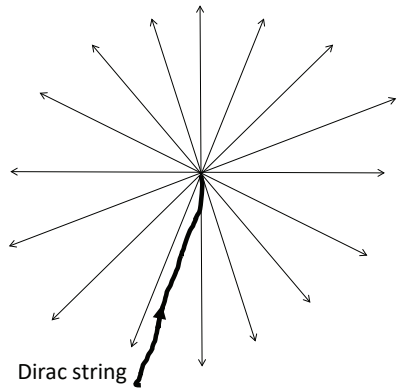}
    \vspace{1cm}
    \caption{Dirac string bringing in the flux to the monopole from infinity along an arbitrary path.  Dirac's quantization condition, Eqn. \ref{diracquant} ensures it is unobservable.}
    \label{dirac}
  \end{figure}

Dirac argued that  an infinitesimally thin  {\em Dirac string} was bringing in the flux $4\pi g$ at  the singularities, (at the north or south pole in our example) and releasing them at the origin in a spherically symmetric manner, as indicated by Figure \ref{dirac}. The location of the string can be changed by a change of gauge.
(For example, by going from $\bA_+$ to $\bA_-$ we can move it from the south pole to the north.) Thus the string should not be observable, not just  where it enters the sphere, but all the way to infinity, where it begins.  A particle going around it would acquire a Bohm-Aharanov phase factor 
\beq
\exp \lt{{i e \Phi \over \hbar}}\rt= \exp \left[ {  4 \pi i e g \over \hbar}\right].
\eeq

Dirac demanded that this be unobservable, i.e.
\beq
{4 \pi eg\over \hbar} = 2 \pi m, \ \ \ m=0, \pm 1, \pm 2..
\eeq
or 
\beq eg = \half m \hbar .\label{diracquant}
\eeq

This is a remarkable result: it implies that the presence of even a single monopole in the universe forces  the electric charges of particles to  be integral multiples  of ${\hbar \over 2g}$. This gives a clue to why the electron and proton, which are so different, have charges of the same magnitude. I for one believe monopoles exist,  if only for this reason.

\subsection{Modern approach to the monopole problem} 

\begin{figure}[h]
    \centering
\includegraphics[width=3in]{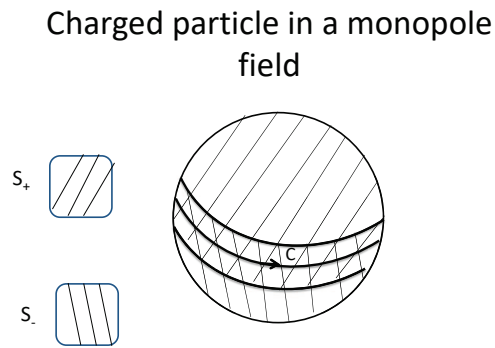}
    \vspace{1cm}
    \caption{One divides the sphere into two overlapping patches $S_+$ and $S_-$, one excluding the south pole and the other the north pole,  with respective vector potentials  $\bA_{\pm}$ and  wavefunctions $\psi_{\pm}$ which are called {\em sections}\index{sections}. The overlap region is doubly shaded.}
    \label{sec}
    \end{figure}

In the modern approach one abandons the notion of a single wave function for the particle or a single  $\bA$ for the particle in the field of a monopole. Let us say the particle is moving on a sphere with the monopole at the center. One divides  the sphere into two overlapping patches $S_+$ and $S_-$, one excluding the south pole and the other the north pole, as shown in Figure \ref{sec}. In the two  patches  there will be vector potentials  $\bA_{\pm}$ and  wavefunctions $\psi_{\pm}$ which are called {\em sections}\index{sections}.

In the overlap region (say in a belt around the equator) the two potentials and corresponding wavefunctions will be related by a gauge transformation. Look at the contour $C$ in Figure \ref{sec}. When traveled as shown we have 
\beq
\oint_C \bA_+\cdot  d \br = \Phi_+,  \ \mbox{the flux intercepted by $S_+$}
\eeq
using Stokes' theorem in a patch where $\bA$ is regular. If we now traverse $C$ the other way, i.e., along $-C$, and integrate $\bA_-$ we obtain
\beq
\oint_{-C }\bA_-\cdot d \br = \Phi_-, \ \mbox{the flux intercepted by $S_-$.}
\eeq
We add the two equations and manipulate as follows
\beq
\oint_C (\bA_+ -\bA_-)\cdot d \br = \ \mbox{total monopole flux = $4\pi g$.}
\eeq

Since $\bA_{\pm}$ describe the same $\bB$ they must be related as follows
\beq
\bA_+- \bA_-= \nablab \chi, 
\eeq
 leading to 
\beq \oint \nablab \chi \cdot  d \br = 4 \pi g.\label{predirac}
\eeq 
(Recall  our concrete example of $\bA_{\pm}$ 
\beqr
\bA_{\pm}&=&  \pm {g\over r \sin \t}( 1 \mp \cos \t)\be_{\phi}\\
\bA_+-\bA_-&=&{2 g \over r \sin \t}\be_{\phi}=2g \nablab \chi\ \ \ \mbox{where $\chi(\t, \phi)= \phi$.)}
\eeqr 
Under this gauge transformation by $\chi$  the wave functions are related by 
\beq
\psi_+ = \exp\left[ {ie\over \hbar}\chi \right] \psi_-.
\eeq
The requirement that $\psi_+$ be single valued assuming $\psi_-$ is, means 
\beqr
{e\over \hbar} \oint \nabla \chi \cdot d \br&=& 2 \pi m, \ \ \mbox{or, using Eqn. \ref{predirac}}\\
eg&=& m {\hbar \over 2}.
\eeqr

\section{The Berry phase}  
Consider a system whose Hamiltonian $H$ is a function of time. Let $|n(t)\rangle$ be an eigenket of $H(t)$:
\beq H(t) |n(t)\rangle= E(t)|n(t)\rangle.
\eeq
What is the solution to 
\beq
 i \hbar {d|\psi(t)\rangle \over dt}=H(t)|\psi(t)\rangle
\eeq
 assuming that the state never jumps to any other eigenstate (labeled by $n'\ne n$) as $t$ is varied, i.e., the evolution is adiabatic?   This is possible if there is a gap $\hbar \omega$ in the spectrum and $H$ changes sufficiently slowly, on a time scale $1/\omega$.  

 A reasonable guess is that if we started with $|\psi (0)\rangle = |n(0)\rangle$ at $t=0$, then at time $t$
\beq
|\psi(t)\rangle = \exp \lt - {i \over \hbar}\int_{0}^{t}E(t')dt'\rt |n(t)\rangle, 
\eeq
where $\int_{0}^{t}E(t')dt'$ is the accumulated phase shift. You can verify that this does not work because $|n(t)\rangle$ has its own time derivative. So we substitute  
\beq
|\psi(t)\rangle =e^{i\g (t)} \exp \lt - {i \over \hbar}\int_{0}^{t}E(t')dt'\rt |n(t)\rangle.
\eeq
into the Schr\"{o}dinger equation, dot both sides with $\langle n(t)|$ to find 
\beq
\dot{\g}= i \langle n|{dn\over dt}\rangle
\eeq
 with a solution 
\beqr
\g (t) &=&  \int_{0}^{t}A(t') dt' \ \ \mbox{where}\\
A(t)&=& i \langle n|{dn\over dt}\rangle.
\eeqr
Thus we have
\beq
|\psi(t)\rangle =\exp \lt {i\int_{0}^{t}A(t') dt'} \rt \exp \lt - {i \over \hbar}\int_{0}^{t}E(t')dt'\rt |n(t)\rangle.
\eeq

One may be tempted to dismiss the extra phase due to $A$, because the phase can always be changed  by  a change of phase of the kets $|n(t)\rangle$ without affecting  their defining property as instantaneous eigenkets of $H(t)$). Under such a change
\beqr
 |n(t)\rangle &\to & |n(t)\rangle e^{i\chi(t)}\ \ \mbox{then}\\
A(t)&\to & A(t) -{d \chi \over dt}.
\eeqr

But suppose  $H$  returns to the original starting point after some time $T$, i.e., if $H(0)=H(T)$, then the relationship 
\beq
\psi (T)\rangle = e^{i\oint_{0}^{T}A(t') dt'}|\psi (0)\rangle
\eeq
 is unaltered by gauge transformations, given that $\chi$ is single-valued: $\chi (T)=\chi (0) 2\pi m$. This is easier to visualize if we think that the space of parameters in $H$ is labeled by a coordinate $R$  and $H$ varies with time because $R$ does:
\beq 
H(t)= H(R(t))\ \ \ \ \ \ \  |n(t)\rangle = |n(R(t)\rangle. 
\eeq
Then 
\beqr
\int_{0}^{t}A(t')dt'&=& i \int_{0}^{t}\left\langle n|{dn\over dR}\right\rangle {dR\over dt'} dt'\\
&=&   \int_{0}^{t}A(R) {dR\over dt'} dt' \ \  \mbox{where}\\
A(R)&=& i\left\langle n|{dn\over dR}\right\rangle.
\eeqr
In this form we see that $A(R)$ couples to the velocity of the fictitious particle with coordinate $R(t)$. Indeed we soon encounter  problems where $R$ is the coordinate of a real particle and $A(R)$ affects its dynamics like a genuine vector potential, except that it is not of electromagnetic origin.  

In this version it is readily seen that 

\beqr
\exp \lt i\oint_{0}^{T}A(t') dt'\rt &=&\exp \lt i\oint_{R(0)}^{R(T)}A(R) dR\rt\\
&=& \exp \lt i\int \!\int {\cal B} \cdot d\bS \rt
\eeqr
 where 
\beq 
{\cal B}= \nablab \times A
\eeq
is called the {\em Berry curvature} \index{Berry curvature ${\cal B}$} and $S$ the surface bounded by the loop traversed by the system in the time $t=0-T$.

Next imagine a huge particle lumbering along in its configuration space with coordinate $R$. Riding on it is a small but fast moving system. The small system experiences a Hamiltonian $H(R)$. For example the small system could be a spin experiencing the magnetic field $\bB(R)$ at the heavy particle's location. (The heavy particle could be electrically neutral and unaware of $\bB$.) We assume the small system stays in one particular state $|n(R)\rangle$ and does not jump to other states with a different $n$.  The Berry phase it accumulates 
$$\exp \lt i\int_{0}^{R(T)}A(R) dR\rt$$
clearly affects the fate of the large particle as any vector potential would. 

Or consider some electrons moving along with their parent nuclei. At a given location $R$ of the 
nuclei, the electrons settle down to some state $|n(R)\rangle$ and stay at fixed $n$ as $R$ slowly moves. The nimble electrons manage to find an eigenstate at each fixed $R$, which is a slowly varying parameter for them. This is the {\em Born-Oppenheimer approximation.} However Born and Oppenheimer  did not consider the  potential $A(R)$ that arises from the fast motion of the electrons.

We will now discuss   a simpler example in which a Berry vector potential appears and modifies the dynamics. 
\subsection{Berry phase affecting slow degree of freedom}

\begin{figure}
    \centering
\includegraphics[width=3in]{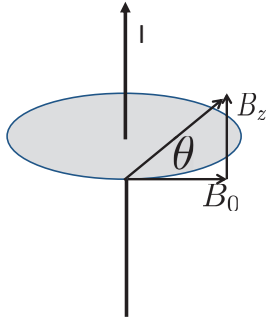}
    \vspace{1cm}
    \caption{The heavy particle with moment of inertia $I$ moves slowly on a circle carrying a spin which evolves rapidly, i.e., is in the instantaneous eigenstate of the local magnetic field $\bB$  which  is the sum of an azimuthal field due the current perpendicular to the plane of motion and a uniform  field  perpendicular to the plane of motion. }
    \label{ady}
  \end{figure}

Consider the situation depicted in Figure \ref{ady}. A massive particle is forced to move along a circle of radius $r$ lying in the $xy$ plane. There is a uniform magnetic field $\bB= \bk B_z$.
In addition a wire passing through the center produces an azimuthal field $\bB_0= B_0 \be_{\phi}$. The particle, assumed neutral, does not feel $\bB$. Riding on the particle is a spin which sees the field  
\beq \bB= \bk B_z+ \bi B_0 \sin \phi + \bj B_0 \cos \phi.
\eeq
(The reason the sines and cosines seem interchanged is because the magnetic field at $\phi$ is  the tangent to the circle.) 
The Hamiltonian for the combined system is 
\beq H= {L^2 \over 2I}-\sigmab \cdot \bB\label{adyH}.
\eeq
We will assume the spin is locked into the instantaneous ground state $|+\rangle$  (parallel to $\bB$). The naive expectation is that energy eigenstate and eigenvalue are 
\beqr
|\Psi \rangle &=& e^{im\phi}|+\rangle\ \ \ \ \  m= 0, \pm 1 \ldots \\
E_m&=& {m^2 \hbar^2 \over 2I}-\sqrt{B_{0}^{2}+ B_{z}^{2}}.
\eeqr
But this result ignores the Berry phase which we will now incorporate. 
We first expand the combined state as 
\beq 
|\Psi \rangle= \int \psi(\phi )|\phi  \otimes n(\phi)\rangle d\phi
\eeq
Notice that $\phi$ is the only real degree of freedom: when the particle is at $\phi$, the spin is forced to be in $|n(\phi)\rangle$. This is why a single wavefunction $\psi(\phi )$ describes both. Just to make sure you got it:
\beq
\langle \phi\otimes n(\phi )|\Psi\rangle= \int \psi(\phi') \underbrace{\langle \phi\otimes n(\phi)|\phi'\otimes n(\phi')\rangle}_{\delta (\phi-\phi')} d\phi'= \psi (\phi).
\eeq

Instead of directly jumping into the eigenvalue problem of $L^2$, let us deal with the eigenvalue problem of $L$ first. 
\beqr
\langle \phi \otimes n(\phi )|L|\Psi\rangle &=&  \int \langle \phi \otimes n(\phi)|L|\phi'\otimes n(\phi')\rangle \langle \phi'\otimes n(\phi' )|\Psi\rangle d\phi' \nonumber\\
&=& \int \langle n(\phi)|n(\phi'\rangle \underbrace{\langle \phi |L|\phi'\rangle}_{-i\hbar \delta (\phi-\phi')d/d\phi'} \psi(\phi')d\phi'\\
&=& -i\hbar \int \delta (\phi -\phi'){d \over d\phi'}\lt \langle n(\phi)|n(\phi'\rangle\psi(\phi')\rt\\
&=& -i\hbar \left\langle n(\phi)|{dn(\phi ) \over d\phi}\right\rangle \psi (\phi)  -i\hbar {d\psi \over d\phi}\\
&=& (-i\hbar {d \over d\phi}-\hbar A)\psi (\phi).\ \ \ \ \ 
\eeqr

Choosing solutions of the form $e^{im\phi}$ we find that the eigenvalues  of $L$ are 
\beq 
l= m\hbar -\hbar A.
\eeq
We see the spectrum is no longer the integers but shifted by $A$.
Let us now compute $A$ for the spin whose  ground state  is 
\beq
|n (\phi) \rangle = \left( \begin{array}{c} \cos {\t\over 2} \\  i \sin {\t \over 2} e^{i\phi}\end{array}\right),
\eeq
where $\tan \t = {B_0 \over B_z}$. (The $i$ in the lower components is there because the field at $\phi$ is tangent to the circle at that point.)
It is easily verified that 
\beq
A= -\sin^2 {\t \over 2}.
\eeq
Thus the spectrum of $L$ is 
\beq
l = \hbar \left( m+ \sin^2  {\t \over 2}\right).
\eeq

Going back to 
\beq H= {L^2 \over 2I}-\sigmab \cdot \bB\label{adyH2}
\eeq
you may be tempted to conclude that the spectrum is 
\beq
E_m = {\hbar^2 \over 2I}\left( m+ \sin^2  {\t \over 2}\right)^2- \sqrt{B_{0}^{2}+ B_{z}^{2}}.
\eeq
This is however incorrect: there is an extra constant ${\hbar^2 \over 4} \sin^2 \t$. The details 
are left to the following exercise.
\begin{ex}
Show that in addition to $A$, a scalar potential 
\beq
\Phi = \hbar^2 \lt \langle dn|dn\rangle - \langle dn|n\rangle \langle n|dn\rangle \rt
\eeq
arises when second derivatives enter $H$. Show that in our problem $\Phi = {\hbar^2 \over 4} \sin^2 \t$. 
In the above I use a compact notation in which $|dn\rangle = {d| n (\phi) \rangle \over d\phi}$ etc. 
\end{ex}    

\subsection{Berry monopole in parameter space}
Where does the Berry flux come from? It has to do with degeneracies in parameter space where two levels $E_1$ and $E_2$ collapse to a common $E$.  Consider a problem where two levels cross. Let us focus on them and ignore the rest. In the two-dimensional vector space the Hamiltonian has to have the following form 
\beq
H=\left(\begin{array}{cc}
E + h_z & h_x-ih_y\\
h_x+ih+y & E-h_z
\end{array} \right) = E I+ \sigmab \cdot \bh\ \ \ \label{hofspin}
\eeq
where 
\beq
 \bh= (\bi h_x+\bj h_y + \bk h_z)
\eeq
are the three free parameters of the $2 \times 2$ traceless Hermitian Hamiltonian. 
We see that $\bh$  is the position vector in parameter space. (We can choose the degenerate energy $E=0$. )

  The energy levels of  the two states  $|n_{\pm}\rangle$ are 
\beq
 E_{\pm}= \pm |\bh | .
\eeq
The levels become degenerate at the origin in $\bh$ space. 
 At every other point the levels are split.

\begin{figure}
    \centering
\includegraphics[width=4in]{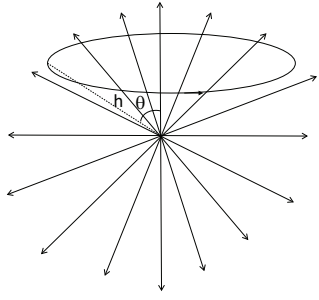}
    \vspace{1cm}
    \caption{The source of Berry flux is a monopole at $\bh =0$, the origin of parameter space and the point of degeneracy. The figure shows a closed path on a sphere of radius $h$. The dynamical phase at time $t$ is $h t$. The Berry phase is the monopole flux penetrating the surface bounded by the closed loop, independent of how long it took to go around.  }
    \label{monoberry}
  \end{figure}

This looks like a spin in  magnetic field $\bB= \bh$ in $\bh$ space, which we previously refereed to as $\bR$ space.
Let us assume the system is in the spin up or $|n_+\rangle$ state. It has a {\em dynamical phase factor }
\beq
e^{i\g_d}= \exp \lt -{i \over \hbar}\int_{0}^{t}E(t')dt' \ \rt= \exp \lt -{i \over \hbar}\int_{0}^{t}|\bh (t')|dt' \ \rt.
\eeq
This phase depends on details of the motion. For example if it is a circle on a sphere of radius $h$ as shown in Figure \ref{monoberry},   the accumulated dynamical phase at time $t$ is  $ht$.  It  can be big or small depending on $t$. 

But there is  also a Berry phase, which is the focus of our discussion.

Let us say the system is in the spin up or $|n_+\rangle$ state. Then 
\beqr
\bA_+\cdot d \bh &=& i \langle n_+|dn_+\rangle\\
&=& i \left( \cos {\t \over 2}, \sin {\t \over 2}e^{-i\phi}\right)
\left(\begin{array}{c} d(\cos {\t \over 2})\\ d(\sin {\t \over 2})e^{i\phi}+i d\phi \sin {\t \over 2}e^{i\phi}\end{array}\right)\nonumber \\
&=& -{(1 - \cos \t )\over 2}d\phi\ \ \mbox{so that}\\
 \bA&=& -{(1 - \cos \t )\over 2}\be_{\phi}
\eeqr

Consider a loop $C$ along a latitude $\t$ on a sphere surrounding the origin, as shown in Figure \ref{monoberry}. We find
\beq
 \oint A_{\phi} d \phi =- \pi ( 1 -\cos \t)
\eeq
 which we recognize as flux of a monopole of strength $g = - \half$ at the origin. (When $\t \to \pi$, the total flux enclosed is $-2\pi$.)

The Berry phase is thus due to a source which has a $\d$-function divergence at the origin. 
Unlike the dynamical phase it does not depend on the "velocity "  with which the path is traversed (as long as the adiabatic approximation is valid.)

{\em Note: Some of you may be worried  about the $\phi$ integral. We normally expect to integrate $A_{\phi} r\sin \t d\phi$ but here we have just  $A_{\phi}d\phi$. This is because we have defined \beq
A_{\phi}= i\langle n|dn\rangle =i\left\langle n \right|\left.{ \p n\over \p \phi }\right \rangle.
\eeq
 We could have instead used (remembering there is no $r$ dependence of $|n\rangle$)

\beqr \bA &=&  i \langle n|\nablab n\rangle\\
  &=& \be_\t \left\langle n \right|\left.{ \p n\over r \p \t }\right \rangle+ \be_{\phi}\left\langle n \right|\left.{ \p n\over r \sin \t  \p \phi }\right \rangle.
\eeqr
  in which case we would have integrated $A_{\phi} r \sin \t d\phi$.

The point is that if we parametrize the curve on which the system moves by the variable  $\eta$ and set
\beq
 A_{\eta}= i \left\langle n \right|\left.{ \p n\over \p \eta }\right \rangle,
\eeq
then $A_{\eta}d \eta$ is the phase change when there is a change $d \eta$, while if we 
then proceed  to re-parametrize the curve  by $\zeta (\eta)$ and  define 
\beq
 A_{\zeta}= i  \left\langle n \right|\left.{ \p n\over \p \zeta }\right \rangle,
\eeq
then 
\beq 
A_{\zeta}d \zeta
\eeq
is the change in phase over the {\em same} segment in the new parametrization. This all works out because of the transformation law 
\beq 
A_\eta = A_{\zeta}{d \zeta \over d \eta}.
\eeq
To understand this better you should learn differential forms. 
}

Back to the flux which seems to be due to a monopole of strength $- \half$ at the origin. We demonstrate this as follows. Let $|n\rangle$ be the state the spin is in. (It was $|+\rangle$ in our example. )

\beqr
A_{\mu}&=& i \langle n | \p_{\mu}n\rangle \ \ \ \  \ \   (\p_{\mu}= {\p \over \p h_{\mu}})\\
F_{\mu \nu}&=& \p_\mu A_{\nu}- \p_\nu A_{\mu}\\
&=& i \lt \langle \p_\mu n| \p_{\nu}n\rangle-\langle \p_\nu n| \p_{\mu}n\rangle \rt\\
&=& \sum_{m\ne n}i \lt \langle \p_\mu n|m\rangle \langle m| \p_{\nu}n\rangle-\langle \p_\nu n|m\rangle \langle m| \p_{\mu}n\rangle \rt,
\eeqr
where  $m\ne n$ because the $m=n$ term vanishes identically.
\begin{ex}
Show that the $m=n$ vanishes identically using  $$0= d\langle m|m\rangle=  \langle d m|m\rangle+ \langle m|dm\rangle.$$
\end{ex}
Next we derive a very useful relation:
\beqr 
\langle m |H|n\rangle &=& 0 \ \ \mbox{ (remember $m\ne n$)}\\
\langle dm|H|n\rangle +\langle m|dH|n\rangle+ \langle m|H|dn\rangle&=&0\\
(E_m - E_n)\langle dm|n\rangle &=& \langle m |dH|n\rangle \ \ \ \mbox{(use $\langle m | dn\rangle = - \langle dm| n\rangle$)}\nonumber \\
\langle dm|n\rangle &=& {\langle m|dH |n\rangle \over E_m-E_n}\\
\langle m|dn\rangle &=& {\langle m|dH |n\rangle \over E_n-E_m}.
\eeqr

Consequently
\beqr
F_{\mu \nu}&=&i \sum_{m \ne n}{ 1 \over (E_n-E_m)^2}\lt \langle n|\p_\mu H|m\rangle 
\langle m|\p_\nu H|n\rangle - (\mu \leftrightarrow \nu)\rt \nonumber\\
&=& i\sum_{m \ne n}{ 1 \over (2h)^2}\lt \langle n|\p_\mu H|m\rangle 
\langle m|\p_\nu H|n\rangle - (\mu \leftrightarrow \nu)\rt
\eeqr
Now put in the $m=n$ term because it vanishes and use completeness and $H= \sigma_{\mu}h_{\mu}$ to obtain
\beqr F_{\mu \nu}&=&{i\over 4h^2}\langle n|\p_\mu H \p_\nu H - \p_{\nu}H \p_\mu H\rangle\\
&=& {i \over 4h^2}\langle n|\sigma_{\mu}\sigma_{\nu}- \sigma_{\nu}\sigma_{\mu}|n \rangle\\
&=& -{1 \over 2h^2}\e_{\mu \nu \l}\langle n|\sigma_\l|n\rangle \\
{\cal B}_\pm&=& \mp {1 \over 2h^2}\hat{\bh}\ \ \ \ \mbox{for $|n\rangle = |\pm\rangle$}.
\eeqr
Since $h^2$ is the distance squared from the origin, this describes a monopole of strength $-\half$ sitting at the origin, the point of degeneracy. 

Notice that the sign of ${\cal B}$ depends on which state ($|n_+\rangle$ or $|n_{-}\rangle$) we are working with. 

So the picture bear in mind is that there is a monopole with $g=\mp \half$ sitting at the origin in parameter space and  as the system  traverses a closed loop in the eigenstate $|n_\pm\rangle$ , the phase change is given by the flux intercepted by a surface bounded by this loop, as shown in Figure \ref{monoberry}. 

\section{Time-reversal symmetry - TRS}

Let us first meet this symmetry  in  classical mechanics. 

\subsection{TRS in classical mechanics} 
\begin{figure}
    \centering
\includegraphics[width=3in]{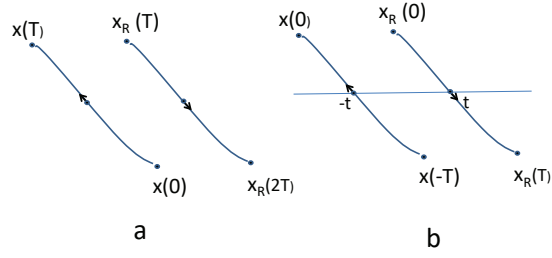}
    \vspace{1cm}
    \caption{(a) A path and its reverse between times $0$ and $2T$. (b) The same paths with initial time changed to $-T$.
     The time reversal operation is done at $t=0$. The original path $x(t)$ and its time-reversed version $x_R(t)$ obeying $x_R(t)=x(-t)$ and $\dot{x}_R(t)= - \dot{x}(-t)$.  A time slice shows this relation between the two paths: at opposite times $x$ and $x_R$ have the same value and opposite derivatives. }
    \label{timerev}
  \end{figure}

Suppose a  planet moves from $x(0)$ to $x(T)$ with  initial and final velocities $\dot{x}(0)$ and $\dot{x}(T)$  as shown in Figure \ref{timerev} a. Say we make a movie of this and play it backwards starting at time $T$. This time-reversed trajectory $x_R(t)$ will have its velocity opposite to that of the original one on the way back and arrive at the starting point at time $2T$  with velocity $\dot{x}_R(2T)=-\dot{x}(0)$. The reversed movie would appear perfectly regular, i.e., in accordance with Newton's Laws,  to a person viewing it. Indeed she will not know if the projector is running forward or backwards. In other words, what she sees can very well be the movie of a real planet obeying Newton's  laws,  traveling along the trajectory $x_R(t)$. This is an example of time-reversal symmetry (TRS) of Newton's Laws and the gravitational force.

For cosmetic reasons let us change the initial time to $t=-T$ so that the reversal takes place at $t=0$ and the clip ends at time $T$ as shown in Figure \ref{timerev}b. In this convention, the time-reversed path $x_R(t)$ is related to the original one as follows:
\beq
x_R(t)= x(-t).\label{xr}
\eeq
You can see this is true from the figure: if you slice it horizontally, you find $x_R(t)= x(-t)$. Differentiating Eqn. \ref{xr} we find
\beq
\dot{x}_R(t)={dx_R(t)\over dt}= {dx(-t)\over dt}=-{dx(-t)\over d(-t)}=-\dot{x}(-t),\label{xdotr}
\eeq
again in accord with the figure.
If you set $t=T$ in Eqns. \ref{xr}, \ref{xdotr}, you  will find, as in Figure \ref{timerev}b,  that after time-reversal the particle returns to its original location with opposite velocity. 

Taking one more derivative we find
\beq
a_R(t)=\ddot{x}_R(t)=\ddot{x}(-t) =a(-t).\label{arev}
\eeq

We are given that the trajectory $x(t)$  obeys 
\beq
 m \ddot{x}(t)= F(x(t)).
\eeq
Using Eqn. \ref{arev}
\beq
m  \ddot{x}_R(t)= m  \ddot{x}(-t)= F(x(-t))=F(x_R(t))
\eeq
 which means $x_R(t)$ also obeys Newton's laws. In the present case this follows from the fact that because the law involves only acceleration, we may change $t$ to $-t$ and get a new solution from an existing one. This is not true if there is friction and there is a term in the equation of motion involving a single time-derivative.
 \subsection{Time-reversal symmetry in quantum mechanics}
 
Wigner showed that there are two kinds of symmetries that leave the {\em magnitude}  of the inner product invariant.

{\em Unitary symmetries $U$} \index{Unitary symmetries $U$} represented by   operators obeying
\beq
U^{\dag}U=I,
\eeq
and whose action is  linear
\beq
 U (\a |\psi\rangle + \b |\chi \rangle)= \a U|\psi\rangle + \b U |\chi \rangle
 \eeq
 and preserve the inner product:
 \beq
 \langle U\phi |U\psi\rangle = \langle \phi |U^{\dag}U|\psi\rangle =  \langle \phi |\psi\rangle.
 \eeq
 If $U$ is a symmetry operator, then 
 \beq
 UHU^{\dag}=H.
 \eeq

Consequently, if  
\beq 
H|E \rangle = E |E\rangle
\eeq
then 
\beqr
UHU^{\dag} U|E \rangle &=& E U|E\rangle\\
H |U\psi \rangle &=& E |U\psi\rangle ,
\eeqr
meaning $|E\rangle$ and $|UE \rangle$ are degenerate.

This is what we normally run into, when we consider symmetries like rotational or translational invariance. But  now need a new beast.

{\em Anti-unitary symmetries}  \index{anti-unitary symmetries}  $\Omega$ act as follows on any two generic kets $|\phi\rangle$ and $|\psi\rangle$:
\beq
\langle \Omega \phi |\Omega \psi\rangle = \langle \psi|\phi\rangle =\langle \phi |\psi\rangle^{*} .\label{antiprop}
\eeq
Thus they preserve only the absolute value  of the inner product.

I bring them up because $\T$, the operator which generates time-reversal is anti-unitary. 

I will now establish one central property of $\T$ (or any anti-unitary operator), returning to others properties  later. It is that they are anti-linear:
\beqr
 \T \a |\psi\rangle &=&\T \a \T^{-1}\cdot \T|\psi\rangle =\a^{*} \T |\psi \rangle \equiv \a^{*} |\T \psi \rangle \ i.e.,\\
 \T \a \T^{-1}&=& \a^{*},  \label{antilin}
 \eeqr
 where $\a$ is a $c$-number. In other words, when $\T$ passes through a $c$-number it complex conjugates it. This property follows from the requirement Eqn. \ref{antiprop}:
 \beq
 \langle \Omega \phi |\Omega \psi\rangle = \langle \psi|\phi\rangle. \label{antiprop2}
 \eeq
 We now establish  Eqn. \ref{antilin} by imposing this condition on   two generic state vectors expanded in an orthonormal  basis as follows:
 \beqr
 |\psi \rangle &=& \sum_n \psi_n |n\rangle\\
 |\phi \rangle &=& \sum_m \phi_m |m\rangle\\
  \T|\psi \rangle &=& \sum_n \T \psi_n \T^{-1} |\T n\rangle\\
 \T|\phi \rangle &=& \sum_m \T \phi_m \T^{-1} |\T m\rangle\\
 \langle \T \phi|\T\psi\rangle &=& \sum_{m,n}(\T \phi_m \T^{-1})^*(\T \psi_n \T^{-1})\underbrace{\langle \T m|\T n\rangle}_{=\langle n|m\rangle=\d_{mn}}\\
 &=& \sum_{n}(\T \phi_n \T^{-1})^*(\T \psi_n \T^{-1})\ \ \ \mbox{which by  ( Eqn. \ref{antiprop2}) equals}\nonumber\\
 \langle \psi|\phi\rangle&=& \sum_{n}\psi_{n}^{*}\phi_n
 \eeqr
 Since $\phi_n$ and $\phi_m$ are arbitrary, this means 
 \beq
 (\T \psi_n \T^{-1})= \psi_{n}^{*}
 \eeq
  and likewise for $\phi_m$.

 If $\T$ is a symmetry operation, it means  
\beq
\T H\T^{-1} = H.
\eeq

In this case 
\beqr
H|E \rangle =&=& E |E\rangle \ \ \mbox{ implies }\\
\T H\T^{-1}\cdot \T |E \rangle &=& E \T|E\rangle\ \ \ \mbox{(remember $E$ is real)}\\
H |\T E \rangle &=& E |\T E\rangle.
\eeqr

Thus $|E\rangle$ and its time-reversed partner $|\T E\rangle$ are degenerate unless $\T|E\rangle = c |E\rangle$. We will consider cases where this possibility is ruled out and a degeneracy is mandatory. 

\subsection{TRS in non-relativistic quantum mechanics}

 Let us consider the action of $\T$  in the non-relativistic quantum mechanics of a spinless particle. I discuss  only one spatial dimension with coordinate $x$, but the extension to higher  dimensions is obvious.   
 
 Consider the eigenkets of position and momentum:
\beqr
X |x\rangle &=& x |x\rangle \label{xt}\\
P |p\rangle &=& p |p\rangle.\label{pt}
\eeqr
The action of an  anti-linear operator (like the linear operator) is fully defined by its action on a basis. We choose  the $|x\rangle$ basis. We demand, based on classical intuition that 
\beq
 \T |x\rangle = |x\rangle.\label{xt2}
 \eeq
 The action of $\T$ on the operator $X$ follows from Eqn. \ref{xt2}:
 \beqr
 \T X \T^{-1}\cdot \T |x\rangle &=& x \T|x\rangle \\
 \T X \T^{-1} |x\rangle &=& x |x\rangle
 \eeqr
 which means 
 \beq
 \T X \T^{-1}=X.
 \eeq
 
 Consider  the expansion
\beqr
|\psi\rangle &=& \int dx |x\rangle \langle x |\psi\rangle   = \int dx  \psi(x)|x\rangle\\
\T|\psi\rangle &=& \int dx \psi^*(x)|\T x\rangle = \int dx \psi^*(x)| x\rangle 
\eeqr
which means the wavefunction get conjugated by $\T$:
\beq
\langle x|\T \psi\rangle = \psi^*(x).\label{psistar}
\eeq
 
 What does $\T$ do to $|p\rangle$? This cannot be answered unless we specify what   $|p\rangle$ stands for. So we furnish its components in the $|x\rangle$ basis:
 \beq
 \langle x |p\rangle =\exp \lt {ipx\over \hbar}\rt.
 \eeq
 
 We may now deduce what $\T$ does to $|p\rangle$
 \beqr
 \langle \T x |\T p\rangle &=& \langle p|x\rangle =\exp \lt -{ipx\over \hbar}\rt\\
 \langle  x |\T p\rangle&=&\exp \lt -{ipx\over \hbar}\rt\\
 &=&\langle  x | -p\rangle\ \ \ \ \mbox{which means}\\
 \T|p\rangle &=& |-p\rangle.
 \eeqr
 It is now easy to show that 
 \beq
 \T P \T^{-1}= -P.\label{minusp}
 \eeq
 \begin{ex} Prove Eqn. \ref{minusp}.
 \end{ex}
 The preceding  results are consistent with the commutation relation:
 \beqr
 XP-PX&=& i \hbar\ \ \ \ \mbox{because}\\
 \T X\T^{-1}\ \T P\T^{-1}-\T P \T^{-1} \T X\T^{-1}&=& \T i \T^{-1}\hbar\\
 -XP+PX &=& -i \hbar.
 \eeqr
 
 \subsection{TRS of  the Schr\"{o}dinger equation}
 
 How does all this apply to the  Schr\"{o}dinger equation?
 Let us begin with the equation in the ket notation and act on both sides with $\T$:
 \beqr
 i \hbar {d |\psi (t)\rangle\over dt}= H(X,P)|\psi (t)\rangle\\
  (-i) \hbar {d |\T \psi \rangle\over dt}= \T H(X,P)\T^{-1}|\T\psi\rangle\ \ \   \\
  i \hbar {d |\T \psi \rangle\over d(-t)}= \T H(X,P)\T^{-1}|\T\psi\rangle=
   H(X,-P)|\T\psi\rangle.\ \ \ \label{trecse}
   \eeqr
   Thus we find that $|\T \psi\rangle$ obeys the equation of motion with $t\to -t$ provided $H(X,-P)= H(X,P)$. This is the case if 
   \beq
   H= {P^2\over 2m}+V(X)
   \eeq
    but not if there is a magnetic field:
    \beq
   H= {(P-eA)^2\over 2m}+V(X).
   \eeq
   
   Now we go to the $x$-basis by projecting Eqn. \ref{trecse} on to the ket $\langle x|$ and recalling Eqn. \ref{psistar}, $\langle x|\T\psi\rangle = \psi^*(x)$:
   \beqr
   i \hbar {\p \psi^*(x,t) \over \p (-t)}&=& \lt \left(i\hbar {\p \over \p x}-eA\right)^2 +V(x)\rt\psi^*(x,t)\\
   &=& \lt \left(-i\hbar {\p \over \p x}-eA\right)^2 +V(x)\rt^*\psi^*(x,t), \ i.e., \\
   i \hbar {\p \psi^*(x,t) \over \p (-t)}&=&H\left(-i\hbar {\p\over \p x}, x\right)^*\psi^*(x,t).
   \eeqr
   This means that {\em $\psi^*(x,t)$ obeys the time-reversed equation if $H=H^*$ in the $x$-representation}. (In particular only if $A=0$.)
   \subsection{The $K$ operator}
    Limiting ourselves to the  Schr\"{o}dinger equation in the $x$-basis, we may set 
    \beq
    \T = K
    \eeq
    where $K$ is the complex conjugation operator on all $c$-numbers.
    
    By definition, 
    \beqr
    K^2&=&I \ \ \ i.e., \\
    K&=&K^{-1}. \ \ \ \
    \eeqr
     Let us see how $K$ acts: 
    \beqr
     i \hbar {\p \psi(x,t) \over \p t}&=&H\left(-i\hbar {\p\over \p x}, x\right)\psi(x,t)\\
     K i \hbar {\p \psi(x,t) \over \p t}K&=&KH\left(-i\hbar {\p\over \p x}, x\right)K \cdot K\psi(x,t)K\\
      -i \hbar {\p \psi^*(x,t) \over \p t}&=&H\left(-i\hbar {\p\over \p x}, x\right)^*\psi^*(x,t)\\
      i \hbar {\p \psi^*(x,t) \over \p(- t) }&=&H\left(-i\hbar {\p\over \p x}, x\right)^*\psi^*(x,t)
     \eeqr
     from which it follows that $\psi^*(x,t)$ is the time-reversed solution provided $H^*=H$.
     
     It is important to note that $\T=K$ only after going to the  $x$-basis, whereupon   only $c$-numbers (constants, functions and their derivatives) enter the picture.

     When we have spin we must modify $\T =K$. Consider 
\beq H_{so} = \bs \cdot (\br \times \bP )
\eeq
 which describes spin-orbit coupling. (Here we are obviously in higher dimensions and $x\to \br$.) This $H_{so}$  should be TRI since both spin and angular momentum get reversed. 
However $\T = K$ does not do the job: it reverses $s_y$ (which is the pure imaginary Pauli matrix) but not the other two. The correct answer is 
\beq
\T = i s_y K\ \ \ \ \ \ \Theta^{-1}= K s_y (-i)
\eeq
under which 
\beqr
\T \bs \T^{-1}&=&  -\bs\\
\T \br  \T^{-1}&=&\br\\
\T \bP \T^{-1}&=&-\bP\\
\T H_{so} \T^{-1}&=&H_{so}.
\eeqr
You may wonder how  $K$ is to act on $\bs$: as a $c$-number or an operator?   The answer is that it   complex conjugates  the  matrix elements of $\bs$  as $c$-numbers that represent, in the $S_z$ basis,  the abstract spin operator $\bS$ of Hilbert space, (just like $\psi (\br)$ represents $|\psi\rangle$ in the $|\br \rangle$ basis). 

Once again $\T = i s_y K$ is true only in the eigenbasis $|\br, s_z\rangle$ of $\bR$ and $S_z$. 

Here is a surprise. You might have expected that $\T^2=I$ since a double reversal should be equal to no reversal. However we find 
\beq
 \T^2 = i s_y K i s_y K = i s_y (-i (s_y)^*)K^2= -1.
\eeq
Indeed there are many problems where $\T^2=-1$, which  was shown by Kramers to imply the degeneracy of states related by $\T$. We will now discuss this at length.

\subsection{General study of $\T$ and Kramers' degeneracy}

{\em Theorem} 
\beq
\T = K U \ \ \ \ \mbox{where $U$ is unitary}.
\eeq
Proof:
 Consider expanding $|\psi\rangle$ in an orthonormal basis and acting with $\T$:
\beqr
|\psi\rangle &=& \sum_{n}\psi_n |n\rangle\\
\T |\psi\rangle &=& \sum_{n}\psi_{n}^{*} |\T n\rangle\label{344}.
\eeqr
But 
\beq
\langle \T n| \T m\rangle = \langle m|n\rangle = \d_{mn}.
\eeq
So $|\T n\rangle$ is also an orthonormal basis which  must therefore be related to the basis $|n\rangle$ by a unitary transformation:
\beq
|\T n \rangle = \sum_{m}U_{nm}|m\rangle.
\eeq
Substituting this in Eqn. \ref{344}, 
\beqr
\T|\psi\rangle &=& \sum_{mn}\psi_{n}^{*}U_{nm}|m\rangle\\
&=& K \sum_{mn}\psi_{n}^{}U_{nm}^{*}|m\rangle\\
&=& K \sum_{mn}\psi_{n}^{}U_{mn}^{\dag}|m\rangle\\
\eeqr
 Dotting both sides with $\langle k|$
\beqr
\langle k|\T \psi\rangle &=& K \sum_{mn}\psi_n U^{\dag}_{mn} \d_{mk}\\
&=& K\sum_{n}U^{\dag}_{kn}\psi_n = K\lt U^{\dag}\psi\right]_k \ \ \ \mbox{which means}\\
|\T \psi \rangle &=& KU^{\dag}|\psi\rangle.
\eeqr
 But $U^{\dag}$ is also unitary and we have 
\beq
 \T = KU.
\eeq
Sometimes I may write $\T = U K$ which is just as good, with the elements of the new $U$ being the complex conjugate of the old $U$. 

{\em Theorem} $\T^2= \pm 1.$

{\em Proof}: 
We know that after two reversals we must get the same physical state:
\beq
\T^2\psi \rangle = c |\psi\rangle.
\eeq
So 
\beqr
KUKU |\psi\rangle &=& c |\psi\rangle\\
U^*U|\psi\rangle &=& c |\psi\rangle \ \ \ \forall |\psi\rangle\\
U^* U&=& c\\
U^*&=& c U^{\dag} =c (U^*)^T= c (cU^{\dag})^T=c^2 U^*\\
c&=&\pm 1.
\eeqr
Note for future use that since $U^*= c (U^*)^T $, 
\beq
U^T = cU.
\eeq

{\em Theorem} If $\T^2=-1$, it can't have an eigenvector.

Proof: If 
\beq
\T |\psi\rangle = \l |\psi\rangle
\eeq
then
\beq
\T^2 |\psi\rangle = \T  \l |\psi\rangle= \l^* \T |\psi\rangle = |\l|^2|\psi\rangle \ne -|\psi \rangle.
\eeq

Thus what we have are (at least)  two degenerate states related by $\T$
\beq
\T |\psi\rangle = |\chi \rangle\ \ \ \ \ \T |\chi\rangle =- |\psi \rangle.
\eeq
This is called a {\em Kramers doublet}.\index{Kramers doublet} We can verify that $|\psi\rangle $ and $|\T \psi\rangle$ are orthogonal if $\T^2 =-1$:
\beq
\langle \psi| \T \psi\rangle = \langle \Theta^2 \psi| \T \psi\rangle= -\langle  \psi| \T \psi\rangle.
\eeq

There is no $\T^{\dag}$ in the usual sense:
\beq
\langle  \chi|\T \psi \rangle \ne \langle \T^{\dag}\chi |\psi\rangle
\eeq
 because the  expression  would be  anti-linear in $\psi$ and  $\chi$ based on the left hand side  but linear in both based on the right. This can be fixed by adding another complex conjugation in the definition, but we will not follow that route. 

\section{Symmetries in momentum space}

Consider a single particle moving in a periodic potential with 
\beqr
{\cal H}  &=& {P^2 \over 2m}+V(x)\\
V(x+a)&=&V(x)
\eeqr
where  $a$ is the lattice spacing. Thus 
\beq
T(a) {\cal H}T^{\dag}(a)={\cal H}
\eeq
where $T(a)$ translates by $a$. Its eigenvalues will have to be of the form $e^{ika}$
\beq
T(a)\psi_k(x) = e^{ika}\psi_k(x)
\eeq
This form respects the unitarity of $T$ and  ensures the group property  that $T(a) T(b)= T(a+b)$. 

Bloch's theorem says that we may choose energy eigenfunctions to be of the form 
\beqr
\psi_k(x)&=& e^{ikx}u_k(x)\ \ \ \mbox{where}
\\
u_k(x+a)&=&u_k(x).
\eeqr
Let us check:
\beqr T(a)\psi_k&=&\psi_k(x+a)= e^{ik(x+a)}u_k(x+a)=e^{ika}e^{ikx}u_k(x+a)\nonumber \\
&=&e^{ika}e^{ikx}u_k(x)=e^{ika}\psi_k(x).
\eeqr

The energy eigenvalue equation obeyed by $u_k(x)$ is
\beqr \underbrace{ \lt{P^2 \over 2m}+V(x)\rt}_{{\cal H}} e^{ikx}u_{k}(x)&=& E_k e^{ikx}u_{k}(x)\\
 \underbrace{\lt{(P + \hbar k)^2 \over 2m}+V(x)\rt}_{H(k)} u_{k}(x) &=& E_k u_{k}(x)\\
H(k)u_k &=&E_k u_k.\label{hofk}
\eeqr
where I have defined {\em in  the $x$-representation}
\beq
H(k) = e^{-ikx}{\cal H}e^{ikx}.\label{eikx}
\eeq
This discussion   assumes that at each $k$ there is just one eigenfunction.  Generally, $H(k)$ can have a tower of eigenfunctions $u_{km}$ labeled by a {\em band index $m$}. In this larger space,   $H(k)$ will  generally  be a matrix with elements $H_{mn}$.  (That is $u_{km}$ and $\psi_{km}$ form a basis but not necessarily an eigenbasis of ${\cal H}$. )

 I remind you of some basic facts about momentum space.
 \begin{itemize}
 \item If the system is periodic with length $L=Na$, the allowed values of $k$ obey
\beq
e^{ikNa}=1 \to k_m = {2 \pi m \over Na}, \ \ \ \ m=0,...N-1.
\eeq
It is more convenient (assuming $N$ is even) to choose the allowed momenta symmetrically around $0$     
\beq
 k=k_m= {2 \pi m  \over Na}\ \ \mbox{where\ } m= \lt 0, \pm 1, \pm 2..., {N\over 2}\rt.\label{allowedm}
\eeq
Note that $0$ and ${N\over 2}$ are equal to minus themselves because $e^{\pm i 0}=1$ and $e^{\pm i \pi }=-1$.
\item
The momenta lie in the interval
\beq
-{\pi \over a} \le k<{\pi \over a}.
\eeq

{\em From now on I will use $a=1.$}  
Thus 
\beq
-{\pi } \le k <{\pi}.
\eeq

When $N\to \infty$,  the allowed values of $k$  get closer and closer, the spacing between adjacent ones (labeled $m$ and $m + 1$) being 
\beq
\D k = {2 \pi \over N}.
\eeq

Then a sum over $m$  of any function of $k_m$ can be turned into an integral over $k$  as follows: 
\beq
\sum_m f(k_m)= {N \over 2 \pi} \sum_m f(k_m) {2 \pi \over N} =N \int {dk \over 2\pi} f(k).
\eeq
\end{itemize}

\subsection{How symmetries of ${\cal H}$ act on $H(k)$.}

I will begin with the toughest example of TRS, implemented by the antilinear operator $\T$. Unitary symmetries will be then be a breeze.

Let us define time-reversed Hamiltonian 
\beq{\cal H}_{\T}=  \T {\cal H} \T^{-1}.
\eeq

If there is TRS, then 
\beq
 \T {\cal H} \T^{-1}= {\cal H}.
 \eeq
  
 Let 
 \beqr
 H_{nm}(k)&=& \langle \psi_{kn}|{\cal H}|\psi_{km}\rangle\\
 &=& \langle u_{kn}|e^{-ikx}{\cal H}e^{ikx}|u_{km}\rangle \equiv  \langle u_{kn}|H(k)|u_{km}\rangle.
 \eeqr
 In the above $k$ labels the conserved momentum index and $n$ and $m$ label  the band. 
We want to know what restrictions ${\cal H}_{\T}={\cal H}$  places on the matrix element of $H(k)$.

In the  abridged notation
\beq
|\psi_{nk}\rangle = |k,n\rangle,
\eeq
\beq
H_{nm}(k)= \langle k, n|{\cal H}|k, m\rangle.
\eeq

Before proceeding we need to define the action of $\T$ on the states $|k, n\rangle$. 
We assume that $\T$ reverses $k$ and then possibly scrambles up the band index by a unitary transformation:
\beq
\T |k, n\rangle= \sum_r U_{rn}|-k, r\rangle.\label{acttheta}
\eeq
Now we  systematically assemble the matrix elements of ${\cal H}_{\T}$:
\beqr
{\cal H}|k, i\rangle &=& \sum_{j} |k, j\rangle               \langle k, j|{\cal H}|k,i\rangle\\
&=&\sum_{j} |k, j\rangle               H_{ji}(k)\\
\T {\cal H}|k, i\rangle &=&\sum_{j}     H_{ji}^{*}(k)\T |k, j\rangle\\
&=& \sum_{j}     H_{ji}^{*}(k)U_{rj} |-k, r\rangle\\
\T {\cal H}\T^{-1} \T|k, i\rangle&=&\sum_{j,r}     H_{ji}^{*}(k)U_{rj} |-k, r\rangle\\
{\cal H}_{\T}\sum_s U_{si}|-k, s\rangle &=&\sum_{j,r}     H_{ji}^{*}(k)U_{rj} |-k, r\rangle\\
\sum_s \langle -k , r|{\cal H}_{\T} |-k, s\rangle U_{si} &=& \sum_{j}     H_{ji}^{*}(k)U_{rj}
\eeqr
But if ${\cal H}_{\T}= {\cal H}$, we have then
\beqr
\lt H(-k) U\rt_{ri}&=& \lt U H^{*}(k)\rt_{ri}\\
H(-k) &=& U H^{*}(k)U^{\dag}.
\eeqr
Using $K$ to produce  the complex conjugation, we rewrite
\beqr
H(-k)&=& U K H(k) K U^{\dag} \equiv \T H(k) \T^{-1}\ \ \ \mbox{where}\\
\T &=&  U K = K\cdot KUK=KU^* \equiv KU'.
\eeqr
Since $U'=u^*$ is also unitary we will refer to it also as $U$ so that $\T = KU$. 
(I denote by $\T$ both the time-reversal operator in Hilbert space and its matrix counterpart that conjugates the matrix $H_{mn}$.)

{\em In words: if $\T$ is a symmetry there must exist a matrix $\T = K U$ such that $\T H(k)\T^{-1}= H(-k)$.}

If parity $\Pi=\Pi^{-1}=\Pi^{\dag}$ is a symmetry, the same arguments lead to 

\beq
\Pi H(k)\Pi= H(-k)\label{436}
\eeq
where $\Pi$ is unitary, there being no need for complex conjugation of $H$ by an anti-linear operator. 
\begin{ex}Derive Eqn. \ref{436} assuming $\Pi |k, n\rangle = \sum_r U_{rn}|-k, r\rangle$.
\end{ex}

\subsection{Symmetry restrictions on Berry phase}
What are the implications for the Berry phase $A(k)$ if the problem is symmetric under the action of  $\T$ or  $\Pi$?    

Consider parity first. We reason as follows:
 \beqr
  H(k) |k\rangle &=& E |k \rangle\\
 \Pi H(k) \Pi \cdot \Pi |k\rangle &=& E \Pi|k \rangle\\
 H(-k) \Pi |k\rangle &=& E \Pi|k \rangle.
 \eeqr
  In the simple case where there is just one state at eack $k$, this means
  \beqr
  \Pi  |k\rangle &=&|-k\rangle \label{pi1} \ \ \\
  |k\rangle &=&\Pi |-k\rangle \\
  A(k)&=& i \langle k|{d \over dk}|k\rangle\\
  &=& i \langle \Pi \cdot  (-k)|{d \over dk}|\Pi \cdot (-k)\rangle\\
  &=& i \langle  (-k)|\Pi {d \over dk}\Pi | (-k)\rangle\\
  &=& i \langle  (-k)| {d \over dk} | (-k)\rangle \ \ \ \mbox{( as $\Pi^2=I$)}\\
  &=& -i \langle  (-k)| {d \over d(-k)} | (-k)\rangle\\
  &=&-A(-k).
  \eeqr
  In general we will have to add a  gradient $-{d\chi /dk}$ to the right hand side:  although  $\Pi$ will always reverse $k$,  the state vector it produces could have a different phase from the one initially chosen in forming the basis:
  \beq \Pi |k\rangle = e^{i \chi (k)}|-k \rangle.
  \eeq
  I will ignore this gradient ${d\chi /dk}$ since it will drop out of gauge invariant quantities. 
   
  With this caveat, we may say   $A(k)$ is an odd function of $k$. This result is true in higher   spatial dimensions where $k \to \bk$. One consequence is that  ${\cal B}$, its curl  in the 2-dimensional BZ with coordinate $\bk$ is an even function:
  \beqr
  \bA(\bk)&=& -\bA(-\bk)\label{tria2}\\
  {\cal B}(\bk)&=& \nablab_{\bk} \times \bA( \bk)=-\nablab_{\bk} \times \bA(- \bk) \nonumber \\
  &=&\nablab_{-\bk} \times \bA(- \bk)={\cal B}(-\bk).\label{trib2}
  \eeqr
  
  If at each $k$ there are many states $|k, m\rangle$  labeled by a band index $m$, we will  write in place of Eqn. \ref{pi1}
  \beq
  \Pi  |km\rangle =\sum_n U_{mn}|-k,n\rangle
  \eeq
   and you may verify that this  $k$-independent unitary operator $U$ will meet its $U^{\dag}$ and disappear in the calculation. 
   \begin{ex} Verify that $U$ drops out as claimed.
   \end{ex}

Consider now TRS. Start with Eqn. \ref{acttheta}:
\beqr
\T |k, n\rangle&=& \sum_r U_{rn}|-k, r\rangle\ \ \ \mbox{}\\
\T^2 |k, n\rangle&=& \sum_r\T U_{rn}|-k,r\rangle\ \ \\
\pm 1 |k, n\rangle &=&\sum_rU^{*}_{rn}|(\T(-k) r\rangle = \sum_rU^{\dag}_{nr}|(\T(-k) r\rangle
\eeqr
which we can write more compactly as 
\beq
|k\rangle =\pm U^{\dag}|(\T(-k) \rangle.
\eeq
(In this notation $|k\rangle$, for example,  is a  column vector whose entries are kets $|k, n\rangle$ while $\langle k|$ is a row vector whose entries are obtained by complex conjugating c-number  coefficients and turning kets  into corresponding bras.)

Now plug this into the expression for $A$, set $(\pm1)^2=1$, and proceed as we did with $\Pi$:
\beqr
A(k)&=& i \langle k|{d \over dk}|k\rangle\\
  &=& i \langle U^{\dag}\T (-k)|{d \over dk}|U^{\dag}\T (-k)\rangle\\
  &=& i \langle \T (-k)|U^{}{d \over dk}U^{\dag}|\T (-k)\rangle\\
  &=& i \langle \T (-k)|{d \over dk}\T (-k)\rangle\\
  &=& i \langle  d_k(-k)|  (-k)\rangle\\
 &=&-i \langle  (-k)| d_k (-k)\rangle\\
  &=&i \langle  (-k)| d_{-k} (-k)\rangle\\
  &=&A(-k).
  \eeqr

Thus if we have TRS
\beq A(\bk) = A(-\bk) \ \ \ \mbox{and} \  {\cal B}(\bk) = - {\cal B}(-\bk).\label{tribo}
\eeq

It follows that if both $\T$ and $\Pi$ are symmetries, 
\beq
{\cal B}(-\bk)\underbrace{=}_{\Pi}  {\cal B}(\bk)\underbrace{=}_{\T}-{\cal B}(-\bk)= 0.\label{pitheta}
\eeq

\subsection{Examples of symmetries of $H(k)$.}
{\em Example 1: Spinless Bernevig-Hughes -Zhang (SBHZ) model}\\
\beq
H(k) = \sigma_1 \sin k_x+\sigma_2 \sin k_y +\sigma_3 (\Delta -\cos k_x -\cos k_y).
\eeq
(In this context $\sigmab$ are matrices in internal space and do not  correspond to actual spin.) 
Let $\T = K$. Then 
\beqr
 \T H(k)\T^{-1} &=& KHK\\
&=& \sigma_1 \sin k_x+\sigma_{2}^{*} \sin k_y +\sigma_3 (\Delta -\cos k_x -\cos k_y)\\
&=&\sigma_1 \sin k_x-\sigma_{2}^{} \sin k_y +\sigma_3 (\Delta -\cos k_x -\cos k_y)\\
&=&\sigma_1 \sin k_x+\sigma_{2}^{} \sin (-k_y) +\sigma_3 (\Delta -\cos (-k_x) -\cos(-k_y))\nonumber\\
&\ne& H(-k).
\eeqr
It seems to be  invariant under $\Pi$  because 
\beq
(i \sigma_3)H(k)(-i \sigma_3)= H(-k).
\eeq
 
However this is just invariance under a $\pi$ rotation around the $z$-axis. In $d=2$ the effect of reflecting both coordinates through the origin coincides with a $\pi$ rotation in the plane.  So by parity one means flipping just one coordinate. This is not  a symmetry of our $H$.  In $d=3$ parity defined as flipping all three coordinates cannot be accomplished by any rotation.

Our $H$  does have an anti-unitary {\em charge conjugation symmetry ${\cal C}$}\index{charge conjugation symmetry ${\cal C}$} where 
\beqr
{\cal C}&=& i \sigma_2 K \ \ \ \mbox{and}\\
{\cal C}H(k){\cal C}^{-1}&=&-H(k).
\eeqr
This means every energy eigenstate $E(k)$ has a partner $-E(k)$. You could have guessed this given that $H(k)$ has the form $\sigmab \cdot \bh$.
\begin{ex} Check the action of ${\cal C}$. Show that levels come in equal and opposite pairs. Verify by explicit computation that   the two eigenspinors  of $H(k)= \sigmab \cdot \bh$ with opposite energies are related by ${\cal C}$. 
\end{ex}

\noindent {\em Example 2:}

Say we drop the  $\sigma_3$ term in the SBHZ model so that 
\beq
H(k) = \sigma_1 \sin k_x+\sigma_2 \sin k_y .
\eeq
Let $\T = i\sigma_2K$. Clearly $\T^2=-1$. 
Then, using $\T \sigmab \T^{-1} = - \sigmab$,  
\beqr
 \T H(k)\T^{-1} &=&  -\sigma_1 \sin k_x-\sigma_{2}^{} \sin k_y \\
&=& H(-k).
\eeqr
Thus our $H$ is  TRS.  But there is another {\em unitary} choice
\beq \T = \s_3,\ \ \ \T^2=+1.
\eeq
In Chapter \ref{six} we will find that the unitary choice places the more relevant restriction on the Hamiltonian. 

{\em Example 3: Kane-Mele model}
\beqr
H(k)&=& \sum_{a=1}^{5} \G_s d_a(k)+ \sum_{a<b=1}^{5}\G_{ab} d_{ab}(k)\ \ \mbox{where}\\
\G_a&=&(\sigma_x \otimes I, \sigma_z \otimes I, \sigma_y \otimes s_x, \sigma_y \otimes s_y, \sigma_y \otimes s_z)\\
\G_{ab}&=& {i \over 2}\lt \G_a, \G_b\rt = i \G_a\G_b\ \ \ \  (\G 's \ anticommute.)
\eeqr
Here the $s$'s are spin matrices. 
The $\G$'s obey the algebra of Dirac matrices in $5$ dimensions and $\G_{ab}$ are generators of rotations acting on the spinors. 

We  choose 
\beq
\T = i I \otimes s_y K.
\eeq
Note that $\T^2=-1$ here. 
Given this definition of $\T$,
\beqr
\T \G_a \T^{-1}&=&\G_a\\
\T \G_{ab} \T^{-1}&=-&\G_{ab}
\eeqr
\begin{ex} Verify this.
\end{ex}

Next, if the functions $d$ are real and obey 
\beqr  d_a(k)&=&  d_a(-k)\\
 d_{ab}(k) &=& - d_{ab}(-k)\ \ \ \mbox{then}\\
 \T H(k)\T^{-1}&=&  
 \sum_{a=1}^{5} \G_s d_a(k)+ \sum_{a<b=1}^{5}(-\G_{ab}) d_{ab}(k)\\
&=& \sum_{a=1}^{5} \G_s d_a(-k)+ \sum_{a<b=1}^{5}(-\G_{ab}) (-d_{ab}(-k))\\
&=&H(-k)
\eeqr
The functions $d$ may be found in the Kane-Mele paper.

\section{Models on a lattice} 
We begin with some preliminaries. 
\subsection{Fourier modes and bands in k-space} 
Consider a fermion (say an electron) that lives on the sites of a lattice labeled by an integer $n$. For now ignore spin or assume it is frozen. In terms of the lattice spacing $a$, the site labeled $n$ is at $x = na$. 

I will  set $a=1$. Thus the allowed values of position   are just $x=n$.

Suppose the lattice has only two sites and that the fermion can stay in either site with energy $E$ or hop to the other with an amplitude $-t$. The Hamiltonian is 
\beq
H= \left( \begin{array}{cc}
E &-t \\
-t & E  
\end{array} \right).
\eeq
The eigenstates and eigen-energies are
\beqr
\psi_+&=& {1 \over \sqrt{2}}\left( \begin{array}{c}1\\ 1 \end{array}\right)\ \ \ \mbox{$E_+=E-t$, bonding state}\\
\psi_-&=& {1 \over \sqrt{2}}\left( \begin{array}{c}1\\-1 \end{array}\right)\ \ \ \mbox{$E_-=E+t$, anti-bonding state.}
\eeqr
Notice that hopping has split a degenerate level at $E$ to two distinct levels $E_{\pm}$. If we now consider a chain of N atoms 
there will be N non-degenerate levels. This is the valence band. 

If there is another atomic level separated by an amount $\D >>t$,  it will form another distinct band. In real life  such bands  may cross.

The second-quantized Hamiltonian  is 
\beq 
H= -t \sum_n (c^{\dag}_{n+1}c_n+ h. c).
\eeq

We now expand 
\beq
c_n = {1 \over \sqrt{N}}\sum_k c_k e^{ikn}
\eeq
and its adjoint in terms of plane wave operators. Using  
\beq
\sum_N e^{ikn}e^{-ik'n}= N\d_{nn'},
\eeq
we find
\beqr
H &=& -t \sum_k c^{\dag}_{k}c_k (e^{-ik}+e^{ik})\\
&=& -2t \sum_k c^{\dag}_{k}c_k (\cos k).
\eeqr

If we assume the system forms a ring of $N$ sites then the allowed momenta are 
\beq
k= {2\pi m \over N}, \ \ m=1, \ldots N \mbox{or \ } m= 0, \pm 1\ldots {N \over 2}.\ \ 
\eeq
The range of energies, going from $-2t$ to $+2t$  has a width $4t$, called the bandwidth.

\begin{figure}
    \centering
\includegraphics[width=3in]{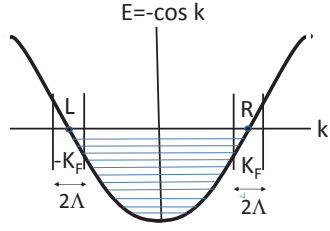}
    \vspace{-.9cm}
    \caption{The energy levels as a function of $k$. The occupied states of negative energy lie in $-K_F <k \le K_F $  where $K_F = {\pi \over  2}$  is the Fermi momentum. The figure also shows a region of low energy excitations near each Fermi point $\pm K_F$ . }
    \label{linearize}
  \end{figure}

In the ground state we want to occupy all negative energy states. These correspond to 

\beq
-K_F <k \le K_F \ \ \ \mbox{where $K_F={\pi \over 2}$}
\eeq
Since only half the states are filled, we are at {\em half-filling}. This is depicted   in Figure \ref{linearize}.

\subsection{ Peierls instability}
So far we imagined the atoms to be at fixed locations $x_n=n$. In reality the atoms can also be found in nearby locations $x_n + \phi_n$  in the course of vibrations about the stable configuration of lowest energy. The elastic energy cost for this deformation is 
\beq
V_{el} = \l \sum_n  (\phi_{n+1} - \phi_n)^2,
\eeq
 which vanishes when all $\phi$'s are equal, as it should, for this just corresponds to translating the whole lattice. 
The continuum form of $V_{el}$, dropping constants, is 
\beq
V_{el}= \int dx \left( {d\phi \over dx}\right)^2.
\eeq

Including $\pi(x)$, the momentum conjugate to $\phi$, we arrive at the {\em phonon Hamiltonian}  (dropping constants): 
\beq
H_{ph} = \int \lt \pi^2(x)+ \phi^2(x)\rt  dx
\eeq
When  quantized, the  excitations at each $k$  are called {\em phonons}. These are to lattice vibrations what photons are to electromagnetism.

The lowest energy configuration has $\phi_n= 0$ for all $n$, i.e., each atom is at its assigned place $x_n = n$. Peierls pointed out that the  system   would choose to be in a new ground state in which there is a non-zero, non-constant $\phi_n$: 
\beq
\phi_n = (-1)^nu
\eeq
or, where
\beq
x_{n+1}-x_n =1+(-1)^n u.
\eeq
In this configuration the distance between atoms would alternate between two values $1 \pm u$. Why would a system do this given that the energy cost per unit volume would go up as $u^2$? 
The answer lies in the electronic sector. In the presence of this Peierls distortion, the drop in the ground state energy of the electrons  exceeds the increase in lattice energy at small $u$. More precisely one finds the energy density for small $u$  is of the form 
\beq E(u) =
 \a  u^2 + \b u^2 \ln u 
 \eeq
 where the first part is from the lattice distortion and the second from the modified electronic energy. Due to the logarithm, which is arbitrarily  large and negative as $u \to 0$, the origin is guaranteed to be a local maximum. As we move away, $E(u)$ will fall symmetrically to two degenerate minima at opposite values of $u$  and then rise up to form a double-well potential. This spontaneous symmetry breaking is called the {\em Peierls instability.} 
\subsection{Computation of Peierls distortion energy}
Let us see how the electrons respond to the lattice distortion and lower their energy. Since the distance between atoms now oscillate between 
two values  in the Peierls state, so would the hopping amplitude oscillate: 

\beq
t= t +(-1)^n u= t+e^{i\pi n}u.
\eeq
 
(The alternating part of $t$ vanishes at $u = 0$. I assuming that it will be linear in $u$ for small $u$.) 
The Hamiltonian is now 
\beq
{H}=-t \sum_n (c^{\dag}_{n+1}c_n + h.c)  
- u \sum_n(c^{\dag}_{n+1}c_ne^{i\pi n}+ h.c) 
\eeq

Since a momentum of $\pi$  (i.e., the factor $e^{i\pi n}$) connects $k$ and $k+\pi$, we pair the operators so connected into a column vector. Since the BZ is a ring of circumference $2\pi$, any interval of width $\pi $ will do. We choose it symmetrically around $k=0$  and write in obvious notation 

\beq
{H \over N}= \int_{-\pi/2}^{\pi/2}{dk\over 2\pi} \lt c^{\dag}_{k}\ c^{\dag}_{k+\pi}\rt 
\left( \begin{array}{cc}
-t \cos k & -u \\
-u & t \cos k  
\end{array} \right)
\lt \begin{array}{c} c_k \\ c_{k+\pi}\end{array}\rt.
\eeq

Though $k$ runs over half the values as before, there are two states at each $k$ with equal and opposite energies:

\begin{figure}
    \centering
\includegraphics[width=3in]{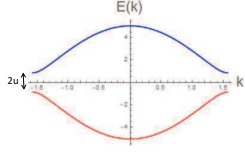}
    \vspace{
.9cm}
    \caption{The energy spectrum of fermions after the Peierls distortion. There is now a gap 2u separating the occupied and empty bands. The occupied states all get pushed down, thereby lowering the electronic energy. }
    \label{peierls}
  \end{figure} 
\beq
{E_{\pm}\over N} =\pm  \sqrt{ t^2 \cos^2 k + u^2} 
\eeq
Notice the gap $2u$ in the spectrum at $\pm {\pi \over 2}$.  At half-filling, the lower branch is totally filled and the upper branch is empty. Without the gap, the system would have been a conductor, while with the gap it would be an insulator.  By this I mean that  a DC voltage   cannot excite an electron from the filled state to an empty state, while in  a gapless ground state, there are  states arbitrarily close the Fermi energy and a DC voltage can produce a DC current. (Of course a an AC voltage of sufficiently high $\omega$ can bridge the gap.) 

Figure \ref{peierls} shows the spectrum after the distortion.  We see that all the occupied levels are pushed down as a result. This is what lowers the electronic energy. Of course to know by how much, we need to  do the integral over $k$ to find  
the ground state energy due to occupied negative energy states. This we do ignoring constants and focusing on just the $u$-dependence:
\beqr
E_g &\simeq& -\int_{0}^{{\pi \over 2}}\sqrt{ t^2 \cos^2 k + u^2} dk\\
{d E_g \over du}&\simeq& -\int_{0}^{{\pi \over 2}}{u \over \sqrt{ t^2 \cos^2 k + u^2}} dk.
\eeqr

For small $u$, the integrand diverges near $k={\pi \over 2}$. Integrating over an interval of width $\L$  near $K_F= {\pi \over 2}$ we find (upon focusing on the leading divergence), 
\beqr
{d E_g \over du}&\simeq&-\int_{0}^{\L}{u \over \sqrt{t^2q^2+u^2}}dq;\ \ \ \ q={\pi \over 2}-k\\
&\simeq &u \ln {u \over \L}+ \mbox{less singular terms, which means}\\
E_g &\simeq &u^2 \ln {u \over \L}+ \mbox{less singular terms.}
\eeqr
as advertised earlier.

\section{Su-Schrieffer-Heeger (SSH) model\label{six}}

The SSH model is a wonderful tool for explaining many features of topological  insulators: a topological index tied to the band structure, edge states, a Berry phase that affects  the dynamics, and charge fractionalization. 
\begin{figure}[h]
    \centering
\includegraphics[width=5in]{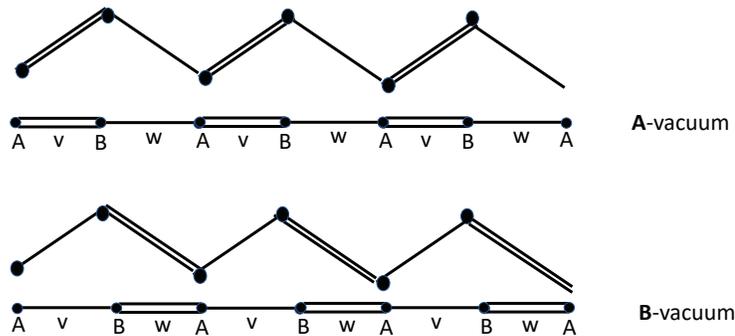}
    \caption{The Peierls distorted lattice. The inequivalent sites are labeled A and B. The degenerate symmetry-breaking vacua are also called A-vacuum and B-vacuum. The single (double) bonds denote atoms which are further (closer) and share one (two) electrons in the covalent bond.  }
    \label{sshsoliton}
  \end{figure}
Consider the following chain depicted in Figure \ref{sshsoliton}. It has undergone Peierls distortion. The inequivalent sites are labeled A and B. The degenerate symmetry-breaking solutions shown one below the other are called the A-vacuum and B-vacuum. In the A-vacuum the bond going from A to B as we move to the right is shown by a double line while the one from B to A is shown by a single line. The saw-tooth and linear depictions are completely equivalent, except in the latter the double (single) bonds are inclined upwards (downwards) in the A-vacuum and oppositely inclined in the B-vacuum. 
In the B-vacuum the pattern is displaced by one lattice unit relative to the A vacuum and corresponds to reversing $u$, the order parameter.

\begin{figure}
    \centering
\includegraphics[width=5in]{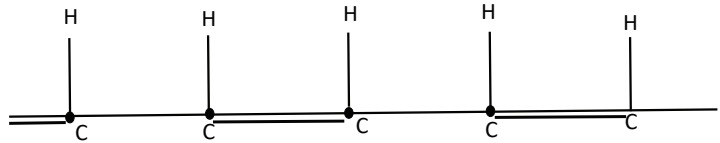}
    \vspace{-.9cm}
    \caption{Each carbon atom wants to form four covalent bonds to fill up its outer shell. One is with a H atom outside the chain, and one with each of its neighbors. These are very strong bonds due to strongly overlapping wavefunctions. They are inert and can be ignored in consideration of excitations.  The last and fourth bond is weaker and is what concerns us. It shown with the neighbor which is closer although the hybridization with the other neighbor is not negligible.  Thus the fourth bond can be with the neighbor on either side, although only  the bond with the closer neighbor is shown. In the $A$ ($B$) vacuum this happens to lie within (between) the unit cells. }
    \label{sshchem}
  \end{figure}

The single and double lines connecting the atoms signify two things.

First, the double line means the atoms are closer and the hopping is stronger. A single line means the atoms are further and the hopping is weaker. The symbols $v$ and $w$ shall denote intracell and intercell amplitudes. In the A-vacuum $v>w$ while it is the other way in the B -vacuum.  

Second, the (single) double lines denote  (single) double covalent bonds, and each bond corresponds to   two electrons in a singlet state  shared by the atoms at either end. To properly interpret the diagram, we need  to dig a little deeper into the   underlying chemistry. Look at  Figure \ref{sshchem}. Each carbon atom has 4 electrons in its outer shell and it would like to have 8 to form a full shell. To this end it reaches out to its neighbors which are also looking for an electron to share. One covalent bond is with a H atom outside the chain, and one with each of its carbon neighbors in the chain. These three bonds are very strong due to the strong overlap of wavefunctions. The bonds are frozen and  form the  inert background as far as the excitations of interest are concerned.  We can forget about them and focus on the  fourth  bond  that this chapter is all about. It is substantially weaker than the other three (due to weaker overlap of orbitals) and preferentially formed with the neighbor which  is closer. In the A  (B) vacuum this happens to be within (between) the unit cells. Although I show only the stronger bond in the figures, they are  to be viewed as an extreme caricatures of the A and B phases in which the weaker bond, smaller but  by no means negligible,  is ignored. (This is like figures that depict the ordered phase of the Ising model with all spins perfectly aligned.) 

For our purposes, we can ignore the  inert bonds,  and view  the single line as denoting a single bond with the more distant neighbor and a weaker hopping amplitude and the double line  as the stronger bond with the closer neighbor and larger hopping. {\em Thus in the following discussions, both single and double lines   represent just one bond (weak or strong) with one pair of  spin-singlet electrons.}

 Since the hopping amplitude is spin-independent and there will be no interactions in the model, we can treat each spin separately. This is what one means by the model with spinless fermions. We will not consider that option here. 

\subsection{The band Hamiltonian}

Consider the A-vacuum. The unit cell has two sites with A and B atoms as shown. We will set $a=1$, where $a$ is  the distance from one  A atom to the next A atom. The Hamiltonian is 

\beq
H=v \sum_n (A^{\dag}_{n}B_n+B^{\dag}_{n}A_n)+w \sum (A^{\dag}_{n+1}B_n+B^{\dag}_{n}A_{n+1}).
\eeq

We can choose $v$ and $w$  to be positive. If they are not, we can make them positive by the following change of operators: 
\beq
\begin{array}{ccc}
\hline 
\mbox{sign of v} & \mbox{sign of w} &\mbox{transformation} \\
\hline \\
+& +&\mbox{do nothing} \\
+&-&A_n \to  (-1)^nA_n, B_n \to  (-1)^nB_n \\
-&+& A_n \to  (-1)^nA_n,\ B_n \to (-1)^{n+1}B_n\\
-&-&A_n \to -A_n \\
\hline
\end{array}
\eeq

Upon Fourier transformation 
\beqr
H (k)&=& \int_{0}^{2\pi} {dk \over 2 \pi} \lt (v+ we^{-i k})A^{\dag}_{k}B_k + 
(v + w e^{i k})B^{\dag}_{k}A_k\rt\\
&=& \int_{0}^{2\pi} {dk \over 2 \pi}  \lt A^{\dag}_{k}, B^{\dag}_{k} \rt
\left[ \begin{array}{cc} 0&v+w e^{-i k}\\
v+w e^{ik}&0\end{array}\right] \lt \begin{array}{c}A_k \\  B_k\end{array}\rt
\\
&\equiv& \int_{0}^{2\pi} {dk \over 2\pi} \Psi^{\dag}_{k}\lt (v + w \cos k)\ \s_1 + w \sin k \ \s_2\rt \Psi_k.
\eeqr

Consider 
\beq
H (k) =(v + w \cos k) \s_1 + w \sin k \s_2\equiv \bh \cdot \sigmab.\label{hsig}
\eeq
It has charge conjugation symmetry:
\beq 
{\cal C} H {\cal C}^{-1}= -H\ \ \ \ \   {\cal C}= \s_3
\eeq
where we have choosen the unitary option with ${\cal C}^2 =+1$. 

There is also  the antiunitary option ${\cal C}= iK\s_2$, with ${\cal C}^2=-1$.

The difference between the two choices is that ${\cal C}=\s_1$ will not allow any term of the form $f(k)\s_3$, by forcing $f(k)=-f(k)$, whereas ${\cal C}= iK\s_2$ (which is just the ``spin-flip'' operator) will allow such a term.

We shall see that it is ${\cal C}=\s_1$ that is more relevant to us because it keeps $\bh$ in the $1-2$ plane, which in turn  is essential in  isolating  different topological sectors. 

This  $H$  also has TRS:
\beq
\T H(k) \T^{-1}= H(-k)\ \ \ \ \ \T =K.
\eeq
TRI does not forbid an $\s_3$ term if it is multiplied by an even function of $k$.

This $H$  also has parity or $ x \to -x$ invariance
\beq
\Pi H(k)\Pi= H(-k)\ \ \ \ \ \Pi = \s_1
\eeq
because  
 $\Pi = \s_1$ exchanges A and B atoms which is what should happen when we reflect through their midpoint.

The ground state of $H$ is  
\beqr
\Psi_{-}&=& { 1\over \sqrt{2}}\lt \begin{array}{c}1\\ -e^{i\phi}\end{array}\rt,\\
\tan \phi &=& {w \sin k\over v + w \cos k}.
\eeqr

\begin{figure}
    \centering
\includegraphics[width=3in]{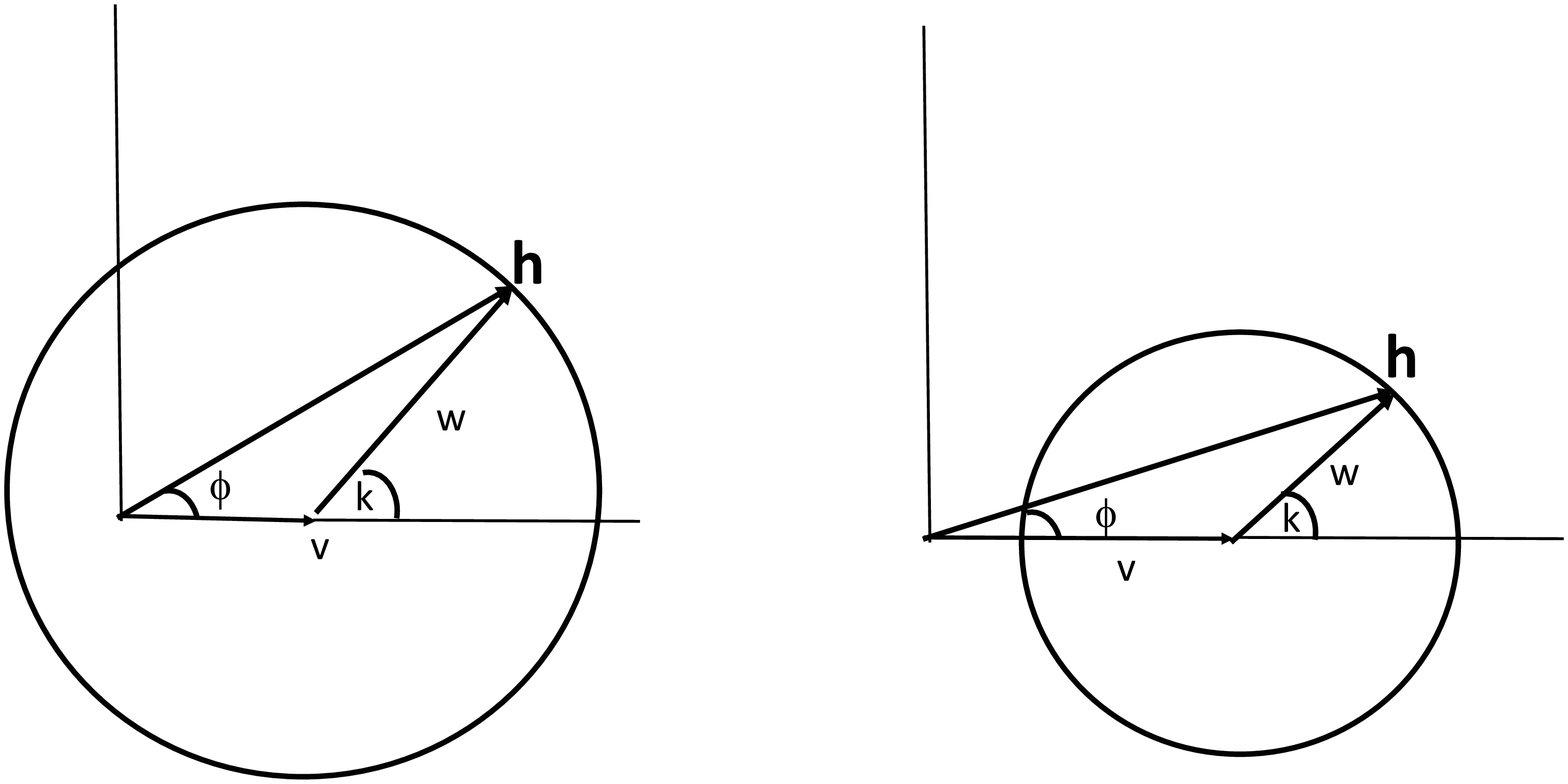}
    \vspace{-.0cm}
    \caption{The ground state spinor at each momentum k follows the 
    field $\bh$ which undergoes a $2\pi$  rotation when $w>v$ (left half) but not so when $w<v$ as in the right half. }
    \label{vw}
  \end{figure}

The ``magnetic field''  $\bh$  that couples to $\sigmab$ in Eqn. \ref{hsig}  lies in the $h_1-h_2$  plane at an angle $\phi$  with respect to the $h_1$-axis. As the momentum $k$  varies from $0$ to $2\pi$,  the tip of $\bh$  moves along a circle of radius $w$  centered at $x = v$. For 
$w>v$, the circle encloses the origin and the tip of the $\bh$  vector rotates by $2\pi$  as shown in the left half of Figure 6.3. On the other hand, as shown in the right half, when $w<v$, the tip of $\bh$  does not go around a full circle. These two possibilities are topologically distinct. 
Since $\hat{\bh}$ is the direction of the expectation value of $\sigmab$  in the ground state, these statements also apply to it. 

When the topology changes, the circle representing $k$ crosses the origin, where $\bh =0$. Here the Hamiltonian is  degenerate, and indeed vanishes entirely. This closing of the band gap is needed for any topology change, which signifies singular behavior.

The word ``topological'' is apt because whether or not $\bh$ encircles the origin is a binary question, not affected by small changes in $v$ or $w$, changes which would affect all other quantity like energies, wavefunctions etc. 

In the presence of a $\s_3$ term, this would no longer be true because the spinor can leave the $h_1-h_2$ plane and there is no real sense to the loop encircling the origin.  Thus the  topological classification relies on charge-conjugation symmetry corresponding to ${\cal C}=\s_3$.

The topology can also be described  in terms of the Berry potential
\beq
A(k) =i \left< \Psi_- |{d \Psi_-\over dk}\right>= - \half {d \phi \over dk}.
\eeq

 The gauge-invariant result is  
\beq
\oint A(k)dk = -\pi \ \ \ \mbox{if $w>v$, else $0$.}
\eeq
This is an example of the Zak phase. 
\begin{ex}
Verify by evaluating $ {d \phi \over dk}$ that $A(k)=A(-k)$, as required by TRS.
\end{ex}


\subsection{Effect of the Berry phase on dynamics }

Say we apply an electric field ${\cal E}$  along our linear system. For this purpose  let us treat the $x$ coordinate as a continuum and imagine a momentum-space wavepacket centered at $k$ slowly (adiabatically) drifting in response. (The adiabatic condition means that the system will never jump to the states in the upper positive-energy band.) The field enters the Hamiltonian   as a potential $-e{\cal E}x$ in real space or as the operator $i \hbar e {\cal E}{d \over dk}$ in 
momentum space. The Schr\"{o}dinger  equation  
\beqr
( i \hbar e {\cal E}{d \over dk}+\e (k)) \psi (k) &=&E\psi(k)\ \ \ \mbox{where}\\
\e(k)&=& -\sqrt{v^2+w^2 +2vw \cos k}
\eeqr
 can be simply integrated to give
 \beq 
\psi(k)= \psi(0) \exp \lt -{i \over \hbar e {\cal E}}\int_{0}^{k}(E - \e (k'))dk'\rt
\eeq
The single valued condition $\psi(0)=\psi (2\pi)$ or 
\beqr
{i \over \hbar e {\cal E}}\lt 2\pi E -\int_{0}^{2\pi}  \e (k'))dk'\rt&=&2\pi i  m\ \ \mbox{implies}\\
E= E_m&=& \bar{\e}+ m \hbar e {\cal E}\ \ \mbox{where }\\
\bar{\e}&=& { 1 \over 2\pi} \int_{0}^{2\pi}\e(k)dk
\eeqr
is  the average energy of the occupied band. This calculation however ignores the Berry phase of $-\pi$  that arises in the slow transport in $k$  and makes a measurable change in the spectrum. Upon including it we find   
\beq
E_m= \bar{\e}+ \left (m+\half \right)\hbar e {\cal E}.
\eeq

\subsection{Topological index $Q$ and edge states}
A salient feature of topological insulators is that gapless or zero energy states appear at the interface of the insulator and the vacuum or another insulator with a different value of the topological index, which is also called the {\em winding number}. The adjective {\em winding}  arises as follows. The Brillouin zone (BZ) is a circle parametrized by $k$ which ranges from $0$ to $2\pi$. The expectation value of $\sigmab$  in the state $
\Psi_-$  also lies on the unit circle parametrized by the angle $\phi$. Thus the the ground state spinor defines map from a circle $S_1$, the BZ, to the circle $\hat{\bh}$. Such a map is indexed by an integer which counts the number of times $\hat{\bh}$ goes around as $k$  goes around once. The winding number is 
\beq
Q=\int_{0}^{2\pi}{dk\over 2\pi} {d\phi (k)\over dk}
\eeq
which meant $ Q = -1$ when $w>v$ and $Q = 0$ when $w< v$. (One can cook up band structures when the non-zero index is +1 or some other integer.)

Suppose we chop off  the $Q\ne 0$  insulator at some point, i.e., there 
 is just the vacuum beyond this. Then $Q$ has to jump from non-zero to zero at the edge. Since $Q$ is restricted to integer values, the jump cannot take place as long as everything is analytic, i.e., as long there is a gap. Hence gapless edge states are mandatory when $Q$ jumps. In our example, instead of the vacuum we could also have a $Q = 0$ insulator i.e., with $w<$ at the interface and a gapless edge state would have to arise there as well. 
 
 Since particle hole symmetry ${\cal C}H {\cal C}^{-1}=-H$ implies that each energy $E$ is accompanied by $-E$, the edge state, if alone, must be at $E=0$. (Without particle-hole symmetry, $E=0$ does not have any significance since $E$ can be moved up and down by adding a constant to $H$.)

Let us now verify the existence of such an edge state when we have the vacuum to the left of the origin and the $w>v$ system to the right. 
It is easier to do this in first-quantization. We rewrite $H$ in Eqn. 6.1 as 
\beq
H =  \sum_{m=1}^{\infty}\left(v\lt |A_m\rangle \langle B_m|+|B_m\rangle \langle A_m| \rt+w  \lt |A_{m+1}\rangle \langle B_m| +|B_{m}\rangle \langle A_{m+1}|\rt\right)
\eeq
where $|A_m\rangle$  and $B_m\rangle$  denote particles of type A or B sitting at site $m$. Notice that there are no sites to the left of $m = 1$. Thus the $m = 1$ term in the sum multiplied by $w$ which invokes $|B_0\rangle$ is actually $0$.

Let us demand that 
\beq
|\Psi \rangle =\sum_n (a_n |A_n\rangle + b_n |B_n\rangle )
\eeq
be a zero-energy eigenket of $H$:
\beqr 
0&=& H|\Psi\rangle \\
&=& v \lt \sum_n (a_n |B_n\rangle + b_n |A_n\rangle )\rt \nonumber \\
&+& w\lt \sum_n (a_n |B_{n-1}\rangle + b_n |A_{n+1}\rangle )\rt
\eeqr

Setting the coefficients of $|A_m\rangle$  and $|B_m\rangle$  to 0, we obtain the recursion relations 
\beqr
a_{n+1}&=& - {v \over w}a_n \\
b_n&=& - {w \over v}b_{n-1}.
\eeqr

While these are the equations away from the ends, there is one equation
\beq
v b_1=- w b_0
\eeq
which, becase there is no $b_0$,  forces $b_1$ and all higher $b's$ to be zero.

  (I suggest you explicitly write out the first few terms to see that $b_1$  has to vanish and hence so must all its descendants.) 

Next,  
\beq
a_n = \left(- {v \over w}\right)^{n-1}a_1.
\eeq
This is a normalizable (exponentially falling) solution if $v<w$, i.e., the system is topologically non-trivial and has a non-zero $Q$. 

\begin{figure}
    \centering
\includegraphics[width=5in]{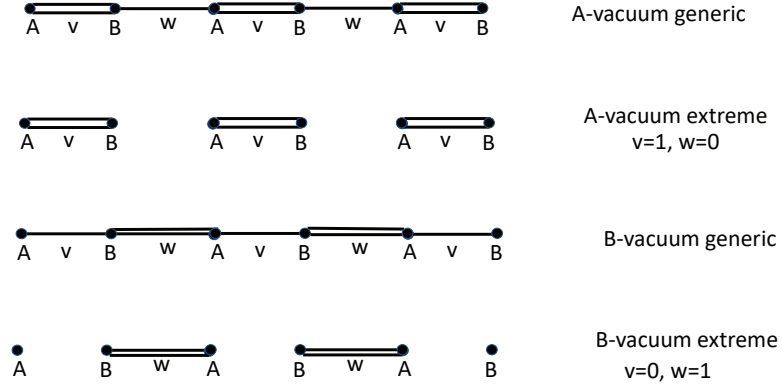}
    \caption{The first and third lines show that $A$ and $B$ vacua while the second and fourth the extreme case when only $v$ or $w$ is non-zero. In the former case all electrons are locked into bonding orbitals and there is a gap to the anti-bonding state. The latter case describes two loose electrons at the ends of the chain not bonding with any other.}
    \label{sshextremes}
  \end{figure}

We can see the distinction between the two phases if we go to the extreme ends of each phase where only $v$ or $w$ alone is   non-zero, as in Figure \ref{sshextremes}. In the former case all electrons are locked into bonding orbitals and there is a gap to the anti-bonding state. The latter case describes two loose electrons at the ends of the chain not bonding with any other. The unpaired spins can point up or down, which implies   degeneracy. 

Notice that in an infinite lattice you would not know if you were in the A or B vacuum because the atoms, also  called $A$ and $B$,  are identical. We need the edges to define the distinction: in the finite system first and last bonds  are  strong (weak) in the $A$ ($B$) vacuum.  Thus edges or boundaries are essential in fully characterizing  topological insualtors.

\subsection{Continuum theory}

When the gap is small or zero, we can approximate the lattice problem by one in the continuum, which may be more tractable mathematically. 
Let us begin with 
\beq 
H(k)=(v + w \cos k) \s_1 +\sin k \s_2. 
\eeq
The gap vanishes when $H = 0$ i.e., 
\beqr
v + w \cos k &=& 0 \\
 w \sin k &=&0. 
 \eeqr
 
The options are 
\beqr
k &=&0 \ \  v + w = 0 \ \ \ \  \mbox{not possible as both are positive} \\
k&=& \pi \ \ v=w \ \ \mbox{possible}.
\eeqr
(Look at Figure \ref{vw} which displays where the circle crosses the origin.) 

So let us go near the second point and set 
\beq
k = \pi + q, \ \ \   \cos k \simeq -1 \ \ \ \  \sin k \simeq  -q
\eeq
so that 

\beq
H =(v - w)\s_1 - w q \s_2 \equiv  m\s_1 - w q \s_2. 
\eeq

 Remember that when $m= v-w>0$, the topological index is $0$ while if $m=v-w<0$, the index is $1$. 

\begin{figure}
    \centering
\includegraphics[width=4in]{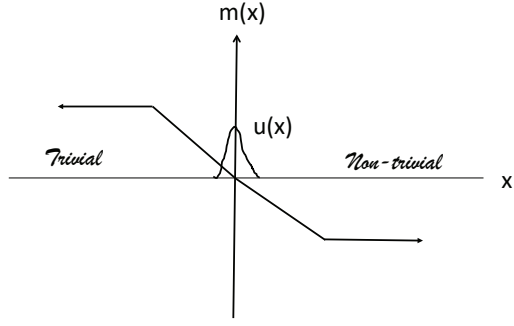}
    \vspace{-.9cm}
    \caption{Localized normalizable zero-energy edge state  separating the topologically trivial state from the non-trivial  one.}
    \label{bhzedge0}
  \end{figure}

Let us set $w = 1$ which merely affects the overall energy scale. In real space we obtain the Dirac equation 
\beq
 i \hbar \s_2 {d \psi \over dx} + m\  \s_1 \psi=E\psi, \label{sshdirac}\ \ \ m=v-1.
\eeq
 In this long-wavelength theory we have the continuum Dirac equation. We replace the sharp edge where $m$ changes sign by a smooth edge in which $m=m(x)$ 
 is a function  that slowly changes from some large positive value $m_0$ at negative $x$  to a large negative value $-m_0$  for large positive $x$, crossing zero linearly at $x = 0$, as shown in Figure \ref{bhzedge0}. Thus to the far left we have the system with $m = v - w >> 0$, the  trivial phase,  and to the far right the system with $v<< w$, the non-trivial phase. The varying $m(x)$ is a smooth interpolation between the phases. 
Multiplying Eqn. \ref{sshdirac}  by $\s_2$  we obtain,  when $E = 0$,  the equation 

\beq
i \hbar {d \psi\over dx}-im \s_3 \psi=0.
\eeq
We cleverly choose  $\s_3 \psi = + \psi$ and solve for $\psi$ by integration: 
\beq
\psi (x) = \psi(0) \exp \lt  {1 \over \hbar} \int_{0}^{x}m(x')dx'\rt.
\eeq

Notice that the $x$ integral falls off in either direction as we move away from $x=0$. (Consider $m(x')=-x'$.) Had we chosen $\s_3 \psi = - \psi$ the wavefunction would have been non-normalizable. 
This solution is a special case of the Jackiw-Rebbi prediction  that fermions always form a zero mode in the presence of a kink or soliton which interpolates between two different  vacua. The kink here is in $m(x)$. Their result in turn is a special case of what are called Index Theorems.

This solution shows a universal feature: there are only half as many states at the $d-1$ dimensional boundary of a $d$ dimensional TI as in an isolated  $d-1$ dimensional system.
Here $d=1$, and the boundary is a point which, in isolation,  could have had both eigenvalues of $\s_3$. 

The fact that the ``spin'' is restricted  $\s_3 = +$ means that the wave function has support only in one sub-lattice (A),  as was the case for our solution on the lattice which had only $a_n \ne 0$.

\subsection{Charge fractionalization}

This model provides one of the simplest examples of charge fractionalization: where the spin and charge of the electron separate. (If we worked with spinless fermions, the excitations would have half-integer charge.) 

\begin{figure}
    \centering
\includegraphics[width=5in]{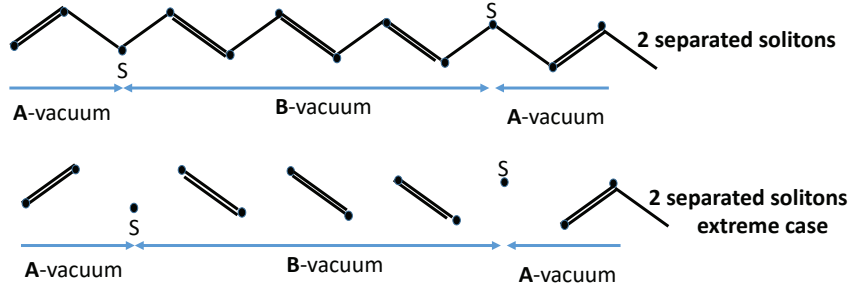}
    \vspace{-.9cm}
    \caption{At the top are a pair of solitons that lead to a change from A to B and back. At the bottom we see an extreme case when the weak bonds are set to zero. Now you can see two atoms  disconnected from everything else. The text and the Figure \ref{sshfractionpptx}  explain the associated charges. }
    \label{frac-extreme}
  \end{figure}

Recall the two Peierls states called A-and B-vacua. A soliton is a configuration that interpolates between the two. Look at Figure \ref{frac-extreme}. We see at the top a switch from A to B vacuum when we run into two adjacent single bonds. This is a soliton. A few sites later we run into another repeated single bond and soliton.The lower line shows the extreme case where weak bonds are set to zero.

\begin{figure}
    \centering
\includegraphics[width=5in]{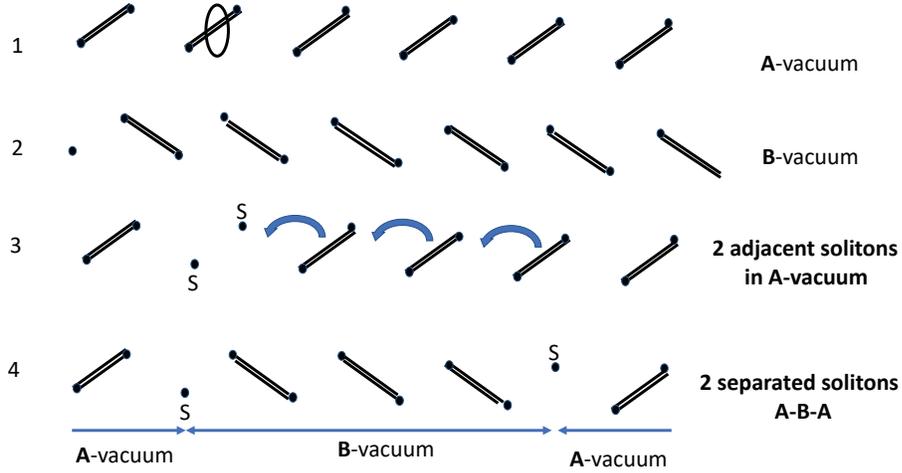}
    \caption{The first two rows show the A and B vacua in the extreme limit where weak bonds are set to zero. In the third row the bond marked by an oval in the top line is eliminated, removing the two electrons. This creates two vacancies.  The curly arrows in the third line show how bonds (and atoms) are moved to the left to reach the fourth line in which the two vacancies and the solitons are separated. The lone sites carry charge $+e$ and no spin. If we put an electron (charge $-e$) there, the soliton would have spin but no charge. These soliton quantum number correspond neither to standard electrons nor holes.  }
    \label{sshfractionpptx}
  \end{figure}

To find the charge associated with these solitionic excitations look at the extreme case shown in Figure \ref{sshfractionpptx}. Now all weak bonds have been set to zero in the two kinds of vacua in the top two lines.  We then pluck out the bond circled  by an oval from the A vacuum as shown in the top line.  The eliminated bond leaves behind  charge $2e$ because it used to contain a pair of electrons in a spin-singlet. This actually creates a double soliton and leaves behind two vacancies which  corresponds to charge 2e as shown in the third line. Now we slowly flip the bonds to the left as shown in the third line so that the double soliton is now split into two, each of charge $e$. The conversion of single to double lines is assumed to be accompanied by the corresponding change in atomic spacing from large to small. Ideally we want to separate the solitons not by just three but  a very large number of sites so they correspond to well defined excitations. 

The charge $e$ excitation   is not the usual hole because it carries no spin, since we removed a spin singlet. 
 If we add an electron at either  soliton location,  the 
soliton would now have spin but no charge. 
In either case the quantum numbers are not those of an electron or hole. This is called {\em spin-charge separation}. 

\section{Chern Bands}
We now extend many of the ideas of the SSH model to two spatial dimensions. 
\subsection{Preview}
Let us recall the highlights of the SSH  topological insulator. It  was defined by a Hamiltonian 
\beq
{ H}(k)=\bh \cdot \sigmab.
\eeq
 The ground-state spinor at each $k$ had an expectation value $\langle \sigmab \rangle= - \hat{\bh}$ which lay on the unit circle. The BZ was also a  circle with coordinate $k $ that ranged from $0$ to $2\pi$. As $k$ varied over the circular BZ, we could follow either $\bh (k)$ or $\langle \sigmab \rangle= - \hat{\bh}$ around the origin.
In the nontrivial case $\langle \sigmab \rangle= - \hat{\bh}$ wound around the origin once while in the trivial case it wound part of the way and then unwound. The topologically non-trivial system had an index $Q$ equal to the winding number alluded to above. 

Equivalently we could define a Berry vector potential 
\beq
A(k) = i \left \langle n | {dn \over dk}\right \rangle
\eeq
 where $|n\rangle$ was the ground state spinor $|\Psi_-\rangle$. Its line integral around the BZ was gauge invariant and yielded $-\pi$ in the non-trivial phase and $0$ in the trivial phase. This difference had measurable consequences.
 
 We now extend these ideas to  two spatial dimension. Here is a preview with details to follow. 
 
 At each $\bk$ there will  a basis of  wavefunctions $\Psi_{\bk\a}$ (where $\a$ is the band index) and the corresponding Bloch functions $u_{\bk\a}$:
 \beq \Psi_{\bk\a}= e^{i \bk \cdot \br} u_{\bk \a}
 \eeq
 with  the usual relation between 
the Hamiltonians $H(\bk)$ and ${\cal H}$:
 \beq
 H(\bk)= e^{-i \bk \cdot \br}{\cal H} e^{i \bk \cdot \br}.
 \eeq
The Berry potential in any occupied band will be 
 \beq
  \bA = i \langle u | \nablab u\rangle 
  \eeq
  where $|u\rangle $ is the eigenspinor at that $\bk$   and $\nablab$ is the gradient in $\bk$ -space. 
  The  curl of $\bA$ is the  {\em Berry flux or Berry curvature} \index{Berry flux or Berry curvature}
  \beq
  {\cal B}=  \nablab \times \bA.
  \eeq
  The {\em Chern number} \index{Chern number ${\cal C}$} will be defined as 
   \beq {\cal C} = { 1 \over 2 \pi}\int  {\cal B} d^2k.\eeq
  
   We will show that ${\cal C}$ is an integer for any filled band with a gap.  A band with ${\cal C}\ne 0 $  will be called a {\em Chern band.}\index{Chern band}
   
   We have shown (Eqn. \ref{tribo}) that 
   \beqr
   {\cal B}(\bk)&=& - {\cal B}(-\bk)\ \ \mbox{if there is TRS. Therefore }\\
   {\cal C} &=& { 1 \over 2 \pi}\int  {\cal B} d^2k=0 \ \ \mbox{if there is TRS}.
   \eeqr
   We can have a non-zero Chern number  only if there is no TRS. 
   
   A Chern band is assured to  have gapless edge states at the boundary with the vacuum or another band with a different ${\cal C}$ for the reason given earlier: an integer like ${\cal C}$ cannot change smoothly from one value to another value unless there is a some non-analyticity due to  gap closing. A surprising result due to Thouless, Kohmoto, Nightingale and den Nijs (TKNN)  is that a Chern band will have a Hall conductance 
   \beq\s_{xy} ={e^2 \over 2 \pi \hbar}{\cal C}.
   \eeq
   This was a surprise for people (like me) who were under the impression  that you needed to put the sample in a perpendicular magnetic field 
   to obtain a non-zero $\s_{xy}$. It turns out you just have to break TRS. A concrete example on the honeycomb lattice was provided by Haldane. 
   
   \subsection{The Kubo formula and the TKNN result}
   The rest of this chapter is dedicated to the derivation of the TKNN formula relating the Chern number to Hall conductance.

   The Hall conductance of a sample in the $x-y$ plane is  the ratio
   \beq
   \sigma_{xy}={ j_y \over E_x}
   \eeq
   of the current density and applied field. 
   
   One way to compute it is  as follows.
   \begin{enumerate}
   \item Consider a system in the distant past in its ground state, $|0\rangle$,  a filled band   wherein  every single-particle state $|\bk, \a \rangle$ is occupied. Imagine there is just one unoccupied band of single-particle states $|\bk,\b \rangle$.
   \item Couple the system to a spatially uniform  external electric field  produced by a spatially uniform, time-dependent transverse vector potential $A_x$:
   \beqr
   E_x&=& - { \p A_x\over\p t}\\
   E_x(\omega)&=& i \omega A_{x}(\omega).
   \eeqr

   (We prefer $\bA$ to a scalar potential with a gradient in order to keep the system translationally invariant. ) 
   \item 
   In the interaction picture let the initial state with $\langle j_y\rangle =0$ evolve, working to first order in the perturbation
   \beq
    H_{I}= -e j_x A_x
    \eeq
    where $e$ is the charge of the particle. (The sign of $e$ will drop out.)
    \item Compute the expectation of $\langle j_y(\omega)\rangle $ in the distant future and evaluate  
    \beqr
    \s_{xy} &=& \lim_{\omega \to 0} {\langle j_y(\omega)\rangle \over E_x(\omega)}.
    \eeqr
     \end{enumerate}
    The result of standard first-order time-dependent perturbation theory is:
     \beqr
     \s_{xy} (\omega)&=& -{e^2 \over i \omega \hbar^2}\sum_{n\ne 0}\lt {\langle 0|j_x|n\rangle\langle n|j_y|0\rangle \over E_n-E_0+\hbar \omega}+ {\langle 0|j_y|n\rangle\langle n|j_x|0\rangle \over E_n-E_0-\hbar \omega}\rt\label{715}
     \eeqr
    where $|n\rangle$ is an exact eigenstate of the unperturbed Hamiltonian.  
    
    It can be shown that the sum in Eqn. \ref{715} vanishes {\em at} $\omega =0$.
    \begin{ex}
    Prove this claim. Be ready for some serious juggling.
    \end{ex}

    Since, since   $\omega\to 0$, we expand
     \beq
     { 1 \over E_n-E_0 \pm \hbar \omega}= {1 \over E_n-E_0}\left( 1 \mp {\hbar \omega \over E_n-E_0}+\ldots\right).
     \eeq
     to arrive at 
     \beq
     \s_{xy}(0)={ie^2 \over \hbar}\sum_{n\ne 0}\lt {\langle 0|j_x|n\rangle\langle n|j_y|0\rangle \over (E_n-E_0)^2}- {\langle 0|j_y|n\rangle\langle n|j_x|0\rangle \over (E_n-E_0)^2}\rt.\label{prekubo}
     \eeq
     
     Here is one way  to approach the matrix elements of $j_x$ and $j_y$. 
     The current operator in second quantization is 
     \beq
     j_x= \int \Psi^{\dag}(\br){ (-i\hbar \nablab )_x\over m}\Psi (\br) d\br
     \eeq
     Now we expand 
     \beq
     \Psi_{} (\br) = \sum_{\a}\int {d^2k \over 4 \pi^2}e^{i \bk \cdot \br}u_{\bk\a}c_{\bk \a}\eeq
     and find, after using the orthogonality properties of $u$,
     \beqr
     j_x&=& \int {d^2k \over 4 \pi^2}\sum_{\b \a}c^{\dag}_{\bk \b}\lt { (-i\hbar \nablab +\hbar \bk)_x\over m}\rt_{\b\a}c_{\bk \a}\label{721}\\
     &=& \int {d^2k \over 4 \pi^2}\sum_{\b \a}c^{\dag}_{\bk \b}\lt { \p H(\bk) \over \p k_x}\rt_{\b \a}c_{\bk \a}\ \ \mbox{where}\\
     \lt { \p H(\bk) \over \p k_x}\rt_{\b \a}&=&\int_{\mbox{unit cell}}d\br u^{*}_{\bk \b}(\br){ (-i\hbar \nablab +\hbar \bk)_x\over m}u_{\bk \a}\label{723}
     \eeqr
      and similarly for $j_y$. 
\begin{ex} Here is a sample exercise in orthogonality in a finite system in one spatial dimension. Consider a system of length $N$ and unit cell size $a=1$ and assume
 that 
\beqr
H(x)&=&H(x+1)\\
u_{k\a}(x+1)&=& u_{k\a}(x)\\
\psi_{k\a}(x)&=& { 1 \over \sqrt{N}}e^{ikx}u_{k\a}\\
\int_{0}^{1}u_{k\a}^{*}(y)u_{k\b}(y)dy &=& \d_{\a\b}\ \ \mbox{ ($u_{k\a}$  orthonormal.)  }
\eeqr
Show that 
\beq
\int_{0}^{N}   \psi_{k'\a}^{*}(x)\psi_{k\b}(x) dx= \d_{kk'}\d_{\a\b}.
\eeq
Hint: Let $x=nx+y, 0\le y\le 1$ and break the integral over $x$ into $N$ integrals over adjacent unit intervals  $n\le x\le n+1$. First do the sum over $n=0,1,2..N-1$ to extract a $\d_{kk'}$. Then do the $y$ integral over $0\le y \le 1$. Remember $u$ is periodic.
\end{ex}
     
\begin{figure}
    \centering
\includegraphics[width=4in]{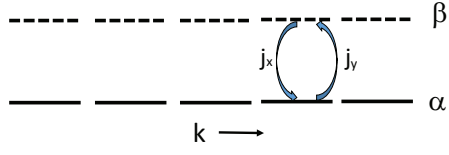}
    \vspace{-.9cm}
    \caption{The transition out of the $\a$ band to the $\b$ band at the same $\bk$ using $j_y$ and back to $\a$ using $j_x$.  }
    \label{jxjy}
  \end{figure}

     In Eqn. \ref{prekubo} $j_y$ will transform the single-particle state $|\bk \a\rangle$ to the state $|\bk \b\rangle$ with matrix element $\langle \bk \b|{\p H \over \p k_y}|\bk\a\rangle$ and then $j_x$ will similarly bring it back to $|\bk \a\rangle$, so that the many-body state returns to the filled band $|0\rangle$. Figure \ref{jxjy} describes this process. The $\bk$ label will be suppressed in the kets from now on. 
     
     Applying Eqn. \ref{721} through \ref{723} to Eqn. \ref{prekubo} we find
     \beqr
     \s_{xy}(0)&=&{ie^2 \over \hbar}\int {d^2k \over 4\pi^2}\sum_{\a \ne \b}\lt {\langle \a|\p_xH|\b \rangle\langle \b |\p_y H|\a\rangle - \langle \a|\p_yH|\b\rangle\langle \b|\p_x H|\a\rangle \over (E_{\b}-E_{\a})^2}\rt.\nonumber\\
     & &
     \eeqr
     Next we call repeatedly on relations like 
     \beq
     \langle \a|\p_xH|\b \rangle =(E_\a-E_\b)\langle \p_x\a|\b \rangle
     \eeq
     to arrive at

     \beqr
     \s_{xy}(0)&=&{ie^2 \over \hbar}\int {d^2k \over 4\pi^2}\sum_{\a \ne \b}\lt {\langle \p_x\a|\b \rangle\langle \b |\p_y\a\rangle - \langle \p_y\a|\b\rangle\langle \b|\p_x \a\rangle }\rt\nonumber \\
     &=& {ie^2 \over \hbar}\int {d^2k \over 4\pi^2}\sum_{\a }\lt {\langle \p_x\a |\p_y\a\rangle - \langle \p_y\a|\p_x \a\rangle }\rt.\label{kubolast}
   \eeqr
  where to get to the last line I have inserted the $\a=\b$ term because its contribution vanishes and then used completeness.
   
   \begin{ex} Verify that the $\a=\b$ term vanishes.
   \end{ex}

I will now show  that the Berry flux ${\cal B}$ enters the formula above:
   \beqr
   A_x&=& i \langle \bk, \a|{d \over dk_x}|\bk, \a\rangle \equiv i \langle \a|\p_x\a\rangle\\
   {\cal B}_{}&=& \p_xA_y -\p_yA_x\\
   &=& i\lt \p_x\langle \a|\p_y \a\rangle - \p_y\langle \a|\p_x \a\rangle \rt\\
   &=& i\lt {\langle \p_x\a |\p_y\a\rangle - \langle \p_y\a|\p_x \a\rangle }\rt.
   \eeqr
   Inserting this into Eqn \ref{kubolast},  finally we arrive at the celebrated TKNN result relating $ \sigma_{xy}$ to the Chern number: 
   \beqr
   \sigma_{xy}&=&{e^2 \over 2\pi \hbar}\int {d^2k \over 2\pi}{\cal B}\\
   &=& {e^2 \over 2\pi \hbar}{\cal C}
   \eeqr
   where the defintion of the Chern number 
   \beq
   {\cal C}= \int{d^2k \over 2\pi}{\cal B}
   \eeq
   has been recalled.
   
    \subsection{Quantization of ${\cal C}$ in the BZ}
\begin{figure}
    \centering
\includegraphics[width=4in]{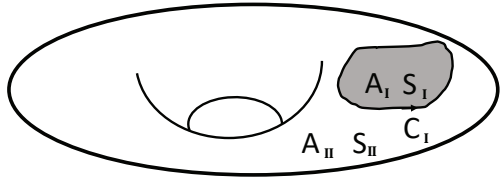}
    \vspace{-.9cm}
    \caption{Two patches $S_I$ and $S_{II}$ on a surface $S$ with boundary $C_I$ enclosing $S_I$.  The vector potentials $\bA_I$ and $\bA_{II}$ reside in the two patches and are related by a gauge transformation $\nablab \chi$. }
    \label{ch}
  \end{figure}

    We prove that ${\cal C}$ is quantized to integral values pretty much along the same lines we used to show Dirac's condition $eg= m \hbar/2$.
    Consider a closed two-dimensional surface $S$ (which represents the BZ)  as shown in Figure 
    \ref{ch}. Let $C_1$ be a loop enclosing an area $S_I$ within which is a well defined non-singular $A_I$. The complimentary region is $S_{II}$ and the potential there is $A_{II}$. (We have seen in the monopole problem that when there is non-trivial topology, we will need more than one patch.) We know that 
    \beq
     \bA_I-\bA_{II}= \nablab \chi.
     \eeq
     
     Then we follow the well known path
     \beqr
     \exp\lt i\oint_{C_1}\bA_I \cdot d \bk\rt &=& \exp\lt i\int_{S_I}{\cal B} \rt\\
     \exp\lt i\oint_{-C_1}\bA_{II} \cdot d \bk\rt &=& \exp\lt i\int_{S_{II}}{\cal B} \rt\\
     \exp\lt i\oint_{C_1}(\bA_I-\bA_{II}) \cdot d \bk\rt &=& \exp\lt i\int_{S}{\cal B} \rt\\
     \exp\lt i\oint_{C_1}\nablab \chi \cdot d \bk\rt &=& \exp\lt i\int_{S}{\cal B} \rt\\
     \exp \lt 2 \pi i m \rt&=& \exp\lt i\int_{S}{\cal B} \rt.\\
     m&=&{1 \over 2 \pi}\int {\cal B}={\cal C}.
     \eeqr

     \subsection{Hall conductance of the filled Dirac sea}
     We will establish that  
     \beq
     \sigma_{xy}= \half {m \over |m|}
     \eeq
     for the filled sea of a Dirac fermion of mass $m$.
     The procedure is once again to apply an electric field 
     \beq
     E_y(\omega)= i \omega A_y(\omega),
     \eeq
     find $\langle j^x\rangle$  and take the ratio.
      
      By defintion 
      \beqr
      \langle j^{\mu}(x)\rangle &=& {e\over Z} \int \lt D\psi D \bar{\psi}\rt \bar{\psi}\g^{\mu}\psi (x)\exp \lt{{iS \over \hbar}}\rt \ \ \mbox{where}\\
      S&=& \int \bar{\psi}(y)(i \p\!\!\! / -e A\!\!\!/ -m)\psi(y)d^3y\\
      Z&=& \int \lt D\psi D \bar{\psi}\rt\exp \lt{{iS \over \hbar}}\rt.
      \eeqr
      
     If we bring down the interaction to first order in $e$ we find
     
     \beq
     \langle j^{\mu}(x)\rangle= -{i e^2 \over \hbar} \left\langle \int d^3 y \bar{\psi}\g^{\mu}\psi (x)\bar{\psi}\g_{\nu}\psi (y) A^{\nu}(y)  \right\rangle.
     \eeq
     where the average is taken with respect  to $Z_0$, the partition function with $A=0$. Notice that 
     $j^x$ (and not $j_x=-j^x$) is the current we want and $A^y$ is the vector potential we want, which is why I have chosen $\g_{\nu}$ with the lower index.

      Let us now go to momentum space 
      \beq
      F(t,\br)= \int F(\omega , \bq) \exp 
      \lt i \omega t - i \bq \cdot \br\rt {d \omega d^2\bq \over 8\pi^3}
      \eeq
      
      and use the Feynman rules to arrive at 
      \beq
      j^{\mu}(q)={-ie^2 \over \hbar}\underbrace{(-1)}_{loop}\int {d \omega d^2\bq \over 8\pi^3}\ Tr \lt \g^{\mu}{i \over p\!\!\!/ - m + i \e}\g_{\nu}{i \over p\!\!\!/ + q\!\!\!/- m + i \e}\rt A^{\nu}(q)
      \eeq
      
We want to set $\mu =x, \nu = y$ and extract just the part of the trace  proportional to $q_0 = \omega$, setting $q=0$ everywhere else:
      
\beqr
      \mbox{``Tr''}&=&Tr \lt \g^{x}{p\!\!\!/+m  \over p^2 - m^2 + i \e}\g_{y}{p\!\!\!/ +q\!\!\!/+m \over (p+q)^2- m^2 + i \e}\rt A^{y}(q)\nonumber\\
      &\underbrace{=}_{q \to 0}&Tr { \g^x m \g_y \g_0 \omega A^y\over (p^2 - m^2 + i \e)^2}+\mbox{terms that do not matter}
      \eeqr
      
      Now we are going to choose our $\g$ matrices in a particular way.
      Suppose our single-particle Hamiltonian is 
      \beq
       H = \sigmab \cdot (\bp - e \bA)\label{sph}
       \eeq
       then we will find, upon following the usual route to the path integral, (see my book on QFT and CMT) that 
       \beq
       \g^0= \sigma_z\ \ \ \  \g^x=i\sigma_y \ \ \ \ \g_y=i\sigma_x=-\g^y.
       \eeq
       and 
       \beq
       \mbox{``Tr''}= { 2im  \omega A^y\over (p^2 - m^2 + i \e)^2}.
       \eeq
       So we need to evaluate
       \beq
       \sigma_{xy}= {-ie^2 (2m)\over \hbar}(-1)\int {d \omega d^2\bq \over 8\pi^3}{1 
       \over (\omega ^2 - |\bp|^2-m^2 + i \e)^2}.
       \eeq
       There are poles at 
       \beq
        \omega = \pm (E-i\e)\ \ \ \mbox{where $E= \sqrt{|\bp|^2+m^2}$}
        \eeq
         which lie just below (above) the real axis for $\omega$ positive (negative). 
         This allows us to do a Wick-rotation of the $\omega$-axis by$\half  \pi $  without encountering any singularities, set 
         \beq
         \omega = i p_0
         \eeq
         and integrate up the $y$-axis from $-\infty$ to $+\infty$ to arrive at the final result
         \beqr
         \sigma_{xy}&=&{e^2 \over \hbar}{2m \over 8 \pi^3}4\pi \int_{0}^{\infty}{|\bp|^2 dp \over (|\bp|^2+m^2)^2}\\
         &=&  {e^2 \over 2\pi \hbar}\cdot \half {m \over |m|}.
         \eeqr
         
        Suppose  in a filled band we have a {\em Dirac point} near which  $H$ takes the form in Eqn. \ref{sph}, and $m$ changes sign.  Then the change in Chern number (due to the Dirac point) will be 
         \beq
         \D {\cal C}= \half \D \left({m \over |m|}\right).
         \eeq
         Thus if $m$ changes from positive to negative values, $\D {\cal C}=-1$.
         
      \subsection{An alternate expression for  the Chern number}
      When there are just two bands, there is a very appealing way to express the Chern number.
      
      Let the one-body Hamiltonian at any $\bk$ be written as 
      \beq
      H = \sigmab \cdot \bh (\bk)
      \eeq
      This follows from hermiticity and the freedom to shift the energies of the two bands so that they add up to zero and $H$ becomes traceless. 
      
      {\em Theorem}:
      \beqr
      {\cal C}&=& { 1 \over 4\pi}\int \hat{\bh} \cdot (\p_x\hat{\bh}\times \p_y \hat{\bh})d^2k\\
      &=& { 1 \over 4\pi}\int \e_{abc}\hat{h}_a (\p_x\hat{h}_b\ \p_y \hat{h}_c)d^2k\\
      \p_j&=& {\p \over \p k_j}\ \ \mbox{etc.}
      \eeqr
      {\em Proof:}
      Let the occupied band be called $n$, the empty one $m$ and let $l$ run over   both. 
      Remember in the ground state
      \beq
      \langle n| \sigmab|n\rangle = - \hat{\bh}.
      \eeq
      Now,
      \beqr
      A_{x}&=& i\langle n|\p_{x}n\rangle \ \ \ \ \ \ \ A_{y}= i\langle n|\p_{y}n\rangle \\
      {\cal B}&=&i\lt \langle \p_{x}n|\p_{y}n\rangle - \langle \p_{y}n|\p_{x}n\rangle\rt\\
      &=&i\sum_{l}\lt \langle \p_{x}n|l\rangle\langle l|\p_{y}n\rangle - \langle \p_{y}n|l\rangle\langle l|\p_{x}n\rangle\rt.
      \eeqr
      The $l=n$ term vanishes (check this). So we may sum over $m\ne n$ and use old results
      like
      \beq
      \langle \p_{x}n|m\rangle = {\langle n|\p_{x}H|m\rangle \over E_n-E_m}\ \ \ \mbox{if $m\ne n$}
      \eeq
      and 
      \beq
      (E_m-E_n)^2= 4h^2,
      \eeq
      there being just one state with $m\ne n$ with a gap $2h$. 
      We now arrive at 
      \beq
      {\cal B}={i \over 4h^2}\sum_{m\ne n}\lt \langle n|\p_{x}H|m\rangle \langle m|\p_{y}H|n\rangle - \langle n|\p_{y}H|m\rangle \langle m|\p_{x}H|n\rangle \rt.
      \eeq
      Now we can put back the term with $m= n$  because its contribution vanishes. (Check this!)
      Then, using completeness,
      \beq
      {\cal B}={i \over 4h^2}\lt \langle n|\p_{x}H\p_{y}H - \p_{y}H\p_{x}H|n\rangle \rt.
      \eeq
      Now we use 
      \beq
      \p_{x}H = \sigma_{a}{\p_{x} h_a }\ \ \ \ \ \ \p_{y}H = \sigma_{b}{\p_{y} h_b }
      \eeq
      to arrive at 
      \beqr
      {\cal B}&=& {i \over 4h^2}\lt \langle n|\p_{x}h_a\p_{y}h_b(\sigma_a\sigma_b - \sigma_b\sigma_a)|n\rangle \rt\\
      &=& -{1 \over 2h^2}\p_{x}h_a \p_{y}h_b \e_{abc}\langle n|\sigma_c|n\rangle\\
      &=& {1 \over 2h^2}\hat{\bh}\cdot (\p_{x}\bh \times \p_{y}\bh) \ \ \mbox{using $\langle n|\sigma_c|n\rangle=-\hat{h}_c$. So }\label{171}\\
      {\cal C}&=& {1 \over 2 \pi}\int d^2k\ {\cal B}={ 1\over 4 \pi}\int {\hat{\bh}\cdot (\p_{x}\bh \times \p_{y}\bh)\over h^2 }\ dk_x dk_y.\label{clast}
      \eeqr
      
      We are now going to interpret Eqn. \ref{171} geometrically.
        \begin{figure}
    \centering
\includegraphics[width=5in]{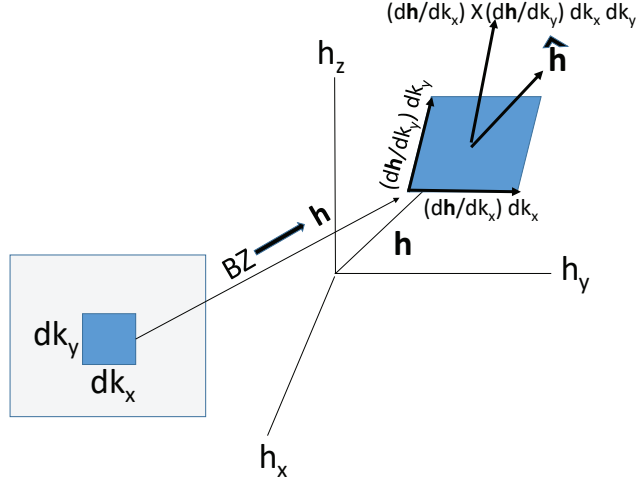}
    \caption{Under the map $BZ \to \bh$ a patch of sides $dk_x \cdot dk_y$ maps into a parallelogram of area ${\p \bh \over \p k_x}\times {\p \bh\over \p k_y} dk_x dk_y$. Projecting along $\hat{\bh}$ and dividing by $h^2$ gives $d \Omega$,the solid angle subtended at the origin of $\bh$ space.}
    \label{hspace}
  \end{figure} 
      
      Consider a patch of size $dk_x dk_y$ in the BZ as shown in Figure \ref{hspace}. Under the map $BZ \to \bh$ this patch  maps into a parallelogram of area ${\p \bh \over \p k_x}\times {\p \bh\over \p k_y} dk_x dk_y$. Projecting the area vector along $\hat{\bh}$ and dividing by $h^2$ gives 
     \beq 
     {1 \over h^2}\hat{\bh}\cdot \p_{x}\bh \times \p_{y}\bh= d \Omega
     \eeq
       the solid angle subtended by the patch at the origin of $\bh$ space. 
      Thus
      \beq
      {\cal B}\ dk_x dk_y= { 1 \over 2 } d\Omega.\label{173}
      \eeq
      If the totality of  patches representing the BZ surround the origin, 
      \beq
      {\cal C}= { 1\over 2 \pi}\int {\cal B}\ dk_x dk_y={ 1\over 4 \pi}\int d \Omega = 1.
      \eeq
      otherwise ${\cal C}=0$
      
       This same solid angle $d\Omega$ can be measured on the unit sphere if we map $BZ \to \bh /h =\hat{\bh}$. So we may rewrite Eqn. 
     \ref{clast} as an integral over the unit sphere $S^2:|\bh|=1$.
     \beqr
      {\cal C}&=& {1 \over 2 \pi}\int_{S^2} d^2k \  {\cal B}={ 1\over 4 \pi}\int \hat{\bh}\cdot \lt \p_{x}\hat{\bh} \times \p_{y}\hat{\bh}\rt \ dk_x dk_y\label{alterc}
      \eeqr
      which establishes  the theorem. 
      \begin{ex}
      Starting with $\bh = h \hat{\bh}$ show by direct computation that Eqn. \ref{clast} leads to  Eqn. \ref{alterc} above.
      \end{ex}

      These result Eqn. \ref{173}  has  another interpretation.  Imagine a monopole of strength $\half$ at the origin of $\bh$ space. It creates a radial field
      \beq
      \bB = {1 \over 2 }{\hat{\bh} \over h^2}.
      \eeq
      
      The flux intercepted by an area $d\bS$ is 
      \beq
      d\Phi = \bB \cdot d\bS={1 \over 2}{\hat{\bh}\cdot d\bS\over h^2}=\half d \Omega
      \eeq
      
      Comparing to  Eqn. \ref{173} we find that:\\
      {\em The Berry flux piercing a small area in the $BZ$ equals the monopole flux piercing its image in $\bh$ space under the map $BZ \to \bh$.}\\

\begin{figure}
    \centering
\includegraphics[width=3in]{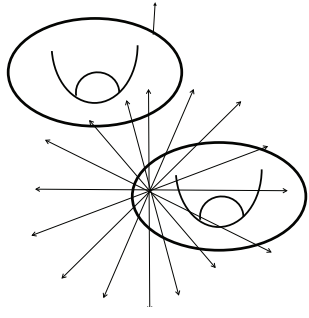}
    \vspace{.9cm}
    \caption{Two possible maps of the BZ to $\bh$ space: one which encloses the monopole  and corresponds to ${\cal C}=1$,  and one which does not, corresponding to ${\cal C}=0$. It is clear from the figure that if ${\cal C}$ is to change, the monopole has to enter or exit the BZ-torus and that will be a Dirac point.}
    \label{mono}
  \end{figure}

     I  depict this situation in Figure \ref{mono}. It shows a monopole sitting at the origin and two possible maps of the BZ to $\bh$ space. In one, the monopole is enclosed by the image of the BZ (also depicted as a torus) and ${\cal C}=1$, while in the other the monopole lies outside and ${\cal C}=0$.
     
     It is clear from the figure that if ${\cal C}$ is to change, the monopole has to enter or exit the BZ-torus. That point of entry or exit will be a point of degeneracy since $\bh =0$ at the location of the monopole.

      We may re-express these ideas  involving Berry  flux into corresponding statements about vector potentials. The Berry potential produces a phase change
      \beq
      d\chi = \bA(\bk) \cdot d \bk
      \eeq
       over an infinitesimal  line segment $d \bk$ in the BZ. Under the map $BZ \to \bh$ the line segment goes into the line segment $d \bh$ and the same phase change is produced there by a vector potential
       $\bA( \bh)$:
       \beq
        \bA(\bk) \cdot d \bk=\bA(\bh) \cdot d \bh.
        \eeq
        The curl of $\bA (\bk)$ is ${\cal B}$ and the curl of $\bA (\bh)$ is $\bB$, the monopole field.

        The line integral of $\bA(\bk) \cdot d \bk$   around a closed loop in the BZ will equal that of $\bA(\bh) \cdot d \bh$ around the image of the loop under the map. 
        (These statements can be made more precise in the language of differential forms.)
        \begin{ex}
Show that the Chern number of the  two  bands adds up to zero. Generalize to many bands. (Consider a  pair  at a time.)
        \end{ex}
        
      \section{The Spinless-Bernevig-Hughes-Zhang (SBHZ)  model.}
      The Hamiltonian is 
      \beqr
      H(\bk)&=& \sigma_x \sin k_x+\sigma_y \sin k_y +\sigma_z (\D - \cos k_x - \cos k_y)\\
      &\equiv & \bh \cdot \sigmab\\
      E&=& \pm \sqrt{\D^2 + 2 (1 - \D (\cos k_x+ \cos k_y)+\cos k_x \cos k_y)}
      \eeqr
      Here are the questions we will ask and answer:
      \begin{itemize}
      \item Where does the gap close?
      \item What is ${\cal C}$ as a function of $\D$?
      \item How do the Dirac points mediate the change in ${\cal C}$?
      \end{itemize}
      
      {\bf Where does the gap close?}
      The gap closes when $\bh =0$. This means we must kill the $\sin$ terms that multiply $\s_x$ and $\s_y$  and then handle the $\s_z$ term. There are four options:
      \beq
      \begin{tabular}{cc|c}
      $k_x$&$k_y$&$\D$ \\
      \hline\\
      0&0&2\\
      $\pi$&$\pi$&-2\\
      0&$\pi$&0\\
      $\pi$&0&0
      \end{tabular}
      \eeq
      
      {\bf What is ${\cal C}$ as a function of $\D$?}
      First consider $\D >>2$. Then $H\simeq \sigma_z \D\ \ \forall \bk$ and 
      the BZ maps into a speck way up the $h_z$ axis. It does not come close to enclosing the monopole, which sits at $\bh =0$. Indeed for any $\D >2$, we know $h_z>0$ and the 
      image torus cannot wrap around the origin. This region must have ${\cal C}=0$.
      
      For the same reason $\D<-2$ also corresponds to ${\cal C}=0$.
      
      Consider the transition at $\D =2$, where $\bk $ vanishes . Near this point let
      \beq
       \bk = \boldmath{0}+ \bp.
       \eeq
       Near the transition, and to first order in $\bp$,
       \beq
       H(\bp)= \sigma_x p_x+\sigma_y p_y +\sigma_z \underbrace{(\D -2)}_{m}.
       \eeq
       When $\D>2$ and $m=0^+0$  the Dirac point contributes (in its immediate neighborhood) 
       \beq
       {\cal C}_{DP}= \half { m\over |m|}= \half.
       \eeq
       Since the band as a whole has ${\cal C}=0$, the rest of the BZ must contribute 
      \beq
       {\cal C}_{band}= -\half { m\over |m|}= -\half.
       \eeq
       When we cross over to $\D<2$ or $m<0^-$, the Dirac contribution to ${\cal C}$ changes by $-1$ while the band contribution, being smooth stays the same. So we have 
       ${\cal C}=-1$ for $\D <2$ till we hit the Dirac point at $\D =0$.

       Before going to the next transition, I will show you that as $m \to 0$,  the Dirac point contribution 
       ${\cal C}_{DP}$ is essentially a $\d$-function in ${\cal B}$ at the origin.
       
       The lower energy eigenspinor $|n\rangle$  satisfies
       \beq
       \left( 
       \begin{tabular}{cc}
       $m+E$ & $p_x-ip_y$\\
       $p_x+ip_y$ &$m-E$
       \end{tabular}
       \right)\ \left(
       \begin{tabular}{c}u \\ v\end{tabular}\right)=0.
       \eeq
       The solution which is regular at the origin as $m\to 0^+$ is 
       \beq
       |n\rangle ={ 1 \over \sqrt{2E(E+m)}}\left( 
       \begin{tabular}{c}$-p\ e^{-i\phi} $\\ $E+m$ \end{tabular}\right).
       \eeq
       where $\phi$ is the angle in the $p_x-p_y$ plane. You can see that as even though $e^{-i\phi}$ is ill defined at the origin, it is multiplied by $p$ and so approaches $0$.The normalization factor becomes $1/(2m)$ as $p \to 0$.
       \begin{ex}
       Show that if $m \to 0^-$, you should multiply both components by $e^{i\phi}$ to get a regular solution.
       \end{ex}
       
       As for the  Berry potential,
       \beqr
       A_{\phi}d\phi &=& i\langle n|\p_{\phi}n\rangle d\phi\\
       &=& {p^2 d\phi \over 2E(E+m)}\\
       &\simeq&\half \left( 1- {m \over E}\right)d\phi  \ \ \mbox{when $m \to 0$}.
       \eeqr
       When integrated over a small circle of radius $p>>m$, when $m\to 0$,
       \beq
       \oint A_{\phi}d\phi = \pi \left( 1- {m \over p}\right)\label{diracC}
       \eeq
       (Remember $A_{\phi}$ is defined such that $A_{\phi}d\phi$ is the phase change.)
       As $m \to 0$, the entire contribution of $\pi$ comes from an arbitrarily small circle surrounding  the origin. The Berry flux density ${\cal B}$ is therefore  a $\d$-function with coefficient $\pi$.
       The Chern number is
       \beq
        {\cal C}= { 1 \over 2 \pi}\oint A_{\phi} d\phi= \half.
        \eeq
        Of course the full band has to have an integral value for ${\cal C}$ and the balance ($-\half$ in this case) comes from the non-singular contribution over the rest of the BZ.

      We can now jump to the transition at $\D = -2$ at $(\pi, \pi)$. Denoting by  $\bp$ the deviation from $(\pi, \pi)$, 
      \beq 
      H(\bp)= -\sigma_x p_x-\sigma_y p_y +\sigma_z \underbrace{(\D +2)}_{m}.
       \eeq
       We can reverse the sign of {\em both}  $p_x$ and $p_y$ (or $k_x$ and $k_y$) or $h_x$ and $h_y$ in formula Eqn. \ref{clast}   for ${\cal C}_{DP}$,  without any effect.

       When $\D<-2$ or $m<0$, the total ${\cal C}=0$ with ${\cal C}_{DP}=-\half$ and ${\cal C}_{band}=\half$. 
       When we cross over to $m>0$, ${\cal C}=1$ due to the jump in ${\cal C}_{DP}$.
       
\begin{figure}
    \centering
\includegraphics[width=4in]{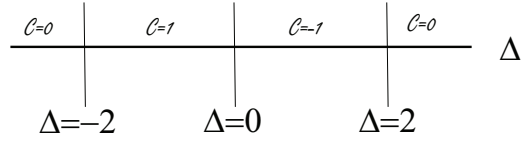}
    \vspace{-.9cm}
    \caption{Phase diagram of the SBHZ model as a function of $\D$. The change in ${\cal C}$ by $\pm 1$ is due a single Dirac point while the change by $\pm 2$ at $\D =0$ is due to two Dirac points, as explained in the text.}
    \label{bhzphases}
  \end{figure}

       The situation is depicted in Figure \ref{bhzphases}.

We finally confront the point  $\D =0$ where ${\cal C}$ changes by $2$. Consider first the Dirac point  
at $(0, \pi)$. Near it
\beq
 H = p_x \sigma_x - p_y \sigma_y + \D \sigma_z.
 \eeq
 If we flip $k_y \to -k_y$  or $p_y \to -p_y$ to bring $H$ to the standard form, we reverse the formula for ${\cal C}$:
 \beq
 {\cal C}_{DP}= - \half {m \over |m|}
 \eeq
 The same applies to the point $(\pi, 0)$. So  when $\D$ goes from positive to negative at the origin, ${\cal C}_{DP}$ goes {\em up} by $+2$. 
 
 How can the monopole hit two points of the torus at the same time? The answer is that 
 the torus is actually twisted and turned inside out before its ends are glued and at $\D=0$ two points $(0, \pi)$ and $(\pi , 0)$ touch. \begin{figure}
    \centering
\includegraphics[width=5in]{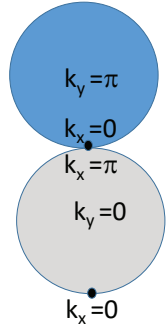}
    \caption{As we scan the BZ at $\D=0$ as a function of $k_x$ at fixed  $k_y$ , we get a slice of the torus in $\bh$ space. As we vary $k_y$ from $0$ to $\pi$ the image starts at the page (lower circle),  goes into the page and then returns at $k_y=\pi$ after twisting and turning inside out.   The points $(0, \pi)$ and $(\pi, 0)$ touch as shown. The orientations of the outward normals are opposite so that when the monopole crosses over it increases the flux entering at both points.  }
    \label{doubledirac}
  \end{figure} 
  As the monopole crosses this point, it  increases the net flux because the  orientation of the surface (for computing the flux with proper sign) is opposite at  these points,  
 as shown in Figure \ref{doubledirac}.


      \begin{ex}
      Start with 
      \beqr
      {\cal C}&=& { 1 \over 4 \pi h^3}\int d^2 k\left(
      \begin{tabular}{ccc}
      $h_x$& $h_y$&$h_z$\\
      ${\p h_x\over \p k_x}$&${\p h_y\over \p k_x}$&${\p h_z\over \p k_x}$\\
      ${\p h_x\over \p k_y}$&${\p h_y\over \p k_y}$&${\p h_z\over \p k_y}$
      \end{tabular}\right)
     \eeqr
      where 
      \beqr
      h&=&(\sin k_x, \sin k_y, \D - \cos k_x - \cos k_y)
      \eeqr
      and show that 
      \beq
      {\cal C}= { 1 \over 4 \pi}\int d^2 k {(\D \cos k_x \cos k_y - \cos k_x -\cos k_y)\over (\D^2 + 2(1 + \cos k_x \cos k_y -\D(\cos k_x + \cos k_y))^{3/2}}.
      \eeq
      Integrate this numerically and show that 
      \beqr {\cal C}&=& 0 \ \ \ |\D|>2\\
      &=& -1 \ \ \ \  0<\D <2\\
      &=&+1\ \ \  -2 < 
      \D <0.
      \eeqr
      \end{ex}

      \subsection{Edge states of the SBHZ model}
      Consider the sample (with negative $x$) terminated by an edge at $x=0$ separating the ${\cal C}=-1$ 
      region with $m= \D -2 <0$ from the ${\cal C}=0$ region with $m = \D -2 >0$.The continuum limit of the Dirac equation for energy   is 
      \beqr
      ( p_x \sigma_x + p_y \sigma_y + m(x) \sigma_z)\psi (x,y)&= &E \psi (x,y)\label{225}\\
      m(x)& \lim_{|x|\to \infty} & =\!\!\mp \!\!M\   
      \eeqr
      where $M$ is large. 
      
We may exploit translation invariance in $y$ to choose 
      \beq 
      \psi (x,y)= e^{iky}u_k(x).
      \eeq
      Then we may write 
      \beq
      ( p_x \sigma_x + k \sigma_y + m(x) \sigma_z)u (x)=E u (x)\label{226}
      \eeq
      Let us first set $k=0$ and look for a solution at $E=0$. The equation to solve is 
      \beq
      (p_x \sigma_x +  m(x) \sigma_z)u (x)=0.
      \eeq
      Multiplying by $\sigma_x$ and rearranging we find
      \beq
      {du \over dx}=-m(x) \sigma_y u.\label{zeroe}
      \eeq
      With some foresight we choose 
      \beq
      \sigma_y u= +u\label{sigy}
      \eeq
      Then Eqn. \ref{zeroe} has a solution
      \beq
      u_k(x)=\exp \lt - \int_{0}^{x}m(x')dx'
      \rt |+\rangle.
      \eeq
       You may verify that the exponential falls like a Gaussian in both  sides of the interface $x=0$. (Consider an approximation $m(x)=x$.) 
       
       Now go back to Eqn. \ref{226} and restore  the $\sigma_y k$. It simply adds to the energy an amount $k$ because of Eqn. \ref{sigy}. Thus   
       
       \beq 
       E=k.
       \eeq
       
\begin{figure}
    \centering
\includegraphics[width=3in]{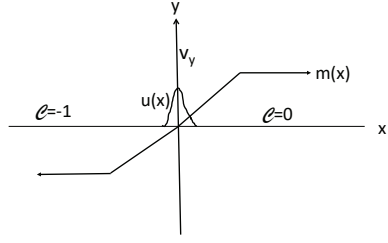}
    \vspace{-.9cm}
    \caption{Localized normalizable zero-energy edge state with indicated velocity $v_y$ separating ${\cal C}=-1$ from ${\cal C}=0$.  }
    \label{bhzedge}
  \end{figure}

        The group velocity is 
        \beq
        v_y = {d E\over dk}= +1
        \eeq
         as indicated in the Figure \ref{bhzedge}.
         
         Notice that the edge carries current only in one direction. It thus differs from a one-dimensional wire in that it supports only half the number of modes as the latter. {\em It is true of all $d$-dimensional topological insulators  that their $d-1$-dimensional boundaries   support only half as many  states as an isolated $d-1$ dimensional system. } 
         
         When we discuss  the Hall effect, we  will find that the edge current in the Figure has the opposite direction to what you in the Hall effect  at the interface of  a ${\cal C}=-1$ sample and the vacuum. The reason is that the Dirac particle is confined by a mass term $m\sigma_z $ whereas the usual Hall sample is confined by a scalar potential $V$. 
         
      
\begin{figure}
    \centering
\includegraphics[width=3in]{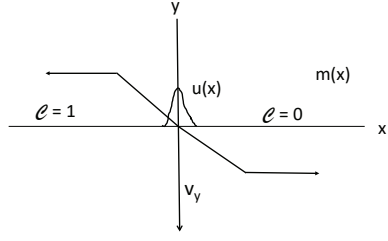}
    \vspace{-.9cm}
    \caption{Localized normalizable zero-energy edge state with indicated velocity $v_y$ separating ${\cal C}=1$ from ${\cal C}=0$.  }
    \label{bhzedge2}
  \end{figure}

      Consider the transition which occurs at $\D=-2$ when  ${\cal C}=1$ and $m=\D+2 >0$ changes to   ${\cal C}=0, m<0$ as we cross the interface in the direction of increasing $x$.  The situation is depicted in Figure \ref{bhzedge2}.
      Now we find that \beqr
      E&=&-k\\
      v&=&=-1
      \eeqr
       as shown in the figure.
       \begin{ex}
       Furnish the details leading to  the result above. Remember that near  $(\pi, \pi)$ the equation to solve is
       \beq
        ( -p_x \sigma_x - p_y \sigma_y + m(x) \sigma_z)\psi (x,y)=E \psi (x,y).
       \eeq
       \end{ex}

      Finally consider the edge states separating ${\cal C}=1$ from ${\cal C}=-1$. That is, we have glued a sample with $m<0$ to one with $m>0$ at $x=0$. The function $m(x)$ rises from negative to positive values, crossing the origin at $x=0$. The Dirac points at $(0, \pi)$ and $(\pi, 0)$ each contribute an edge state. We just need to find $E$ as a function of $k$ and determine the direction of the edge currents.
      
      The equations are 
      \beqr
      ( p_x \sigma_x - p_y \sigma_y + m(x) \sigma_z)\psi (x,y)&=&E \psi (x,y)\ \ \ \ \ (0, \pi )\\
      ( -p_x \sigma_x + p_y \sigma_y + m(x) \sigma_z)\psi (x,y)&=&E \psi (x,y)\ \ \ \ \ (\pi , 0 )
      \eeqr
      
\begin{figure}
    \centering
\includegraphics[width=3in]{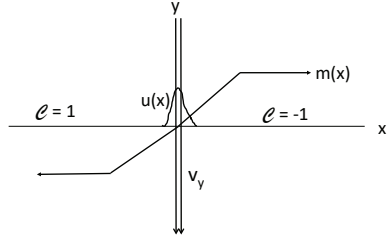}
    \vspace{-.9cm}
    \caption{Localized normalizable zero-energy edge states with indicated velocities $v_y$ separating ${\cal C}=1$ from ${\cal C}=-1$.  The edges are due to the Dirac points at $(0, \pi)$ and $(\pi , 0)$ which are responsible for  the transition at $\D=0$.}
    \label{bhzedge3}
  \end{figure}

      The profile of $m(x)$ is the same as in Figure \ref{bhzedge} and so is the normalizable solution
      which must again obey $\sigma_y u = +u$ . This is an eigenfunction also of $- p_y \sigma_y=-k\sigma_y$ withe eigenvalue $-k$. Thus, for the $(0, \pi)$ Dirac point
       \beqr
      E&=&-k\\
      v&=&-1.
      \eeqr
      Now consider the case $(\pi , 0)$. This differs from the standard form by reversal of $x$. The normalizable solution must   now have $\sigma_y =-1$. This is an eigenfunction also of $ p_y \sigma_y=k\sigma_y$ withe eigenvalue $-k$. So once again $E=-k, v =-1$. Consequently there will be two edge states running down the $y-$axis as shown in Figure \ref{bhzedge3}.

      \section{Graphene}
      Graphene is the two-dimensional version off graphite. 
      \begin{figure}[b]
    \centering
\includegraphics[width=3in]{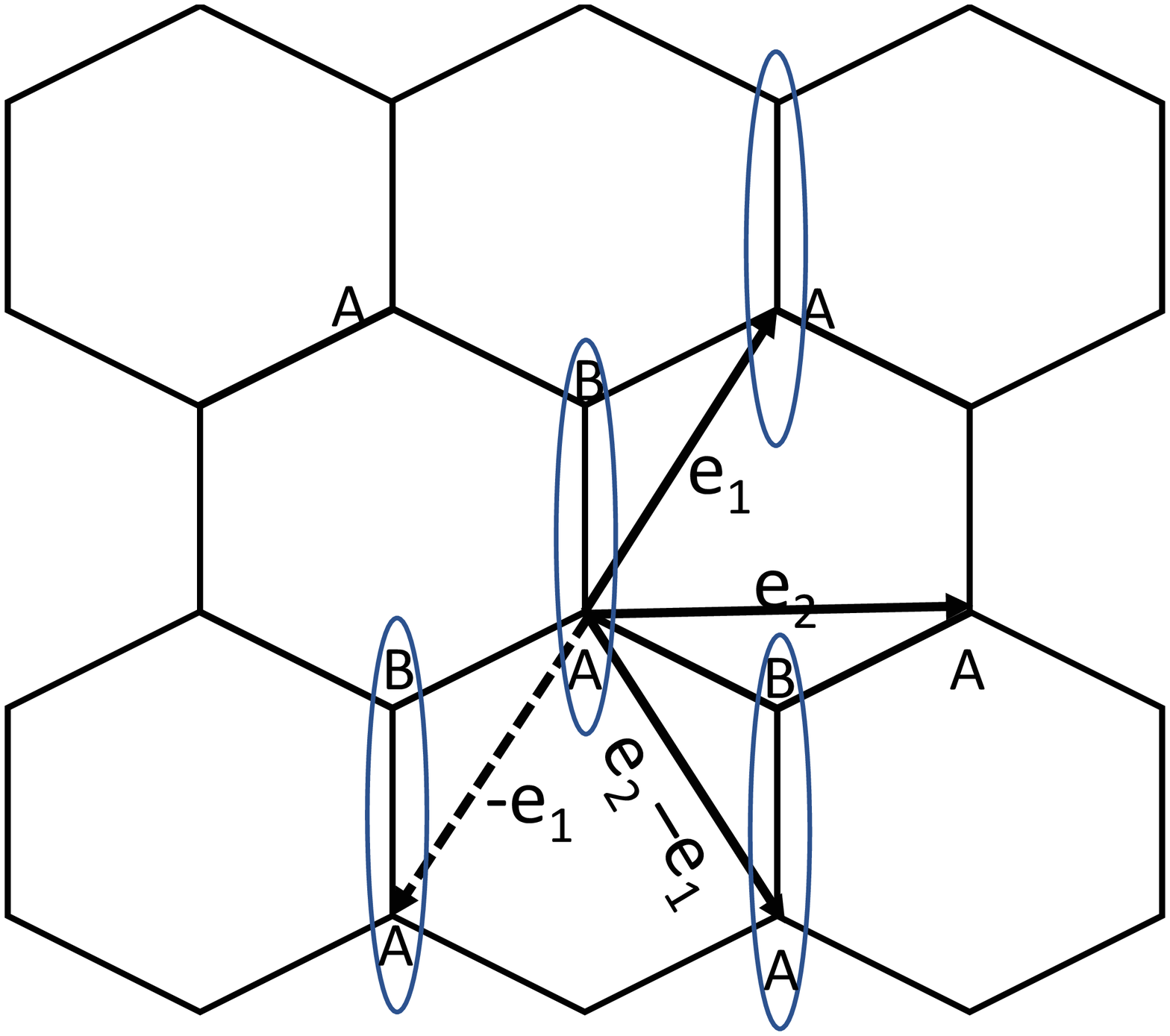}
    \vspace{-.1cm}
    \caption{The graphene lattice. The basis  vectors are $\be_1$ and $\be_2$ and their length is set to $1$. The unit cell (marked by an oval) has two atoms labeled $A$ and $B$.   }
    \label{graphene}
  \end{figure}   
      It has the lattice  structure shown in Figure \ref{graphene}.  The unit cell (enclosed  by an oval) has two atoms labeled $A$ and $B$.

      Let hopping be allowed only between nearest neighbors, i.e., members of opposite  sublattices. The Hamiltonian is 
      \beqr
      {\cal H}&=& -t \sum_{<A,B>}c^{\dag}_{B}(\br')c_A(\br)+ h.c. \label{graphH}
     \eeqr

      The coordinates $\br$ and $\br'$ refer to that of the $A$ atom in that unit cell. The $B$ atom in each unit cell is assigned the same spatial coordinate as the $A$ even though it is off by an amount in the vertical direction. That information is contained in our calling it a $B$ atom. 
      
      The basis  vectors are $\be_1$ and $\be_2$ and their length is set to unity. The $A$ atom from which these are measured in the figure will be called the central $A$ atom.
      We now perform a Fourier transform and obtain
      \beqr
      H(\bk)&=&\!\!\!  -t \sum_{\bk}\lt(c^{\dag}_{B}(\bk)c_A(\bk))(1 +e^{i \bk \cdot \be_1}+e^{i \bk \cdot (\be_1 -\be_2)})\rt \!+ \! h. c.\nonumber\\
      &&
      \eeqr
    In matrix notation
       \beqr
        H(\bk)&=&\sum_{i,j=A,B}c^{\dag}_{i}(\bk )H_{ij}c_j(\bk)\\
       H_{ij}(\bk)&=&-t \left(
       \begin{tabular}{cc}
       0&$1 +e^{-i \bk \cdot \be_1}+e^{-i \bk \cdot (\be_1 -\be_2)}$\\
       $1 +e^{i \bk \cdot \be_1}+e^{i \bk \cdot (\be_1 -\be_2)}$&0
       \end{tabular}\right)\nonumber\\
       && \label{grapham}\\
       E_{\pm}&=  &\pm |1 +e^{i \bk \cdot \be_1}+e^{i \bk \cdot (\be_1 -\be_2)}|
       \eeqr
       The matrix elements need some explanation. Consider all the jumps out of the central $A$ site (at the origin $\br =\boldmath{0}$) to its three neighbors. If they are at the cell with coordinate  $\br'$, the matrix elements in momentum space will be as follows
       \beq
       c^{\dag}_{B} c_A e^{i\bk \cdot (\br- \br')}=c^{\dag}_{B} c_A e^{i\bk \cdot (- \br')}
       \eeq
       The central atom jumps to its companion in the same cell ($\br'=0$ ) or to its neighbors in the southeast and southwest in cell located at $\br'= -\be_1+\be_2$ and $\br'= -\be_1$ respectively. The corresponding $e^{-i\bk \cdot \br'}$ appear in the second row first column. The conjugates appears in the transposed location. 
       
       From Eqn \ref{grapham} we see the problem has TRS with $\T =K$:
       \beq
        K H(k)K= H(-k).
        \eeq

We need to find the zero's of $E$ to locate the Dirac points. We want three unimodular numbers to add to zero. As one of them is $+1$, the other two must be complex conjugates and add up to $-1$. They must be at angles $2 \pi /3$ and $4\pi/3$.

Here are the relevant equations for finding $K$ and $K'$ the two Dirac points. (All others we find will differ by a reciprocal lattice vector and thus equivalent.)

      \beqr
      \be_1 &=& \left(\half, {\sqrt{3}\over 2}\right)\  \ \ \  \ \ \   \be_2= (1, {0})\\
      \bk\cdot \be_1 &=& {4 \pi \over 3}\  \ \ \  \ \ \  \bk\cdot (\be_1-\be_2) = {2 \pi \over 3}\ \ \ \mbox{K point}\\
      \bK&=& \left(  {2 \pi \over 3}, {2 \pi \over \sqrt{3}}\right) \ \ \ \mbox{K point}\\
      \bk\cdot \be_1 &=& {2 \pi \over 3}\  \ \ \  \ \ \  \bk\cdot (\be_1-\be_2) = {4 \pi \over 3}\ \ \ \mbox{K' point}\\
      \bK'&=& \left(  -{2 \pi \over 3}, {2 \pi \over \sqrt{3}}\right) \ \ \ 
      \eeqr
\begin{ex}
Provide the steps leading to the determination of $K$ and $K'$.
\end{ex}

      \subsection{The BZ}
      This is a two-step process.
      \begin{itemize}
      \item Find the reciprocal lattice vectors $\bG_1$ and $\bG_2$ such that 
      \beq
      \bG_i \be_j= 2 \pi \d_{ij}
      \eeq
      \item Find the unit cell by drawing perpendicular bisectors of reciprocal lattice vectors.
      \end{itemize}
      The reciprocal vectors are are given by 
      \beqr
      \bG_1&=& 2 \pi {\be_2 \times \hat{\bz} \over \hat{\bz} \cdot (\be_1 \times \be_2)}=\left(0, {4 \pi \over \sqrt{3}}\right)\\
      \bG_2&=& 2 \pi {\hat{\bz}\times \be_1 \over \hat{\bz} \cdot (\be_1 \times \be_2)}=\left(2\pi , -{2 \pi \over \sqrt{3}}\right)
      \eeqr
      As for the perpendicular bisectors, the one for $\bG_1$, which is purely along the $y$-direction   is just a horizontal line passing through $(0, {2\pi \over \sqrt{3}})$. 
      
\begin{figure}
    \centering
\includegraphics[width=3in]{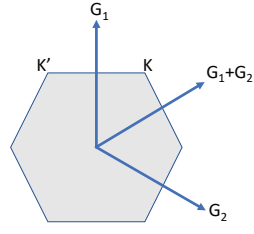}
    \vspace{-.1cm}
    \caption{The graphene Brillouin zone. Note $K$ and $K'$ are at the corners. The other corners are related by lattice vectors and physically equivalent.   }
    \label{bz}
  \end{figure}

      Next consider 
      \beq
       \bG_1+\bG_2= 2\pi\left(1, {1\over \sqrt{3}}\right)
       \eeq
       The equation of the perpendicular bisector is 
       \beq 
       y= -\sqrt{3}x + { 4 \pi \over \sqrt{3}}
       \eeq
       and it cuts the bisector of $\bG_1$ at 
       \beq
       \left( {2\pi \over 3}, {2 \pi \over \sqrt{3}}\right)= \bK.
       \eeq
       Thus one Dirac point lies at a corner of the BZ. The bisector of $\bG_2$ cuts bisector of $\bG_1+\bG_2$ at $\left( {4 \pi \over 3}, 0 \right)$.
       The rest of the BZ may be deduced by symmetry. It too is hexagonal  and the $K'$ Dirac point lies at the corner obtained  by reflecting $K$ on the $y$-axis as shown in figure \ref{bz}.
       \begin{ex}
       Derive the coordinates of the corners of the BZ.
       \end{ex}
       
      For a square lattice the BZ has area $(2\pi)^2$. The unit cell in real space has unit area (upon setting  the lattice constant $a=1$). So we have the result
      \beq
      \mbox{Area of cell in real space} \times \mbox{Area of BZ}= 4\pi^2.\label{volumes}
      \eeq
      This is also true in the graphene problem.
     \begin{ex}
     Verify Eqn. \ref{volumes}.
     \end{ex}
  \subsection{Dirac points of graphene}
  Let us consider 
  $H_{ab}$ 
  near $K$. Writing 
  \beq
  \bk = \bK + \bp
  \eeq
   and expanding to first order in $\bp$ we find
   \beqr
  {H_{ba}\over -t}&=&  1 +e^{i (\bK+\bp) \cdot \be_1}+e^{i (\bK+\bp) \cdot (\be_1 -\be_2)}\\
  &=& 1+ e^{i \bK \cdot \be_1}(1+i\bp \cdot \be_1)+e^{i (\bK\cdot (\be_1-\be_2)}(1+i\bp  \cdot (\be_1 -\be_2))\ \ \ \mbox{to order $\bp$}\nonumber \\
  &=& {\sqrt{3}\over 2}(p_x-ip_y).
 \eeqr
 \begin{ex}
Furnish the steps between the last two equations.
 \end{ex}
 This means 
 \beq 
 H(\bK+\bp) = -t {\sqrt{3} \over 2}(\sigma_x p_x- \sigma_y p_y).
 \eeq
 If we repeat the calculation at $K'$ we find
 
        \beq 
 H(\bK'+\bp) = -t {\sqrt{3} \over 2}(-\sigma_x p_x- \sigma_y p_y).
 \eeq

       The Chern densities are not defined for this gapless system. To produce a gap we need to add
       \beq
       \D H = m \sum_n(c^{\dag}_{An}c_{An} -c^{\dag}_{Bn}c_{Bn})=m\sum_n \s_z (n)
       \eeq
       which in turn will add an $m \s_z$ term to the Dirac Hamiltonian. 
       This term denotes     a chemical potential that alternates with the sub-lattice. The model still has TRS with $\T = K$. However it does not have {\em inversion symmetry}, or symmetry under reflection with respect to the horizontal line that bisects the the bond joining $A$ and $B$ sites of a unit cell. If we call this operation ${\cal I}$, with 
        
       \beqr
       {\cal I}&=& \sigma_x, \ \ \mbox{we want} \\
      {\cal I}H(k_x, k_y){\cal I}^{-1}&=& H(k_x, -k_y)
        \eeqr
        This is true true without the $\sigma_z$ term but not with it. 
        If we open a gap using a non-zero $m$, ${\cal C}$ is defined and the two Dirac points contribute oppositely to it because they have opposite coefficients of $\sigma_x p_x$.   This had to be so because of TRS of the Graphene Hamiltonian (Eqn. \ref{grapham}:
       \beq 
       \Theta H(k) \Theta^{-1}= H(-k)\ \ \ \ \ \ 
       \eeq
       where $\Theta $, is the complex conjugation operator, which I do not want to refer to as $K$ for obvious reasons.

    \section{Quantum Hall State as a Topological Insulator}
    This is the most studied  example of a TI. I will limit myself to introducing the Integer Quantum Hall problem, focusing on  how $\sigma_{xy}$ is computed. 
    The Hamiltonian is 
    \beqr 
    H&=& {P_{x}^{2} \over 2m}+{(P_y - e B_0 x)^2\over 2m}\ \ \ \mbox{where}\\
    \bA&=& B_0(0,x)\ \ \ \mbox{Landau gauge}\\
    \bB &=& \hat{\bz}B_0\eeqr
    We choose the Landau gauge because $P_y$ is conserved. Consequently
    \beqr
   \psi_k(x,y)&=& e^{iky}u(x)\\
   \lt {P_{x}^{2} \over 2m}+{(\hbar k - e B_0 x)^2\over 2m}\rt u_n(x)&=&E_n\\
  \lt {P_{x}^{2} \over 2m}+{e^2B_{0}^{2} (x - kl^2)^2\over 2m }\rt u_n(x)&=&E_n\ \ \mbox{where}\\
   l^2&=& {\hbar \over eB_0}
   \eeqr
   is the square of the {\em magnetic length}. \index{magnetic length}
   
   This is just a harmonic oscillator at the {\em cyclotron frequency} \index{cyclotron frequency}
   \beq 
   \omega_0 = \sqrt{{``k''\over ``m''}}= \sqrt{{\hbar^2 \over m^2l^4}}= {eB\over m}
   \eeq
   centered at 
   \beq
   x_0=kl^2.\label{109}
   \eeq
   In the Lowest Landau Level (LLL) which we shall focus on, we have $n=0$ and the solution is 
   \beq
   \psi = e^{iky}\phi_0(x-kl^2)
   \eeq
   where $\phi_0$ is the familiar Gaussian.

   You should visualize the wave functions as strips along $y$ of width  $\simeq l$ centered at $x_0$.
    There is  a large degeneracy because $E$ is independent of $k$. Let us compute this for an $L_x \cdot L_y$ sample where the $y$ direction is periodic i.e., $y=0$ and $y=L_y$ are joined. 
    
    The allowed momenta are
    \beq
    k_y = {2 \pi m\over L_y}\ \ \ \ m= 1, 2,    \ldots
    \eeq
    The largest allowed value for $m$ is determined by the sample width $L_y$. Since the solution is centered at $x= kl^2$ and $x\le L_x$ we demand that $M$, the largest value of $m$, satisfies 
    \beqr
    {2 \pi M\over L_y}l^2&=&    L_x\\
    M&=& {L_xL_y \over 2\pi l^2}={L_xL_y eB_0 \over 2 \pi \hbar}= {\Phi \over \Phi_0}
    \eeqr
    which states that the degeneracy of the LLL (or any LL) is the flux penetrating the sample in units of the flux quantum. 
    \subsection{Hall conductance computation}
\begin{figure}
    \centering
\includegraphics[width=3in]{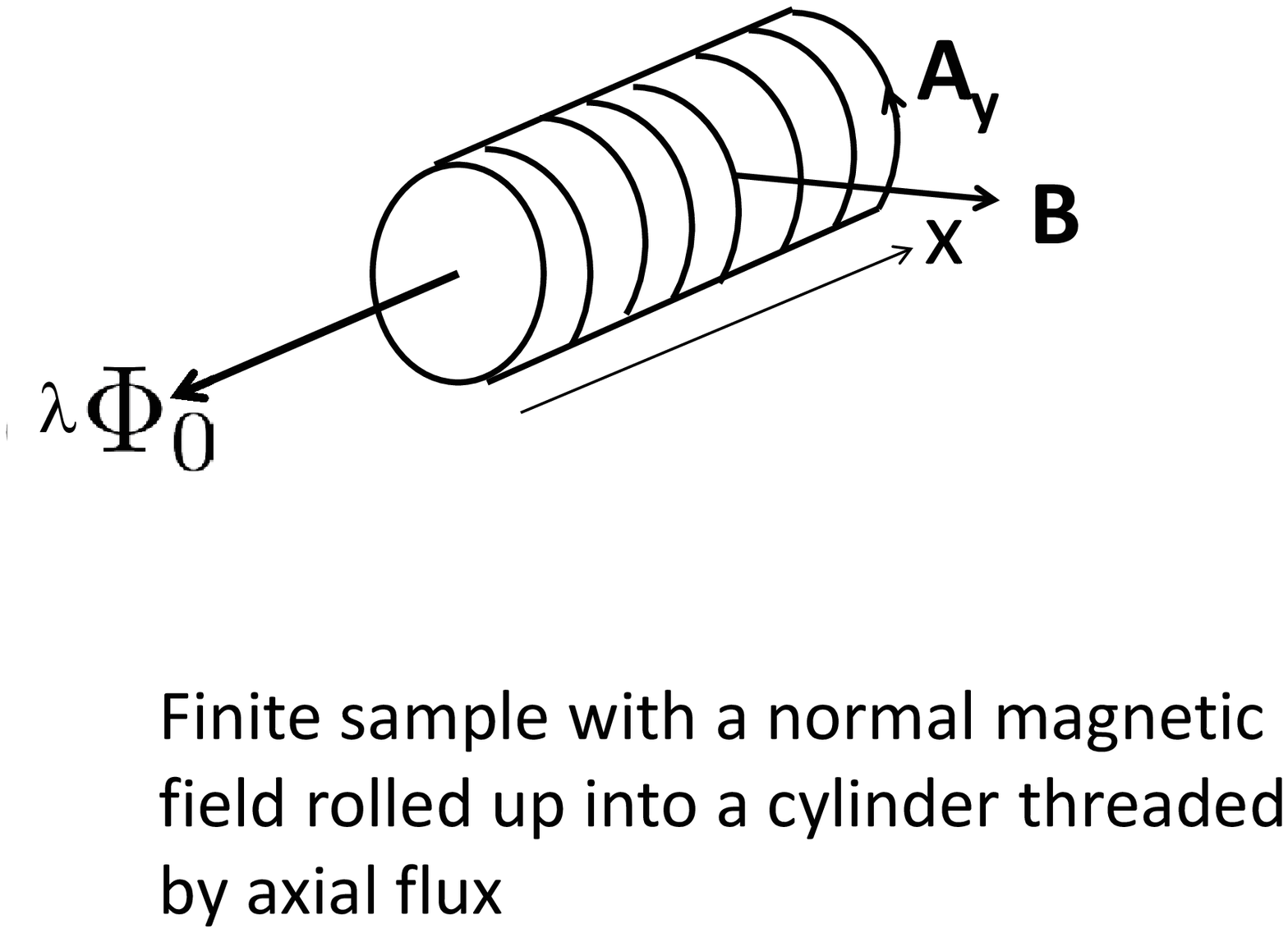}
    \caption{A finite sample rolled up into a cylinder and threaded by a flux that grows from $0$ to $\Phi_0$ adiabatically. At the end one particle of charge $e$ is transported from one end to the other.
    Relating the integral of the current to $e$ and the changing $A_y$ to an electric field along $y$
    one finds $\sigma_{xy}= {e^2 /2 \pi \hbar}$. }
    \label{pump}
  \end{figure}

    Following Laughlin we roll up the system into a cylinder of circumference $L_y$ in the $y$-direction. Now we thread some flux along the axis of the cylinder as shown in Figure \ref{pump}. 
    We choose a vector potential 
    \beq
    A_y = - {2 \pi \hbar  \over eL_y}\l
    \eeq
    where $\l$ is a parameter that will be slowly raised from $\l=0$ to $\l =1$.
    The flux along the cylinder follows from Stokes's theorem
\beq
    \Phi= \oint A_y dy = A_y L_y = -{2\pi \hbar \over e}\l = -\Phi_0 \l.
    \eeq
    At some intermediate value of $\l$ 
    \beq
     H = {P_{x}^{2}\over 2m} +{( \hbar k - eB_0 x+{2\pi \hbar \l \over L_y})\over 2m}.
     \eeq
     When $\l$ rises from $0$ to $1$, 
     \beq
     k= {2 \pi m \over L_y}\to {2 \pi (m+1) \over L_y}
     \eeq
     The particles follow the states they are in (in the adiabatic limit) and at the end of the process, the
     particles have moved over by one strip: the leftmost is empty and there is an extra particle at the right end. Thus charge $e$ has been transported from one end of the sample to the other. From this we can deduce the Hall conductance as follows. 
     When $A_y$ grows there is an electric field 
     \beq E_y = - {\p A_y \over \p t}=- {\p A_y \over \p \l}{d \l \over dt}={2\pi \hbar \over eL_y}{d \l \over dt}.
     \eeq
     If $\sigma_{xy}$ is the Hall conductance, it will produce a current density 
     \beqr
     j_x&=& \sigma_{xy}E_y \ \ \mbox{and a current} \\
     I_x&=& \sigma_{xy}E_yL_y =\sigma_{xy}{2\pi \hbar \over e}{d \l \over dt} 
     \eeqr
     Since the time-integral of $I_x$ is $e$, the charge transported across the sample, we may write
     \beqr
     e&=& \int I(x)dt =\sigma_{xy} \int_{0}^{1} {2\pi \hbar \over e}d\l= \sigma_{xy}{2\pi \hbar \over e}
     \eeqr
      which leads to 
      \beq
      \sigma_{xy}= {e^2 \over 2 \pi \hbar}.\eeq
      
      \subsection{Hall conductance: another look}
      If you take a  wire with just one conducting channel it will have the  conductance $\sigma_{xy}= {e^2 \over 2 \pi \hbar}$. How can a sheet have the same conductance? This is what we want to understand now by computing the Hall current when a voltage is applied. 
      
      Consider any one strip with the wave function peaked at $x_0(k)=kl^2$. By applying a tiny $A$ and taking the derivative at $A=0$, we find 
      \beq
      j_y= - {\p H \over \p (e A_y)}= {\hbar k -eB_0x \over m}\simeq (x-x_0(k))\ \ \mbox{see (Eqn. \ref{109}.)}
      \eeq
      The current density changes sign as we cross the strip. The oppositely moving currents from two adjacent strips cancel and we are left with just  the two uncancelled edge currents. Thus all the action (conduction)  is at the boundary in the Hall state. If there is no voltage across the sample, these edge currents will be equal and opposite. Let us  see what happens when a bias is applied. 
      
      Suppose we apply a tiny potential  difference $\D V= V_R-V_L$ between the two edges. The Landau level which used to be flat now get a little tilt. Starting with 
      \beq
      \langle v\rangle= \left\langle{d E \over \hbar dk}\right\rangle
      \eeq
      we find the current due to the wave function at each $k$:
      \beq
      I_y(k) = {e \over \hbar} \int |\psi_k(x)|^2 {d E \over dk} dx
      \eeq
      In the presence of the potential $V(x)=V(kl^2)$, ${d E \over dk}$ acquires a non-cancelling part $e {d V \over dk}$. Continuing, 
      \beq
      \psi_k(x) = {1 \over \sqrt{L_y}}e^{iky}u_k(x)
      \eeq
       so that 
       \beqr
       I_y (k)&=&{e \over\hbar L_y}{d E \over dk}\ \ \ \ \mbox{using $\int |\psi_k (x)|^2dx = {1 \over L_y}$}
       \eeqr
       assuming $dE/dk$ is constant over the narrow Gaussian wavefunction.
       
       The contributions from all $k$ values is obtained by integrating  with a measure ${L \over 2\pi} dk$:
       \beqr
       I_y&=& \int {dk \over 2 \pi}I_y(k)\\
       &=& {e \over 2\pi \hbar}{(E(R)-E(L))}={e^2 \D V \over 2\pi \hbar}.\\
       \sigma_{xy}&=& {I_y \over \D V}={e^2  \over 2\pi \hbar}.
       \eeqr

      \subsection{Chern number of the LLL}
      We know  ${\cal C}=1$ from $\sigma_{xy}$. However we cannot readily compute it because we do not have wavefunctions $u(\bk)$ in a BZ. In the Landau gauge $k_y\equiv k$ is a good quantum number but there is no $k_x$. The states are localized in $x$ near $x=kl^2$.

      The problem with finding $\bk$ states is that the translation group acts differently here because of  this fact:  {\em even if $\bB$ is uniform $\bA$ is not.} For example $A_y \propto x$ in the Landau gauge where 
        \beq
      H= {P_{x}^{2} \over 2m}+{(P_y - e B_0 x)^2\over 2m}.
      \eeq
       Whereas  translation of $a$ along $y$ implemented by 
       \beq
      T_y(a)= e^{a\p_y}
      \eeq
      is a symmetry of $H$ because it  has no $y$-dependence,  
 a translation in $x$  by  $a$   (in this gauge) has to be accompanied by a ($y$-dependent)  gauge transformation that compensates for the change in $A_y$.

      Here is a $T_x(a)$ that does the job:
      \beqr
      T_x(a)&=& \exp \lt  {a\p_x}\rt \exp \lt -{i a eB_0 y \over \hbar}\rt\\
      T_x(a)H(x,y)T^{\dag}_{x}(a)&=&H(x,y).
      \eeqr
      This means that in general  $\lt T_x, T_y\rt \ne 0$. Instead 
      \beq
      T_yT_x T^{-1}_{y}T^{-1}_{x}= \exp \lt 2 \pi i {\Phi \over \Phi_0}\rt\label{txty}
      \eeq
      where $\Phi=a^2B_0$ is the flux enclosed in the square cell of side $a$. This flux has to be an integral  multiple of $\Phi_0$ for the two translations to commute and for us to define $\bk$ as their simultaneous eigenvalue. In the simplest case this multiple is $1$. 
      \begin{ex}
      Verify that $T_x(a)H(x,y)T^{\dag}_{x}(a)=H(x,y)$.  Then fill in the steps leading to Eqn. \ref{txty}. (Hint: $e^A e^B= e^{A+B}e^{\lt A,B\rt}$ if the commutator is a $c$-number.)
      \end{ex}
      So we take the  planar sample in the continuum and divide it into unit cells, which are squares of size $a^2$ such that 
      
      \beq
      B_0 a^2= \Phi_0 \ \ \ \mbox{or}\  a^2= 2 \pi l^2.
      \eeq
      
      Consider now a rectangle  of width $a$ and height $L_y = N a$ with its left edge at $x=0$. It can be shown that 
      it contains $L_y/a$ strip states with momenta 
      \beq
       k_y = {2 \pi m \over L_y}\ \ \ \  m= 1, .. N=(L_y/a)
       \eeq
       \begin{ex} Show this.
       \end{ex}
       Pick a strip state at some $k$ (or at $x= kl^2$ ) in the chosen rectangle  and form superpositions with its counterparts at the same location in the   rectangles $j a $ away  with phase factor $\exp \lt {i k_x ja}\rt$.
      
       \begin{ex} Show  (ignoring normalization) the states obtained by the prescription above are 
       \beq
       \psi_{k_x, k_y}(x,y)= \sum_{j} \exp \lt iy (k_y +ja /l^2)\rt \exp \lt i k_x ja\rt \phi_0 (x - k_y l^2 - ja).
       \eeq
       Confirm that they respond to $T_x(a)$ and $T_y(a)$ as they should. Remember $a^2 = 2 \pi l^2$ if there is one flux quantum per unit cell. 
       \end{ex}
        Both $k_x$ and $k_y$  will lie within a BZ of sides $2\pi /a$. It is in this BZ that one must compute 
       the Berry flux and from it the Chern number. The fact that ${\cal C}\ne 0$ happens  to imply that one cannot find Bloch functions $u(\bk)$ defined in all of the $BZ$;  instead we will need at least two patches, exactly as in the monopole problem and for the same reason. At the end we will find ${\cal C}=1$. 
      
      \section{Time-reversal symmetric (TRS) models}
      So far we have considered topological insulators which violate TR. A non-zero Chern number is possible only if TR is violated. Now we consider two models which respect TR and yet are topologically distinct from trivial insulators. They also have gapless modes at the edge. They  involve spin in an essential way.
      
   \subsection{BHZ model}
   Consider the following Hamiltonian:
   \beqr
   H&=& \left( \begin{tabular}{cc}
   $h(k)$& 0\\
   0& $h^*(-k)$
   \end{tabular}
   \right)\ \ \ \mbox{where}\\
   h(k)&=& \sigma_x \sin k_x+ \sigma_y \sin   k_y +\sigma_z (\D - \cos k_x -\cos k_y).
   \eeqr
   We define $\tau$ matrices acting on the $2 \times 2$ blocks
   \beq
   \tau_y =  \left( \begin{tabular}{cc}
   0& -iI\\
   iI& 0
   \end{tabular}
   \right)
   \ \ \ \ \ \tau_z =\left( \begin{tabular}{cc}
   I& 0\\
   0& -I
   \end{tabular}\right)
   \eeq
   where $I$ is the $2 \times 2 $ identity in $\sigma$ space. 
   Let us note that 
   \beqr
   h(k)&=&\sigma_x \sin k_x+ \sigma_y \sin   k_y +\sigma_z (\D - \cos k_x -\cos k_y)\\
   h^{*}(-k)&=& -\sigma_x \sin k_x+ \sigma_y \sin   k_y +\sigma_z (\D - \cos k_x -\cos k_y)
   \eeqr
   
  This means 
   \beq
   H= ( \sigma_y \sin   k_y +\sigma_z (\D - \cos k_x -\cos k_y) \otimes I + \sigma_x \otimes \tau_z \sin k_x .
   \eeq
   This problem has charge conjugation  symmetry 
   \beqr
   C H(k)C^{-1}&=& -H(k)\ \ \ \ \mbox{where}\\
   C&=&i \sigma_y \ \cdot K
   \eeqr
   and TR symmetry
    \beqr
   \Theta  H(k)\Theta ^{-1}&=& H(-k)\ \ \ \ \mbox{where}\\
   \T&=&i \tau_y \ \cdot K.
   \eeqr
   We also have Kramers' degeneracy because 
   \beq
   \T^2=-1.
   \eeq
   
   \begin{ex}
   Verify these two symmetries.
   \end{ex}
   
\begin{figure}
    \centering
\includegraphics[width=4in]{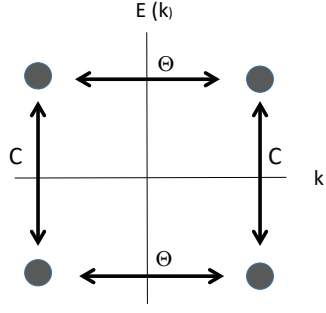}
    \caption{The effect of charge conjugation ($C$) and time-reversal $\T$ on energy eigenstates. States of opposite energy at the same $k$ are related by $C$ and states at same energy and opposite $k$ are Kramers' pairs related by $\T$.}
    \label{bhze}
  \end{figure}

   TRS implies that if 
   \beqr
   H(k)|u_k\rangle &=& E(k)|u_k\rangle \ \ \ \mbox{then}\\
   H(-k)|\T u_k\rangle &=& E(k)|\T u_k\rangle.
   \eeqr
   Thus every energy eigenket  $|u_k\rangle$ has a degenerate Kramers partner $|\T u_k\rangle$.
   The situation is summarized in Figure \ref{bhze}. 
   
     \subsection{Edge states of the BHZ model}
     
    We can look at edges of 
     \beq
   H= ( \sigma_y \sin   k_y +\sigma_z (\D - \cos k_x -\cos k_y) \otimes I + \sigma_x \otimes \tau_z \sin k_x
   \eeq
    in the sectors with $\tau_z=\pm 1$. If we are to use the continuum Dirac theory we have to be near a gapless state. Let us stay near the transition at $\D =2, \bk =(0,0)$ and $m= \D -2$.
    The system goes from a ${\cal C}=-1$ state to the vacuum with ${\cal C}=0$ as we increase $x$.

    I ask you to   verify that when $\tau_z=+1$, 
    \beqr
    |u_+(k)\rangle &=& \exp \lt -\int _{0}^{x}m(x')dx'\rt e^{iky} \chi_+ \ \ \mbox{where}\\
    \sigma_y \chi_+&=& +\chi_+\\
    E(k)&=&k
    \eeqr
   and  that when $\tau_z=-1$, 
    \beqr
    |u_-(k)\rangle &=& \exp \lt -\int _{0}^{x}m(x')dx'\rt e^{iky} \chi_- \ \ \mbox{where}\\
    \sigma_y \chi_-&=& -\chi_-\\
    E(k)&=&-k.
    \eeqr
    \begin{ex}
    Provide the proof  of the above claims.
    \end{ex}
    
    Notice that the spectrum has the features of Figure \ref{bhze}: at each $k$ there are states of opposite energy  and at each $E$ there are states of opposite $k$. 
    
    \begin{ex}
    Show by explicit computation that $\T |u_k\rangle$ with $E=k$ is the eigenstate of $H(-k)$ with same energy. 
    \end{ex}
    
 \section{Kane-Mele model}
 This is a celebrated example of a TI with TRS. The idea is to include spin in a TRS manner, say by using the spin-orbit interaction. 
 
 Let me first write down the model and then describe the origin of the various terms. 
 
 \beqr
 H&=& t \sum_{\langle ij\rangle}c_{i}^{\dag}c_j + i\l_{so}\sum_{\langle\langle ij\rangle \rangle }c_{i}^{\dag}\nu_{ij}c_j s_z +i\l_R \sum_{\langle ij\rangle}c_{i}^{\dag}(\bs\times \hat{\bd}_{ij})_z c_j\nonumber \\
 &+&\l_v \sum_i c^{\dag}_{i}c_i \sigma_z
 \eeqr
 
\begin{figure}
    \centering
\includegraphics[width=3in]{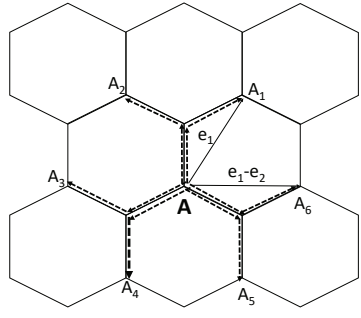}
    \vspace{.9cm}
    \caption{The spin-orbit interaction couples $s_z$ to the orbital angular momentum whose sign depends whether the particle turns clockwise or anticlockwise at the neighboring site as it goes to its second neighbor. For example the journey $\bA \to A_2$ comes with a $+$ sign.  }
    \label{kane-mele}
  \end{figure}

 Look at the honeycomb lattice in Figure \ref{kane-mele}. 
 
 The first term is the usual nearest-neighbor hopping between $A$ and $B$ sublattices with hopping amplitude $t$. 
 
 The second describes {\em spin orbit coupling}\index{spin-orbit coupling} with strength $\l_{so}$. Here the particle goes from the central site shown by boldface $\bA$ to the $6$ second-neighbor $A$ sites numbered $A_1 \ldots A_6$. The orbital angular momentum (along the $z$-axis which is normal to the plane of the lattice) has a sign  depending whether the particle swings clockwise or anticlockwise. {\em Moving with the particle}   we assign a $+$ sign to a left turn and a $-$ sign to a right turn. For example the journey $\bA \to A_2$ comes with a $+$ sign. The physical spin $s_z$ couples to this angular momentum $L_z$. 
 
 The {\em Rashba coupling $\l_R$}\index{Rashba coupling $\l_R$} describes the interaction of the spin with the surface electric field $\bE$ normal to the plane.
 A particle with velocity $\bv$ will see a magnetic field $\bB=\bv \times \bE$ in its rest frame and couple with a Zeeman term 
 \beq
 \bs \cdot \bB= \bs \cdot (\bv \times \bE) =\bE \cdot (\bs \times \bv)\propto \hat{\bz}\cdot (\bs \times \bv).
 \eeq
 We represent the particle velocity $\bv$ with $\hat{\bd}_{ij}$ which is  unit vector separating the nearest neighbors. 
 The coupling $\l_v$ is an alternating on-site potential with opposite signs on the two sub-lattices.

 Here is the final answer with details  left as an exercise.
 \beqr
 H(k)&=& t (\sigma_x (1 + 2 \cos x \cos y)+ 2 \sigma_y \sin y \cos x )\nonumber\\
 &+& 2 \l_{so}(\sin 2x - 2 \sin x \cos y)\sigma_z s_z + \l_v \sigma_z\nonumber \\
 &+& \l_R(\sigma_x s_x \cos x \sin y -\sqrt{3} \sigma_x s_y \sin x \cos y +\sigma_y s_x(1 - \cos x \cos y)- \sqrt{3}\sigma_y s_y \sin x\sin y).\nonumber \\
 & &\label{kmham}
 \eeqr
 In the above
 \beq
 x= \half k_x \ \ \ \ \ y={\sqrt{3}\over 2}k_y.\label{124}
 \eeq
 
 \begin{ex}
 Derive Eqn. \ref{kmham}.
 \end{ex} 
  All these terms are invariant under TR. We expect this because spin and velocity change sign under TR and their product is invariant. You may  verify that 
 \beq
 \T = i s_y K
 \eeq
 does the job. 
 \begin{ex}
 Verify that 
 \beq
 \T H(k)\T^{-1}= H(-k).
 \eeq
 \end{ex}
 
 \subsection{Dirac points of the KM model}
 We are going to handle this in the limited case of $\l_R=0$ which is subject to analysis very similar to what we have encountered. In this case 
 \beqr
 H(k)&=& t (\sigma_x (1 + 2 \cos x \cos y)+ 2 \sigma_y \sin y \cos x\nonumber\\
 &+& 2 \l_{so}(\sin 2x - 2 \sin x \cos y)\sigma_z s_z + \l_v \sigma_z
 \eeqr
 We are free to work with $s_z=+1$ since it is diagonal. We will also set $t=1$.

 To get $E=0$ we need 
 \beqr
 0&=& 1 + 2 \cos x \cos y\\
 0&=& \cos x \sin y\\
 0&=& 2 \l_{so}(\sin 2x - 2 \sin x \cos y) + \l_v 
 \eeqr
 We can kill the middle term in two ways. If we kill $\cos x$, the first condition cannot be satisfied by any $y$. 
 So we kill the $\sin y$ :
 \beq
 y=0, \pi.
 \eeq
 Here is the first  option.
 \beqr
 y&=&0\\
 1 + 2 \cos x&=&0 \ \ \to x={2\pi \over 3}\label{kmd1}\\
 2\l_{so}(\sin 2x - 2 \sin x \cos y) + \l_v &=& 0 \ \ \to {\l_v\over \l_{so}}=3 \sqrt{3}\label{kmd2}
 \eeqr
 We could also choose $(x={4 \pi \over 3}, y=0)$, but the results  will coincide with the option considered below.
 
 \beqr
 y&=&\pi\\
 1 - 2 \cos x&=&0 \ \ \to x={\pi \over 3}\label{kmd1}\\
 2\l_{so}(\sin 2x - 2 \sin x \cos y )+ \l_v &=& 0 \ \ \to {\l_v\over \l_{so}}=-3 \sqrt{3}\label{kmd2}
 \eeqr
 These two Dirac points ${\l_v\over \l_{so}}=\mp 3 \sqrt{3}$ are the boundaries of the non-trivial phase.
 
 Remember we had chosen $s_z=+1$. If we repeat with $s_z=-1$ we will find the Dirac points get exchanged. 
 
 If we turn on $\l_R$, there will be  a two-dimensional
 region of the non-trivial phase in the $\l_R/\l_{so}$ versus $\l_v/\l_{so}$ plane bounded by  ${\l_v\over \l_{so}}=\mp 3 \sqrt{3}$ when $\l_R=0$.

 \subsection{Edge states of the KM model when $\l_R=0$}
 Start with 
 \beqr
  H&=&(\sigma_x (1 + 2 \cos x \cos y)+ 2 \sigma_y \sin y \cos x\nonumber\\
 &+& 2 \l_{so}(\sin 2x - 2 \sin x \cos y)\sigma_z s_z + \l_v \sigma_z
 \eeqr
 and choose $s_z=+1$. Near the Dirac point 
 \beq
 x= {2 \pi \over 3}+{p_x\over 2} \ \ \ \  y=0+{\sqrt{3}\over 2}p_y.
 \eeq
 The factors of $\half$ and ${\sqrt{3}\over 2}$ appear in $p_x$ and $p_y$ because of the definitions in Eqn. \ref{124}:
 \beqr
 x&=& \half k_x,   \ \ \    y=   {\sqrt{3}\over 2}k_y\\
  \d x&=&\half  \d k_x=\half  p_x,   \ \ \    \d y=   {\sqrt{3}\over 2}\d k_y={\sqrt{3}\over 2}p_y
 \eeqr
  We have to first order in $\bp$,
 \beq
 H=  \lt {\sqrt{3}\over 2}(\sigma_x p_x+ \sigma_y p_y) + \sigma_z (\l_v - 3 \sqrt{3}\l_{so})\rt.
 \eeq
 
 \begin{ex}
Derive the $H$ above.
 \end{ex}
 Let is introduce an $x$-dependent mass term
 \beq
 m(x)= {2 \over \sqrt{3}}\lt\l_v(x)-3 \sqrt{3}\l_{so}\rt
 \eeq
 which goes from very negative values (nontrivial insulator) to very positive (trivial insulator) and changes sign at $x=0$. 
 
 If we let $p_y=0$ in $\psi (x)= e^{iky}u(x)$, we find by quadrature
 \beq
 u(x)= u(0) \exp\lt - \int_{0}^{x}  m(x')dx'\rt |+\rangle \ \ \ \mbox{where}  \ \s_y|+\rangle =  |+\rangle.
 \eeq
  This solution is normalizable because we chose  $\sigma_y u=u$. Putting back the $\sigma_y p_y$ term  we find,
  \beq
   E= {\sqrt{3}\over 2} p_y\ \ \ \ \ \ v={\sqrt{3}\over 2}.
   \eeq
   Thus this edge has $s_z=+1$ and runs up the interface at $x=0$.

   The solution with $s_z=-1$  leads to 
    \beq
 H=\lt {\sqrt{3}\over 2}(-\sigma_x p_x+ \sigma_y p_y) + \sigma_z (\l_v - 3 \sqrt{3}\l_{so})\rt..
 \eeq
 and the edge state which  runs down the interface is the time-reversed partner of the $s_z=+1$ state. 
   \begin{ex}
   Verify both solutions at $s_z =\pm 1$.
   \end{ex}

 \subsection{$Z_2$ nature of edge states}
 Let the edge run long $y$ so that $k_y =k$ is a good quantum number. Let us consider a hybrid version in which $x$ remains $x$ and $y$ gets Fourier transformed to $k_y=k$. Since $x$ is invariant umder time-reversal, 
 we have 
 \beq \T H(x, k)\T^{-1}=H(x,-k).
 \eeq

\begin{figure}
    \centering
\includegraphics[width=5in]{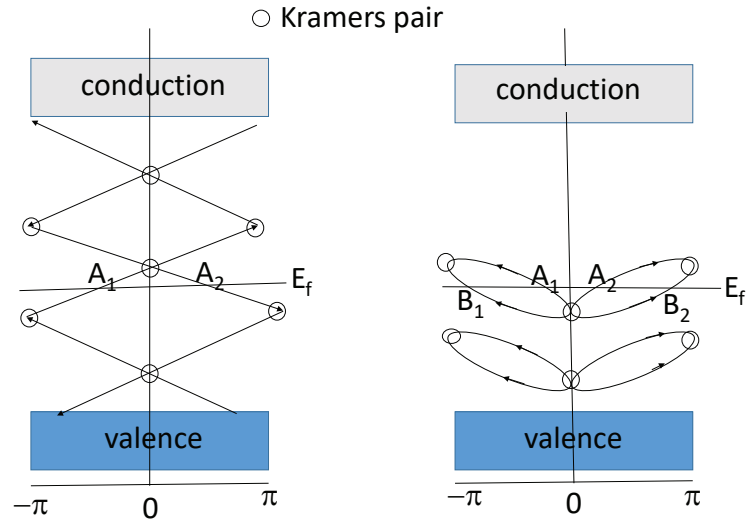}
    \vspace{-.0cm}
    \caption{At $k=0, \pi$, we have  $H(-k)=H(k)$,  so that the levels must come in degenerate pairs. The trivial (right) and non-trivial (left) cases are differentiated by how the lines join up as they connect the TRS points. In the non-trivial case one cannot make a slice at any energy without encountering an {\em odd} number ($=1$  in the figure) of Kramers pairs  in the band gap. These are protected by TRS from gapping out upon scattering from any TRS impurity. In the trivial case a slice will meet an even number ($=0$ or $2$) of Kramers pairs which could pair with the oppositely moving member from the another pair and gap out.}
    \label{kramers}
  \end{figure}

 Consider the points $k_y = 0 , \pi$. Since these  are their own negatives, 
 
 \beq \T H(k)\T^{-1}=H(k).
 \eeq
 So $\T$ is a symmetry and we must have Kramers' pairs. Consider the edge states as a function of $k$ as we go from $-\pi$ to $\pi$ as shown in Figure \ref{kramers}. Each of these TRS points must have a doublet that splits up as we move away. How do these lines join with their TR counterparts?

 I show in Figure \ref{kramers} two simple cases that illustrate the choices.

 In the option on the left  we encounter just one Kramers pair ($A_1$ and $A_2$) and they cannot mix due to TRS and they will remain gapless. Think of a $2 \times 2$ matrix with equal diagonal matrix element  to which we add an off-diagonal term from scattering. This opens up a gap unless symmetry forbids this element, which is the case here.  We cannot find an energy where there is no edge states. This is the non-trivial insulator.
 
 Why can't $A_1=|u_1\rangle$ scatter off its Kramers partner $A_2=|u_2\rangle$? 
 Let $V=\T^{-1}V\T$ be the TRS  scattering amplitude. Then  
 \beqr
 \langle u|V|\T u\rangle&=& \langle V  u|\T u\rangle\\
 &=& \langle \T^2 u|\T Vu\rangle \ \ \ \ \mbox{using $\langle \phi|\psi\rangle=\langle \T \psi|\Theta \phi \rangle$}\\
 &=& (-1)\langle u|\T Vu\rangle\\
 &=& (-1)\langle u|V |\T u\rangle.
 \eeqr
 
 In other option  option, on the right half,   a slice at some energy  encounters no mid-gap states or two Kramers pairs. Suppose there are two pairs. State $A_1$ can scatter off $B_2$ from the {\em other pair} and a gap can open up.  So again there can be energies at which we encounter no states. This is in the same family as the trivial insulator which has no states in the gap.

 The number of Kramers states {\em modulo $2$} is the topological invariant. The sector with even numbers is the trivial one since a band insulator, which  has no states in the gap is trivial. 
 
 \newpage
\centerline{ References}
\noindent R. Jackiw and C. Rebbi, Phys. Rev. D 13, 3398,  (1976). \\
R. B. Laughlin, Phys. Rev. B 23, R5632 (1981).\\
D. J. Thouless, M. Kohmoto, M. P. Nightingale, and
M. den Nijs, Phys. Rev. Lett. 49, 405 (1982).\\
 F. D. M. Haldane, Phys. Rev. Lett. 61, 2015 (1988).\\
 C. L. Kane and E.J Mele PRL 95, 226801 (2005)\\
J. Zak,  Phys. Rev. Lett. 62(23) 2747, (1989).\\
Asbóth, János K. - Oroszlány, László - Pályi, András, A Short Course on Topological Insulators - Band-Structure and Edge States in One and Two Dimensions, 
Springer International Publishing, (2016).\\
B. A. Bernevig, T. L. Hughes, and S. C. Zhang, Science,314, 1757, (2006).\\
B. A. Bernevig, Toplogical Insulators and Topological Superconductors, Princeton Press, (2013).\\
R. Shankar, Quantum Field Theory and Condensed Matter, Cambridge Press, (2017).

\end{document}